\documentclass[prb,twocolumn,amsmath,a4paper]{revtex4-1}
\pdfoutput=1

\usepackage{graphicx}
\usepackage{xcolor} 
\usepackage{dcolumn}
\usepackage{bm}
\usepackage{braket}

\usepackage{sidecap}
\usepackage{natbib}
\usepackage{float}

\usepackage{amssymb}
\usepackage{wrapfig}

\newcommand{\ua}{\uparrow}
\newcommand{\da}{\downarrow}
\newcommand{\la}{\lambda}
\newcommand{\al}{\alpha}

\providecommand{\e}{\varepsilon}
\providecommand{\h}{\hbar}
\providecommand{\p}{\partial}

\usepackage{lipsum}

\makeatletter
\renewcommand\@make@capt@title[2]{%
 \@ifx@empty\float@link{\@firstofone}{\expandafter\href\expandafter{\float@link}}%
  {\textbf{#1}}\@caption@fignum@sep#2\quad
}%
\makeatother

\begin{document}

\title{Generic helical edge states due to Rashba spin-orbit coupling in a topological insulator}

\author{Laura Ortiz}
\affiliation{Departamento de F\'isica Te\'orica I, Universidad Complutense de Madrid, Spain}

\author{Rafael A. Molina}
\affiliation{Instituto de Estructura de la Materia, IEM-CSIC, Serrano 123, Madrid 28006, Spain}

\author{Gloria Platero}
\affiliation{Instituto de Ciencia de Materiales de Madrid, ICMM-CSIC, Madrid 28049, Spain}

\author{Anders Mathias Lunde}
\affiliation{Center for Quantum Devices, Niels Bohr Institute, University of Copenhagen, Denmark}

\date{\today}

\begin{abstract}
We study the helical edge states of a two-dimensional topological insulator without axial spin symmetry due to the Rashba spin-orbit interaction. Lack of axial spin symmetry can lead to so-called \emph{generic} helical edge states, which have energy-dependent spin orientation. This opens the possibility of inelastic backscattering and thereby non-quantized transport.  Here we find analytically the new dispersion relations and the energy dependent spin orientation of the generic helical edge states in the presence of Rashba spin-orbit coupling within the Bernevig-Hughes-Zhang model, for both a single isolated edge and for a finite width ribbon. In the single-edge case, we analytically quantify the energy dependence of the spin orientation, which turns out to be weak for a realistic HgTe quantum well. Nevertheless, finite size effects combined with Rashba spin-orbit coupling result in two avoided crossings in the energy dispersions, where the spin orientation variation of the edge states is very significantly increased for realistic parameters. Finally, our analytical results are found to compare well to a numerical tight-binding regularization of the model.

\end{abstract}

\maketitle

\section{Introduction}\label{sec:introduction}

An insulating bulk energy gap along with gapless edge states is a hallmark of a two-dimensional (2D) topological insulator (TI).\cite{Kane-PRL-2005a,Kane-PRL-2005b,Qi-RMP-2011,Ando-JPSJ-2013} At each boundary, two counterpropagating edge states with opposite spin-polarization and wave numbers form Kramers pairs, i.e.~two distinct degenerate states connected by time-reversal symmetry. These states are denoted helical edge states due to their connection between spin and propagation direction. Due to time-reversal symmetry, elastic backscattering of a single electron from a helical edge state (HES) to its Kramers partner is not possible by any time-reversal invariant potential, e.g. disorder.\cite{Xu-PRB-2006} Thereby an important mechanism to hinder ballistic transport is absent and quantized conductance of $2e^2/h$ per pair of HESs is within reach. The first experimentally realized 2D TI in a HgTe quantum well (QW) indeed found quantized conductance in micrometer-sized samples,\cite{Konig-Science-2007,Roth-Science-2009,Brune-nature-phys-2012,Konig-PRX-2013} along with evidence of edge state transport in both two-terminal \cite{Konig-Science-2007} and multiterminal \cite{Roth-Science-2009} configurations. Prior to the experiments, HgTe QWs were in fact predicted to be 2D TIs beyond a certain QW thickness.\cite{Bernevig-Science-2006} These efforts also resulted in a rather generic Dirac-like model describing the essential physics of some 2D TIs, which is now know as the Bernevig-Hughes-Zhang (BHZ) model. Recently, also InAs/GaSb double QWs were suggested theoretically to be 2D TIs described using the BHZ model,\cite{Liu-PRL-2008} which afterwards have been tested experimentally.\cite{Knez-PRL-2011,Suzuki-PRB-2013,Knez-PRL-2014,Du-PRL-2015,Nichele-et-al-2015}

\begin{figure}
\includegraphics[width=0.96\linewidth]{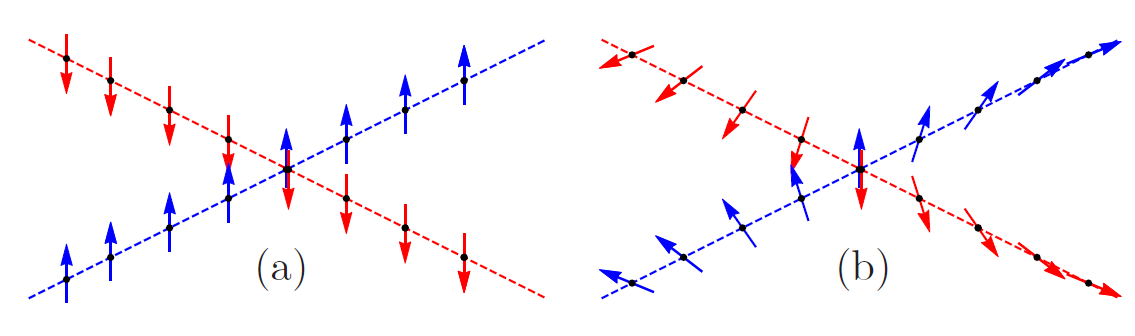}
\caption{\label{fig:GHES_sketch} Illustration of the dispersion relation and spin orientation for (a) helical edge states with constant spin orientation and (b) generic helical edge states with energy-dependent spin orientation. In this paper, we analyse the generic helical edge states and their spin orientation variation due to the Rashba spin-orbit coupling within the BHZ model.}
\end{figure}

Deviations from the quantized conductance have also been found experimentally for longer edges in both HgTe\cite{Konig-Science-2007,Roth-Science-2009,Gusev-PRB-2011,Grabecki-PRB-2013,Konig-PRX-2013,Gusev-et-al-PRB-2014} and InAs/GaSb\cite{Spanton-PRL-2014} QW TIs. Conduction reduction due to \emph{inelastic} backscattering has been studied theoretically,\cite{Schmidt-PRL-2012,Strom-PRL-2010,Budich-PRL-2012,Crepin-PRB-2012,Lezmy-PRB-2012,Geissler-PRB-2014,Kainaris-PRB-2014,Vayrynen-PRL-2013} since it is not a priorly ruled out by time-reversal invariance. Most studies of inelastic backscattering combine some energy-exchange mechanism (e.g. phonons \cite{Crepin-PRB-2012} or electron-electron interactions \cite{Schmidt-PRL-2012,Strom-PRL-2010,Crepin-PRB-2012,Kainaris-PRB-2014}) with a way to manipulate the spin (often some form of spin-orbit coupling \cite{Crepin-PRB-2012,Geissler-PRB-2014,Kainaris-PRB-2014}). Scattering of localized spins \cite{Tanaka-PRL-2011,Lunde-PRB-2012,Eriksson-PRB-2012,Eriksson-PRB-2013,Probst-arxiv-2014} such as magnetic impurities or nuclear spins \cite{Lunde-PRB-2013} has also been analyzed. 

In a particularly interesting proposal for inelastic backscattering, Schmidt et al. \cite{Schmidt-PRL-2012} considered HESs without axial spin symmetry. The Rashba spin-orbit coupling (RSOC)\cite{Rothe-NJP-2010,Virtanen-PRB-2012,Rothe-PRB-2014} and bulk inversion asymmetry (BIA)\cite{Konig-JPSJ-2008,Liu-PRL-2008,Ostrovsky-PRB-2012} can break the axial spin symmetry of the HESs. In this case, a pair of HESs acquire a more generic and intriguing spin-structure than merely having opposite and constant spin-orientations independently of energy. Time-reversal symmetry still dictates that the two counterpropagating Kramers partners have orthogonal spinors, but it does not require equal spinors at different energies as illustrated in Fig. 1. These states were named \emph{generic} helical edge states (GHESs).\cite{Schmidt-PRL-2012} Recently, Kainaris et al. \cite{Kainaris-PRB-2014} extended the original work\cite{Schmidt-PRL-2012} on transport in short GHESs with electronic interaction and disorder to longer ones. Furthermore, the spin-structure of the GHESs was also shown to change the noninteracting transport properties of a point contact and of disordered 2D TI strips. \cite{Orth-PRB-2013} Moreover, the spin-structure of GHESs plays a role in the umklapp-scattering-induced energy gap suggested to host parafermions, a generalisation of Majorana Fermions.\cite{Orth-PRB-2015} 

These studies show that it is worthwhile to analyse the GHESs and their microscopic origin further, which is the purpose of this paper. Very recently, Rod \emph{et al.}\cite{Rod-PRB-2015} studied the spin texture of GHESs due to BIA within the BHZ model and also numerically for the Kane-Mele model.\cite{Kane-PRL-2005a,Kane-PRL-2005b} In contrast, we consider how RSOC\cite{Rothe-NJP-2010} can produce GHESs within the BHZ model. We develop analytical models for the GHESs appearing at an isolated boundary and in the case of a finite width ribbon, where the overlap of the edge states on different boundaries plays an important role. For an isolated edge, we are able to give an analytical formula for the so-called spin-structure parameter, which describes how much the spin-orientation of the GHESs change. This parameter was originally introduced phenomenologically.\cite{Schmidt-PRL-2012} Using realistic numbers for a HgTe QW, we find that an isolated edge is in fact rather robust against spin rotation produced by the RSOC. In contrast, we discover that the combination of RSOC and finite width enhanced significantly the spin rotation versus energy of the GHESs. Throughout the paper, spin rotation refers to the spin orientation variations of the GHES.  Furthermore, we show that our analytical models  compare well to full numerical tight-binding calculations. 

We organise the paper as follows: First, we outline the phenomenology of the GHESs  (Sec.~\ref{sec:helical_edge}) and the BHZ model including the RSOC that breaks the axial spin symmetry (Sec.~\ref{sec:BHZ}). Then, we consider an isolated pair of GHESs at a single boundary in Sec.~\ref{subsec:Semi} and finally analyze the case of a finite width ribbon both analytically and numerically (Sec.~\ref{subsec:twoedges}). Sec.~\ref{conclusions} summarizes the paper and  the Appendices give various technical details.

\section{Phenomenology of\\ the generic helical edge states}\label{sec:helical_edge}

In this section, we discuss the GHESs phenomenologically. GHESs can be modelled as two counterpropagating one-dimensional (1D) states with linear dispersion relations $\e_{k,\pm}=\pm \hbar v k$, i.e.
\begin{equation}
 H_0=\sum_{k,\eta=\pm1} \eta \hbar v k c^{\dagger}_{k \eta} c^{\phantom{\dagger}}_{k \eta}
\end{equation}
 as in Refs.~\onlinecite{Schmidt-PRL-2012,Orth-PRB-2013,Kainaris-PRB-2014,Orth-PRB-2015}. Here $c^{\dagger}_{k \eta}$ ($c^{\phantom{\dagger}}_{k \eta}$) creates (annihilates) a state $\ket{k,\eta}$ with momentum $k$ and propagating direction $\eta$. The states of opposite $k$ and $\eta$ are Kramers partners such that elastic scattering  due to e.g. impurities is still absent. The spin $s_j$ ($j=x,y,z$) expectation value of the Kramers partners are also opposite, i.e.
\begin{equation}
 \bra{k,+} s_j \ket{ k,+} = - \bra{-k,-} s_j \ket{ -k,-},\ \  j=x,y,z.
\end{equation}
The counterpropagating states at each $k$ can be related to the spin states $\sigma=\ua,\da$ along a definite direction by a momentum dependent (and thereby energy dependent) SU(2) matrix $B_k$ as\cite{Schmidt-PRL-2012}
\begin{equation}\label{eq:Bk-full}
 \left(\begin{array}{c} c_{k\ua} \\
 c_{k\da} \end{array}\right) = B_k \left(\begin{array}{c} c_{k +} \\ c_{k-} \end{array} \right).
\end{equation}
Time-reversal symmetry and $B_k \in$ SU(2) lead to $B_k=B_{-k}$. Consistent with these facts, Schmidt {\em et al.} \cite{Schmidt-PRL-2012} introduced the following expansion for small $|k| \ll k_0$: 
\begin{equation}
\label{eq:bk}
  B_k=\left(
  \begin{array}{cc} 
  1-k^4/(2k_0^4) & -k^2/k_0^2 \\ 
  k^2/k_0^2 & 1-k^4/(2k_0^4) 
  \end{array} \right),
\end{equation}
where the spin-quantization axis is chosen such that at the band crossing point $k=0$, we have $c_{k=0,+}=c_{k=0\ua}$ and $c_{k=0,-}=c_{k=0\da}$ as in Fig.~\ref{fig:GHES_sketch}(b). In other words, a constant rotation of all the spins regardlessly of $k$ has been removed from $B_k$ in Eq.(\ref{eq:bk}) following Ref.~\onlinecite{Schmidt-PRL-2012}. Such a $k$-independent rotation corresponds to a constant rotation matrix and can be removed by choosing a rotated basis for the spin. Importantly, a phenomenological spin-structure parameter $k_0$ has been introduced in the expansion (\ref{eq:bk}), which measures the velocity of spin rotation in momentum space. Schmidt {\em et al.}\cite{Schmidt-PRL-2012} showed using perturbation theory that the correction to the quantized conductance due to backscattering processes possible within a pair of GHESs scales as temperature to the forth power with a prefactor depending on $k_0$. In this paper, we find $k_0$ analytically within the BHZ model including the RSOC for an isolated edge.

To gain more insights into the spin structure of the GHESs, we also evaluate the total spin rotation of the edge states, which we define as
\begin{equation}
T_s=\int
dk \left(\left|\left< k_1,\ua | k_1, + \right> \right|^2-\left|\left<k, \ua | k, + \right> \right|^2 \right). 
\label{eq:ts}
\end{equation}
Here $k_1$ is a fixed reference momentum and the integration is over the range of $k$-space, where the edge states exist. The idea behind $T_s$ is to quantify the total variation of the spin orientation of the edge state $\ket{k,+}$ over all relevant $k$. We have constructed $T_s$ such that if $\ket{k,+}$ is a HES (i.e.~$\ket{k,+}=\ket{k,\ua}$), then $T_s=0$. Likewise, if the spin of $\ket{k,+}$ is rotated by the same amount for all $k$, then we still get $T_s=0$.  This is due to the reference term $|\langle k_1,\ua | k_1, + \rangle|^2$ with an arbitrary, but fixed, momentum $k_1$. In our calculations,  we choose  $k_1$ to be the momentum where the edge state dispersion cross the upper bulk band gap edge. In the cases we have analyzed, the reference term $|\langle k_1,\ua\! | k_1, + \rangle|^2$ is very close to one and quite unaffected by small changes in $k_1$. However, generally the choice of $k_1$ does affect the numerical value of $T_s$, but not its variation versus some physical parameter. The behaviour of the spin rotation is more complex for a ribbon than for a single edge, especially for narrower ribbons, as the edge state wave function can have components on both edges. The quantity $T_s$ is useful in that case as this kind of behaviour is difficult to capture with the parameter $k_0$, which quantifies the rotation close to $k=0$. The unit of both $T_s$ and $k_0$ is inverse length.

Before proceeding, we consider a simple 1D model Hamiltonian for a pair of HESs with a generic linear spin-orbit coupling, i.e.
\begin{equation} \label{eq:simplest-case}
 H=\hbar v k\sigma_z+(a_x\sigma_x+a_y\sigma_y)k.
\end{equation}
Here $\sigma_i, i=x,y,z$, are the Pauli matrices and $a_x,a_y$ are the spin-orbit coupling strengths. By diagonalization, we see that this often used \cite{Eriksson-PRB-2012,Eriksson-PRB-2013} simple model does not introduce a $k$-dependent $B_k$, since all matrix elements are linear in momentum $k$. Thereby, it does not give rise to energy-dependent spin-orientation and to GHESs, i.e.~$T_s=0$. This is consistent with the lack of the lowest order inelastic backscattering due to a linear spin-orbit coupling combined with a phonon exchange.\cite{Budich-PRL-2012} In order to get non-trivial GHESs, we resort to calculations for the realistic BHZ model with RSOC.

\section{The BHZ model with Rashba spin-orbit coupling}
\label{sec:BHZ}

The BHZ model is an effective four band model describing the basic physics of a 2D TI.\cite{Bernevig-Science-2006} It was derived using $\mathbf{k \cdot p}$ theory for the band structure of a HgTe QW and therefore valid for small wavevectors $\mathbf{k}=(k_x,k_y)$, i.e.~close to the $\Gamma$ point. It accounts correctly for the physics of HgTe QWs close to the critical well thickness, which marks the transition between a normal semiconductor band structure and an inverted band structure with topologically protected edge states.\cite{Bernevig-Science-2006} The BHZ Hamiltonian consists of two disconnected blocks connected by time reversal symmetry. Each block has the form of a massive Dirac model in 2D in addition to quadratic terms crucial for the band inversion and thereby the topological properties of the material. In fact, the Dirac-like nature makes the BHZ model rather generic for 2D TIs --- even though it grew out of a specific material choice. The BHZ basis states consist of two Kramer pairs of electron-like, $\ket{E\pm}$, and hole-like, $\ket{H\pm}$, states, respectively. The states labeled with $+$ ($-$) are often referred to as the spin-up (spin-down), since they have positive (negative) total angular momentum projection.\cite{Lunde-PRB-2013} In this sense, the time-reversed blocks of the BHZ model have opposite spin. In the basis $\{\ket{E+},\ket{H+},\ket{E-},\ket{H-}\}$, the BHZ Hamiltonian is  
\begin{equation}\label{eq:BHZ-H}
H_0=
\left(\begin{array}{cccc}
 \e_k+M_k & Ak_+ & 0 & 0 \\
 Ak_- & \e_k-M_k & 0 & 0 \\
 0 & 0 & \e_k+M_k & -Ak_- \\
 0 & 0 & -Ak_+ & \e_k-M_k
\end{array}\right),
\end{equation}
where $k_\pm=k_x\pm ik_y$, $\e_k=-Dk^2$, $M_k=M_0-Bk^2$ and $k^2=k_x^2+k_y^2$.  The sign of $M_0/B$ determines the existence of the HESs\cite{Zhou-PRL-2008} and $D\neq0$ induces particle-hole asymmetry in $H_0$. Table \ref{table:parameters} gives the parameters for two different systems modelled by the BHZ model, namely HgTe QWs\cite{Bernevig-Science-2006} and InAs/GaSb double QWs.\cite{Liu-PRL-2008} 
 
In this paper, we utilize an extension of the BHZ model derived in Ref.~\onlinecite{Rothe-NJP-2010} for the inclusion of structural inversion asymmetry (SIA) terms including the RSOC. Importantly, the RSOC couples the two blocks of $H_0$ such that the axial spin symmetry is broken. Here, we include only the most important RSOC linear in momentum, i.e.
\begin{equation}\label{Rashba_hamil}
H_R =
\left(\begin{array}{cccc}
 0 & 0 & -iR_0k_- & 0 \\
 0 & 0 & 0 & 0 \\
 iR_0k_+ & 0 & 0 & 0 \\
 0 & 0 & 0 & 0
\end{array}\right),
\end{equation}
and therefore our full Hamiltonian is $H=H_0 + H_R$. Interestingly, the Rashba term in $H_R$ only couples the electron-like bands, which makes our model more complex than the simple $2\times2$ model Hamiltonian in Eq.(\ref{eq:simplest-case}). Moreover, GHES are now possible as we shall see below. The strength of the RSOC, $R_0$, depends of the amount of SIA, which is often related to an internal or external electric field. For a HgTe QW one can control the RSOC with an external field,\cite{Rothe-NJP-2010} whereas it is an internal field for InAs/GaSb double QWs.\cite{Liu-PRL-2008} Rothe \emph{et al.}\cite{Rothe-NJP-2010} also derives higher order RSOC terms in momentum as we briefly discuss in Appendix \ref{app:higher-order-Rashba}.

The HgTe has a zincblende crystal structure such that inversion symmetry of the crystal is lacking. Therefore, bulk-inversion-asymmetric terms can in principle be included,\cite{Konig-JPSJ-2008} but are often disregarded due to their small size.\cite{Konig-JPSJ-2008,Qi-RMP-2011} However, in InAs/GaSb double QWs BIA terms are in fact significant,\cite{Liu-PRL-2008} so our calculations for this system without BIA terms are not an attempt to model the real system in detail.

\begin {table}[H]
  \begin{center}
   \begin{tabular}{| l | l | l |}
    \hline
    Material & HgTe & InAs/GaSb \\ \hline
    $A$ (meV nm) & 365.0 & 37.0 \\ \hline
    $B$ (meV nm$^2$) & -686.0 & -66.0 \\ \hline
    $D$ (meV nm$^2$) & -512.0 & -58.0 \\ \hline
    $M_0$ (meV)      &  -10.0 & -7.8 \\ \hline
    $R_0$ (meV nm) & 15.6 $\epsilon_z$ & -7.0 \\ \hline
   \end{tabular}
  \end{center} 
\caption{The BHZ model parameters for two QW systems in the topological regime.\cite{ZHANG-BOOK-2013} The parameters for the HgTe QW correspond to a well width of 7 nm, while the values for the InAs/GaSb double QWs are for equal widths of 10nm for both wells. The RSOC constant $R_0$ includes a value for the external electric field $\epsilon_z$ (in meV) in the case of HgTe.}  
 \label{table:parameters}
\end{table}

\begin{figure}
\includegraphics[width=\linewidth]{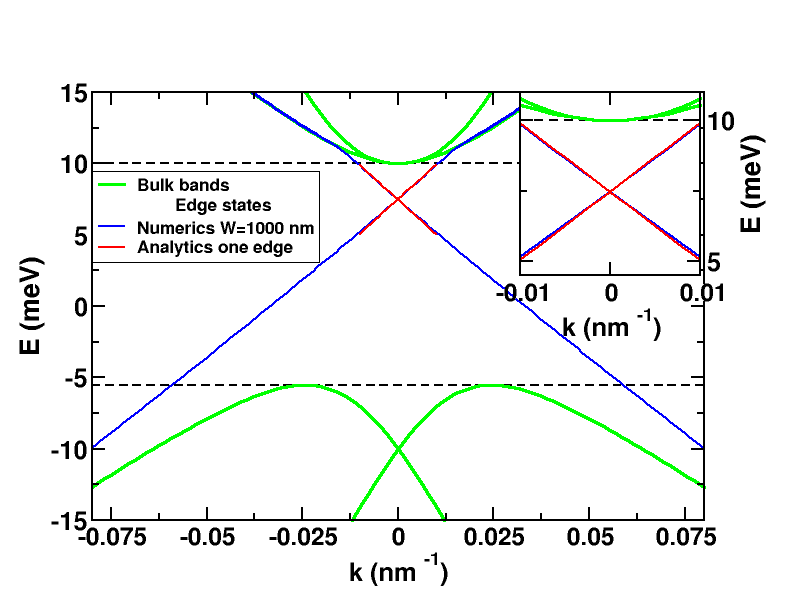}
\includegraphics[width=\linewidth]{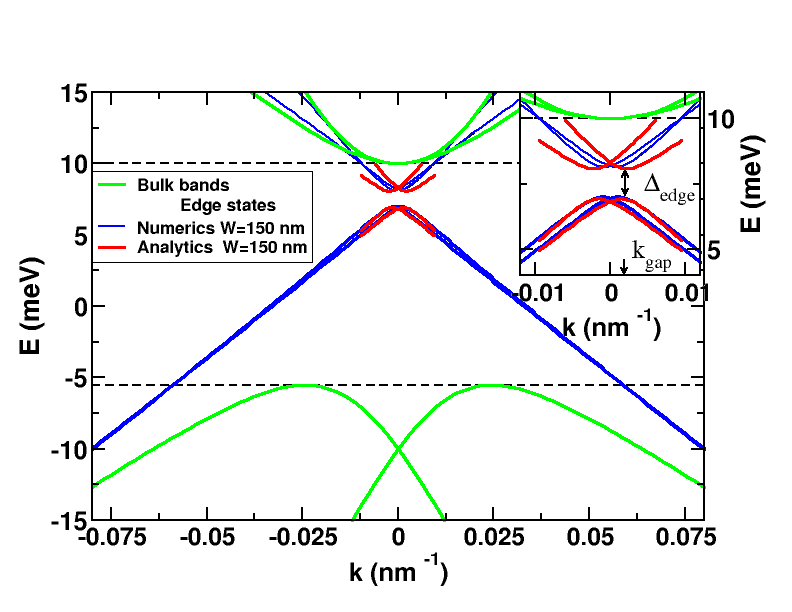}
\caption{\label{fig:dispersions} Bulk and edge state dispersion with the RSOC for a single edge (top panel) and a ribbon (bottom panel) in a HgTe QW. The parameters are given in Table \ref{table:parameters}  and the electric field $\epsilon_z$ is such that $R_0=A$. The bulk gap is marked by dashed horizontal lines and the bulk bands Eq.(\ref{eq:bulk-bands}) are for $H_0+H_R$ with periodic boundary conditions in both directions. Analytical results are not shown in the entire bulk energy gap, because our method requires the existence of both edge states at the same time. The insets show the edge state dispersions close to $k=0$.}
\end{figure}

Before we proceed to the GHESs, we briefly comment on the bulk bands including the RSOC. By diagonalizing $H=H_0 + H_R$, we find
\begin{subequations}\label{eq:bulk-bands}
\begin{align}
E^{n}_{1,2}&=
-D k^2
\pm\frac{R_0 k}{2}-\frac{\sqrt{J_k+K^{\pm}_k}}{2},
\\
E^{p}_{3,4}&=
-D k^2
\pm\frac{R_0 k}{2}+\frac{\sqrt{J_k+K^{\pm}_k}}{2},
\end{align}
\end{subequations}
where $p$ ($n$) is the bulk band with positive (negative) energy and we define  $J_k=4 A^2 k^2+4B^2 k^4+R_0^2 k^2+4M_0^2$  and $K^{\pm}_k=-4 B k^2 (2 M_0 \pm R_0k)\pm4 M_0 R_0 k$ and $k=\sqrt{k_x^2+k_y^2}>0$. Fig.~\ref{fig:dispersions} shows the bulk energy bands with RSOC (together with the edge state dispersions that we consider below). The bulk bands shift due to the RSOC such that the band gap for HgTe becomes indirect. Moreover, the size of the bulk band gap is changed slightly, but not enough to change the topology of the system.

\section{Generic helical edge states}\label{sec:GHES}

The RSOC breaks the spin degeneracy of the BHZ model in such a way that GHESs with energy-dependent spin orientation now becomes feasible. We treat the GHESs below in two cases: (i) A single isolated edge and (ii)  a finite width ribbon with two parallel edges. The isolated edge case offers more analytical insights and we are able to extract the spin-structure parameter $k_0$ defined in Eq.(\ref{eq:bk}).

Without the RSOC, it is possible to obtain the HESs analytically for both an isolated edge and a ribbon with two edges.\cite{Zhou-PRL-2008}  In the appendix \ref{appendix:HES}, we give the details of the analytical wave functions and dispersion relations of the HESs in both cases.  Including the RSOC, it becomes much more challenging to obtain exact analytical forms by the same method, since it is now a problem with four coupled differential equations,  see Appendix \ref{subsec:HES-derivation}. Nevertheless, we are able to obtain analytical results by assuming that the GHESs with RSOC are combinations of the HESs without RSOC, neglecting the possible contribution of the bulk bands. This is a good approximation, since the edge states naturally have a small spatial overlap with the bulk states as long as the edge states are well-localized at the boundary. This is well satisfied especially for momenta close to zero and well into the bulk gap. We find that the bulk gap is reduced as $R_0$ increases and so does the range of applicability of the analytical results. Moreover, we also compare our analytical results with the solution via exact diagonalization of a tight-binding regularization of the BHZ Hamiltonian for a ribbon of width $W$ with periodic boundary conditions in the $x$ direction and edges at $y=-W/2$ and $y=W/2$. The details of the tight-binding formulation is discussed in Appendix \ref{app:tight-binding} and follows Ref.~\onlinecite{book_Shen}. This calculation allows us to unambiguously check the validity of our analytical model.

In the next subsection, we find the GHESs in the presence of RSOC for an isolated edge. We obtain explicit expressions for the spin orientation versus energy and find good agreement with the large-width limit of the numerics. We show that for a single edge the spin orientation is only weakly dependent on energy for a real HgTe sample, i.e. the spin orientation is actually quite robust against RSOC. The following subsection is devoted to a ribbon. Now the expressions become much more complicated but the results as a function of the width of the sample show more interesting patterns, where spin rotation versus energy cannot be neglected.

\subsection{The case of a single isolated edge}
\label{subsec:Semi}

Now we find the pair of GHESs appearing at an isolated boundary of a 2D TI described by the BHZ model including the RSOC. As mentioned above, the starting point is the exact HESs without the RSOC. The HES dispersions are linear,\cite{Wada-PRB-2011} i.e. $E_{\sigma k_y}=E_0+\mathfrak{s}\h vk_y$, where $\mathfrak{s}=+(-)$ for $\sigma=\ua (\da)$, $v$ is the constant velocity and $E_0$ an energy shift. The HESs located at the boundary of the half-plane $x>0$ are given by 
\begin{align}\label{eq:HES-R0-zero}
\psi_{k_y\sigma}(x,y)=\frac{1}{\sqrt{L}}e^{ik_yy}g_{\mathfrak{s}k_y}(x)\hat{\phi}_\sigma,   
\end{align}
 i.e.~a plane-wave running along the $y$-axis combined with a transverse wave function $g_{\mathfrak{s}k_y}(x)$ determining the width of the HES and a $k_y$-independent four-component spinor $\hat{\phi}_\sigma$. There is one spinor from each time-reversed block of the BHZ model, i.e. $\hat{\phi}_{\ua}$ ($\hat{\phi}_{\da}$) only has non-zero components on the two first (last) entries with positive (negative) total angular momentum projection. Periodic boundary conditions are used along the edge of length $L$. The HES wave functions and dispersions are given explicitly using the BHZ parameters in Appendix \ref{subsec-Appendix:HES-isolated-boundary}. 

To include the RSOC analytically, we write the full Hamiltonian $H=H_0+H_R$ in a basis of the HESs for $R_0=0$ given in Eq.(\ref{eq:HES-R0-zero}), i.e.
\begin{align}\label{eq:2_quantization}
\mathcal{H}_0 + \mathcal{H}_R=&\sum_{k_y,\sigma\in\{\ua,\da \}}
E_{\sigma k_y}  c_{\sigma k_y}^\dag c^{}_{\sigma k_y}
\nonumber\\
&+
\sum_{k_y,k_y'}
\sum_{\sigma\sigma'} 
\langle\psi_{k_y\sigma}|H_{R}|\psi_{k'_y\sigma'}\rangle 
c_{\sigma k_y}^\dag c_{\sigma'k_y'}^{},
\end{align}
where  $c_{\sigma k_y}^\dag$ ($c_{\sigma k_y}^{}$) creates (annihilates) a particle in the HES $\psi_{k_y\sigma}$.  In this approach, we neglect the matrix elements between the edge and bulk states. These are presumably very small, since bulk and edge states to a very large extend are spatially separated. This is an excellent assumption well within the bulk gap close to the $\Gamma$ point, whereas the bulk states begin to play a role close to the bulk band gap edge as our numerics show. The full Hamiltonian (\ref{eq:2_quantization}) simplifies by noting that the matrix elements of $H_R$ are diagonal in $k_y$ due to the plane-wave part of the HESs (\ref{eq:HES-R0-zero}). Moreover,  $H_R$ only couples opposite spins, so we find
\begin{align}
\mathcal{H}=
\sum_{k_y}\!
\big(
  c_{\ua k_y}^\dag, c_{\da k_y}^\dag
\big)\!
\left(
 \begin{array}{cc}
  E_0+\h vk_y & k_y\alpha_{k_y}\\
  k_y\alpha_{k_y} & E_0-\h vk_y\\
 \end{array}
\right)\!
\left(
 \begin{array}{c}
  c_{\ua k_y}\\
  c_{\da k_y}\\
 \end{array}
\right)\!,
\label{eq:full-non-diagonal-H}
\end{align} 
 where an effective RSOC $\alpha_{k_y}\equiv\langle\psi_{k_y\ua}|H_{R}|\psi_{k_y\da}\rangle /k_y$ is introduced. In terms of the BHZ parameters, we find
\begin{align} \label{eq:alpha-ky}
\alpha_{k_y}&=
R_0\frac{B-D}{2Bk_y}
\int_{0}^{\infty}\!\! dx
g_{k_y}(x)\big[\p_x+k_y\big] g_{-k_y}(x)
\nonumber\\
&= 
R_0\frac{(B-D)^2}{2B^2}\Big(1-ak_y^2\Big)+\mathcal{O}\Big[k_y^4\Big].
\end{align}
where $a=\frac{D^2 [A^2 B+2 (B^2-D^2) M_0]}{2 B (B^2-D^2) M_0^2}$ and we expanded in $k_y$ in the last step. The exact result for $\al_{k_y}$ and details of the calculation are found in  Appendix \ref{subsec:details-RSOC-isolated}.  

The effective RSOC (\ref{eq:alpha-ky}) only includes the first order RSOC in the BHZ basis given in Eq.(\ref{Rashba_hamil}). In Appendix \ref{app:higher-order-Rashba}, we discuss the effects of higher order RSOC terms. We show that the second order term does not contribute to $\alpha_{k_y}$, while the third order term in principle could contribute even though we face technical difficulties in this case due to the hard wall boundary condition used to find the HESs analytically. 
However, the third order RSOC term cannot introduce terms of a different order in $k_y$ in $\al_{k_y}$ than the ones found here. Therefore it cannot change the physics of the GHESs discussed below. Moreover,  the magnitude of the effects of the third order term can partly be incorporated into the prefactor $R_0$.

The form of $\mathcal{H}$ in Eq.(\ref{eq:full-non-diagonal-H}) is clearly very similar to the simple 1D Hamiltonian for a pair of HESs with a generic spin-orbit coupling Eq.(\ref{eq:simplest-case}), since the effective spin-orbit term $\al_{k_y}\sigma_xk_y$ resembles $a_{x}\sigma_xk$. The important difference is that our effective RSOC $\alpha_{k_y}$ depends on $k_y$ and therefore gives rise to GHESs with $k_y$-\emph{dependent} (or equivalently energy-dependent) spin orientation as we shall see shortly. In contrast, the spin-orbit coupling in Eq.(\ref{eq:simplest-case}) only leads to a constant wavevector-independent spin rotation. In other words, the effective spin-orbit term $\al_{k_y}\sigma_xk_y$ has to be nonlinear in $k_y$ for GHESs to arise.

By diagonalizing $\mathcal{H}$ in Eq.(\ref{eq:full-non-diagonal-H}), we get the dispersion relations including the RSOC 
\begin{subequations}
\begin{align} 
E^{\textsc{rsoc}}_{k_y,\pm}
&=E_0\pm \h v_{\al_{k_y}} k_y , 
\label{eq:dispersions_single_edge_including_R0}
\end{align}
and the eigenstates in $k_y$-space
\begin{align}\label{eq:eigenstate-single-edge-with_R0}
\Psi_{k_y,\pm}&= \frac{1}{\sqrt{2}}
\left(
\begin{array}{c}
\pm\sqrt{1\pm\frac{v}{v_{\al_{k_y}}}}\\ 
\sqrt{1\mp\frac{v}{v_{\al_{k_y}}}}
\end{array} 
\right),
\end{align}
\end{subequations}
where $\pm$ corresponds to two different edge states with the  renormalized velocity $v_{\al_{k_y}}=\sqrt{v^2+(\alpha_{k_y}/\h)^2}$. For $R_0=0$, the states are simply $\psi_{k_y\ua}$ and $\psi_{k_y\da}$, whereas for $R_0\neq0$ they become a superposition of both spins. Moreover, they are GHESs due to their $k_y$-dependent spin orientation. The case described by the model Hamiltonian in Eq.(\ref{eq:simplest-case}) is included here: if $\al_{k_y}$ is \emph{in}dependent of $k_y$, then so are $\Psi_{k_y,\pm}$ and no GHESs appear. Due to time-reversal symmetry, the eigenstates  constitute a Kramers pair with opposite spin orientations (i.e.~orthogonal spinors). This is seen by applying the time-reversal operator $\Theta$ to $\Psi_{k_y,\pm}$ and using $\alpha_{-k_y}{=}\alpha_{k_y}$, i.e.~$\Theta\Psi_{k_y,\pm}= \mp\Psi_{-k_y,\mp}$ (see Appendix \ref{app:TR}). Finally, we observe that the RSOC does not open a gap in the spectrum in accordance with time-reversal symmetry, but merely renormalizes the velocity close to $k_y=0$ and creates a slight nonlinearity for larger $k_y$.

\begin{figure}
   \includegraphics[width=0.95\linewidth]{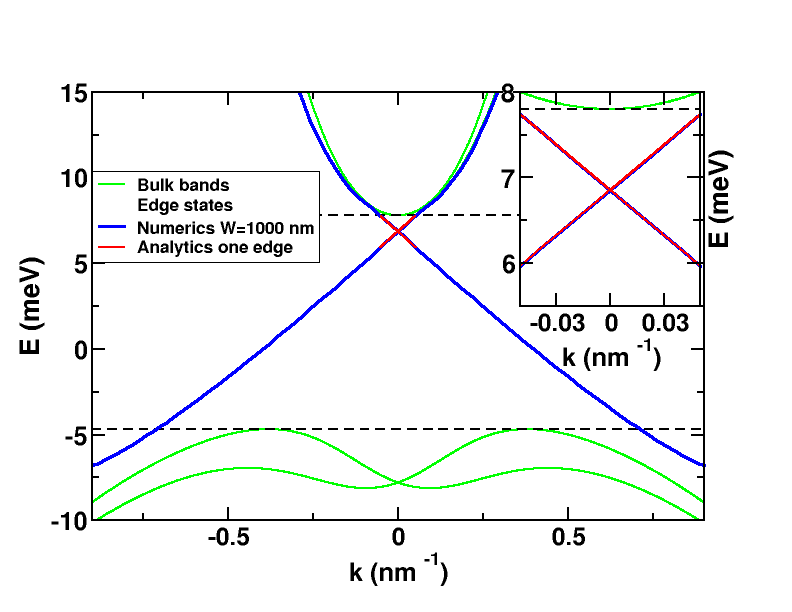} 
\caption{\label{fig:InAs} Dispersion relation for InAs/GaSb QWs using the BHZ Hamiltonian with the parameters given by table \ref{table:parameters}  without taking the BIA terms into account. Analytical results for an isolated edge and numerical results for $W=1000$ nm for the edge states almost coincide.}
\end{figure}

The GHES dispersions (\ref{eq:dispersions_single_edge_including_R0}) for a HgTe QW with $R_0=A$ are shown in the top panel of Fig.~\ref{fig:dispersions} along with a comparison to our numerical results using the tight-binding regularization for $W=1000$nm. We find that the effect of the RSOC on the dispersions is rather weak for a HgTe sample. We also present similar calculations for a InAs/GaSb double QW in Fig.~\ref{fig:InAs}. Our analytical method only works if both HESs without RSOC exist simultaneously, hence the dispersions do not cover the entire bulk gap as seen in Figs.~\ref{fig:dispersions} and \ref{fig:InAs}. Although the bulk bands are quite different for the InAs/GaSb and HgTe QWs, we find that the behavior of the GHESs is very similar for similar values of $R_0$ --- both in the single edge case and for the ribbon discussed in the next section. Therefore, we do not show more figures for InAS/GaSb with the understanding that the results for the latter are similar to our results for HgTe in the presence of an electric field such that $R_0 \approx 0.2A$.

\begin{figure}
 \includegraphics[width=0.9\linewidth]{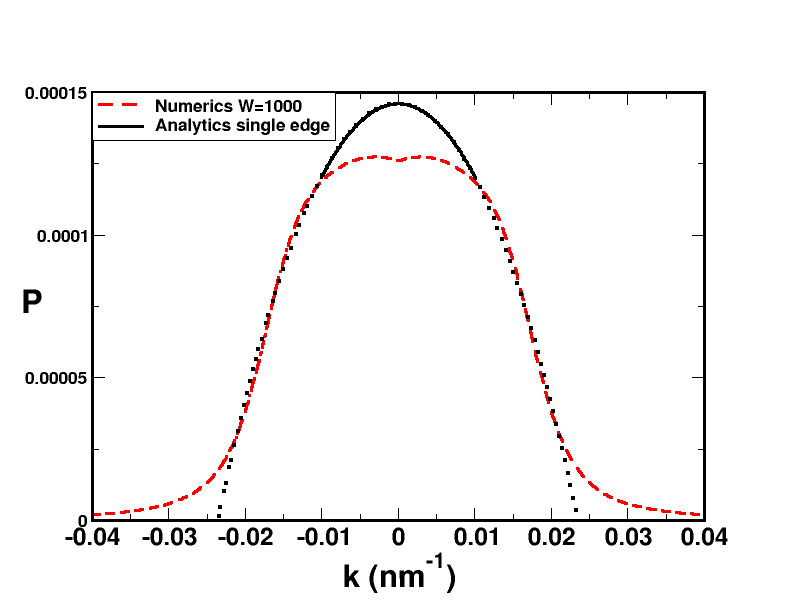}
 \caption{\label{fig:proj_single_edge} The projection $P=|\langle\psi_{k_y\da}|\Psi_{k_y,+}^{}\rangle|^2$ of the GHES $\Psi_{k_y,+}$ with $R_0=0.5A$ into the $R_0=0$ spin-$\da$ state as a function of $k_y$ using the parameters of a HgTe QW in table \ref{table:parameters}. The figure shows a comparison of  the analytical results with the numerics with $W=1000$ nm. The analytical projection is seen in both the bulk energy band gap (full black curve) and in the regime of coexistence of bulk and edge states (dotted black curve). Moreover, we have manually removed the numerical results close $k=0$ where a very narrow peak appears due to finite size effect, see Fig.~\ref{fig:projW} and the discussion in Sec.~\ref{subsec:twoedges}.}
\end{figure}

Next, we consider the $k_y$-dependence of the spin orientation of the GHESs in the case of an HgTe TI. In Fig.~\ref{fig:proj_single_edge}, we show the amount of spin $\da$ in the state $\Psi_{k_y,+}$, which is a spin $\ua$ state for $R_0=0$, i.e. the projection $P=|\langle k_y,\da\! | k_y,+\rangle|^2=|\langle\psi_{k_y\da} |\Psi_{k_y,+}^{}\rangle|^2$. We find a reasonably good comparison between the analytical results for the isolated edge and the numerical results for a large width of $W=1000$nm.  The small discrepancy between the analytical and numerical projections could be due to the truncation of the Hilbert space in the analytical calculation.  As seen in the figure, the spin rotation is rather small in a realistic HgTe QW even for relatively large values of $R_0$, i.e.~the spin orientation of the edge states is rather robust against large external electric field. The analytical projection is found from the GHESs in Eq.(\ref{eq:eigenstate-single-edge-with_R0}) using the exact RSOC $\alpha_{k_y}$ in Eq.(\ref{eq:alpha-single-edge-exact})  in Appendix \ref{subsec:details-RSOC-isolated}. The analytical theory requires simultaneous existence of both HESs without RSOC. The analytical projection in Fig.~\ref{fig:proj_single_edge} is shown in both the bulk band gap region (full curve) and in the region of coexistence between edge and bulk states (dotted curve). In the coexistence regime, the HESs gradually widen and finally the penetration length divergences well within the bulk states as seen in Fig.~\ref{fig:penetration-length} in Appendix \ref{subsec-Appendix:HES-isolated-boundary}. By using the projection, we obtain the total spin rotation $T_s$ Eq.(\ref{eq:ts}). From the numerical results for the entire $k$-space, we find that $T_s$ is proportional to $R_0^2$ to a good approximation. 

Now, we find the analytical form of the spin structure parameter\cite{Schmidt-PRL-2012} $k_0$ in Eq.(\ref{eq:bk}) for the BHZ model including the RSOC. We do this by introducing two unitary transformations, which together diagonalize $\mathcal{H}$ in Eq.(\ref{eq:full-non-diagonal-H}). The first transformation is $k_y$-independent and rotate the spin basis such that it removes the $k_y$-independent part of $\alpha_{k_y}$. This part does not lead to GHESs as discussed above. This rotation is convenient such that we use the same choice of spin-quantization axis as in Ref.~\onlinecite{Schmidt-PRL-2012}, i.e. the spins point along the new rotated spin-quantization axis at $k=0$. The second unitary transformation is $k_y$-dependent and transforms between the eigenstates and the new rotated spin basis. In other words, it is the matrix $B_k$ in Eq.(\ref{eq:Bk-full}). Now we perform the steps explicitly. First, we define $\delta\alpha_{k_y} \equiv \alpha_{k_y}-\alpha_0$, where $\alpha_0=\alpha_{k_y=0}^{}$ is $k_y$-independent. Thereby, we can diagonalize the $\alpha_0$ part of $\mathcal{H}$, i.e.
\begin{align}
\mathcal{H}
&=
\sum_{k_y}
C^{\dagger}_{k_y}
\left(
 \begin{array}{cc}
  E_0+\h vk_y & k_y(\alpha_0+\delta\alpha_{k_y})\\
  k_y(\alpha_0+\delta\alpha_{k_y}) & E_0-\h vk_y\\
 \end{array}
\right)
C_{k_y}
\nonumber\\
&=
\sum_{k_y}
C^{\dagger}_{k_y}
U
\left(
 \begin{array}{cc}
  E_0+ \h v_{\al_0}k_y & 0\\
  0 & E_0-\h v_{\al_0}k_y\\
 \end{array}
\right)U^{\dag}C^{}_{k_y}
\nonumber\\
&\  +
\sum_{k_y}
C^{\dagger}_{k_y}U
k_y\delta\alpha_{k_y}
\left(
 \begin{array}{cc}
  \sin(\theta) & \cos(\theta)\\
  \cos(\theta) & -\sin(\theta)\\
 \end{array}
\right)
U^{\dag}
C^{}_{k_y}.
\label{eq:trivial-rotation-performed}
\end{align} 
Here $C^{\dagger}_{k_y}=\big(c_{\ua k_y}^\dag, c_{\da k_y}^\dag\big)$, $\h v_{\al_0}=\sqrt{(\h v)^2+\alpha_0^2}$ and the first $k_y$-independent unitary transformation $U$ is 
\begin{align}
U=\left(
 \begin{array}{cc}
  \cos(\theta/2) & -\sin(\theta/2)\\
  \sin(\theta/2) & \cos(\theta/2)\\
 \end{array}
\right)
\end{align}
where $\cos(\theta)\equiv v/v_{\al_0}$ and $\sin(\theta)\equiv \alpha_0/(\h v_{\al_0})$. This transformation is simply a $k_y$-independent rotation to a new spin basis, 
\begin{align} 
\left(
 \begin{array}{c}
  c_{\ua' k_y}^{}\\
  c_{\da' k_y}^{}\\
 \end{array}
\right)
=
U^{\dagger}
\left(
 \begin{array}{c}
  c_{\ua k_y}^{}\\
  c_{\da k_y}^{}\\
 \end{array}
\right),
\end{align} 
where $\ua'$ and $\da'$ are the eigenstates of $\mathcal{H}$ at $k_y=0$. Finally, we diagonalize the Hamiltonian completely by a second unitary transformation and obtain 
\begin{align}
\mathcal{H}
&{=}
\sum_{k_y}\!
C_{k_y}^{'\dag}
\mathcal{V}_{k_y}^{}
\!\left(\!
 \begin{array}{cc}
  E_0\!+\!k_y \h v_{\al_{k_y}} & 0\\
  0 & E_0\!-\!k_y\h v_{\al_{k_y}} \\
 \end{array}
\!\right)\!
\mathcal{V}_{k_y}^{\dag}
C_{k_y}^{'},
\nonumber 
\end{align} 
where $C_{k_y}^{'\dagger}=\big(c_{\ua'k_y}^\dag, c_{\da' k_y}^\dag\big)$ and $\h v_{\al_{k_y}}=\sqrt{(\h v)^2+\alpha_{k_y}^2}$. As expected, we find the same dispersions as in Eq.(\ref{eq:dispersions_single_edge_including_R0}). More importantly, we acquire an analytical form of the unitary transformation $\mathcal{V}_{k_y}$, which by construction is exactly $B_{k_y}$ from Eq.(\ref{eq:Bk-full}), i.e.
\begin{align}
B_{k_y}=
\mathcal{V}_{k_y}=
\left(
 \begin{array}{cc}
  \cos(\phi/2) & -\sin(\phi/2)\\
  \sin(\phi/2) & \cos(\phi/2)\\
 \end{array}
\right),
\end{align} 
where $\cos(\phi)\equiv[(\h v)^2+\alpha_{k_y}\alpha_0]/(\h^2 v_{\alpha_0}v_{\al_{k_y}})$ and $\sin(\phi)\equiv \delta\alpha_{k_y} v /(v_{\alpha_0}\h v_{\al_{k_y}})$. Therefore, we have now found the $k_y$-dependent matrix $B_{k_y}$ relating the GHESs to the HESs with a fixed spin-axis for a specific model, namely the BHZ model including the RSOC. We remark that $\delta\alpha_{k_y}=0$ at $k_y=0$ by definition, so $\mathcal{V}_{k_y=0}$ is the unity matrix and therefore $\ua'$ and $\da'$ become eigenstates at $k_y=0$. 

We can now find the spin structure parameter $k_0$ in Eq.(\ref{eq:bk}) controlling the amount of spin-rotation for small $k_y$. By expanding $B_{k_y}=\mathcal{V}_{k_y}$ around $k_y=0$, we obtain
\begin{align}\label{eq:k0-general}
\!\frac{1}{k_0^2}
&=
\frac{D^2 |R_0 A (B{-}D)| \left|A^2 B+2 M_0(B^2-D^2)\right|}{2 \sqrt{B^2{-}D^2} M_0^2 \left| 4 A^2 B^2 (B{+}D) {+}(B{-}D)^3 R_0^2\right|}.
\end{align}
Thereby, we have an explicit expression for $k_0$ --- a parameter originally introduced based on symmetry arguments.\cite{Schmidt-PRL-2012} Such an expression in terms of the BHZ parameters is valuable beyond the case of HgTe QWs due to the generic Dirac-like nature of the BHZ model. Interestingly, we observe that the particle-hole asymmetry parameter $D$ plays an essential role for $k_0$, i.e. for $D=0$ no spin rotation appears and therefore no GHESs come out in the case studied here. This is valid beyond the expansion of $B_{k_y}$ in $k_y$, since the effective RSOC in Eq.(\ref{eq:alpha-ky}) is $k_y$-independent to all orders, $\alpha_{k_y}^{(D=0)}=R_0/2$,  for $D=0$ such that $B_{k_y}$ is the unity matrix (see Eq.(\ref{eq:alpha-D_equal_0}) in Appendix \ref{subsec:details-RSOC-isolated}). Curiously, the parameter $D$ is often removed in many theoretical discussions of topology\cite{book_BH} and thereby the interesting physics of GHESs might be missed. Furthermore, Eq.(\ref{eq:k0-general}) also reveals that $k_0$ depends on the Dirac mass $M_0$ and the RSOC strength $R_0$ in rather non-trivial ways.   

Before proceeding, we briefly discuss the effect of the lowest order BIA terms given by\cite{Konig-JPSJ-2008,Qi-RMP-2011}
\begin{equation}\label{eq:BIA}
H_{BIA}=
\left(\begin{array}{cccc}
 0 & 0 & 0 & -\Delta \\
 0 & 0 & \Delta & 0 \\
 0 & \Delta & 0 & 0 \\
 -\Delta & 0 & 0 & 0
\end{array}\right),
\end{equation}
where $\Delta$ is a constant. Including the $H_{BIA}$ in the basis of the HESs for an isolated edge Eq.(\ref{eq:HES-R0-zero}) as we did for $H_R$ in Eq.(\ref{eq:2_quantization}), we find $\langle\psi_{k_y\sigma}|H_{BIA}|\psi_{k'_y\sigma'}\rangle=0$ for all $k_y\sigma$ and $k'_y\sigma'$. Hence, within our analytic approach, the lowest order BIA terms does not affect the HESs nor their spin orientation for an isolated edge. The second order RSOC terms has the same structure in the anti-diagonal as $H_{BIA}$ and therefore also has zero matrix elements, see Appendix \ref{app:higher-order-Rashba}. Including the small overlaps between the bulk and edge states, a modest effect on the energy dispersions is found due to $H_{BIA}$ close to the bulk gap edge, where these overlaps matter the most.\cite{Michetti-Semicond-Sci-Tech-2012} For a ribbon, the $H_{BIA}$ was found to couple opposite edges.\cite{Krueckl-PRL-2011} Very recently, Rod \emph{et al.}\cite{Rod-PRB-2015} found GHESs for both ribbon and disk geometries due to $H_{BIA}$ in the BHZ model. For both geometries, a finite $k_0^{-2}$ was extracted numerically in the limit of a particle-hole symmetric BHZ model (i.e.~$D=0$), where both edge and bulk states were included in their calculations.    

\subsection{The case of a ribbon with two parallel edges} \label{subsec:twoedges}

In this section, we consider the GHESs for a ribbon with two parallel edges using the BHZ model including the RSOC. Thereby, four edge states come into play, since a pair of GHESs exist on each edge for well-separated edges. We pay special attention to how the finite size effects can enhance spin orientation variation of the GHESs as the width of the ribbon $W$  gets smaller and the edge states on opposite sides begin to overlap.

Before including the RSOC, we briefly summarize the HESs and their dispersions without RSOC for a ribbon.\cite{Zhou-PRL-2008} We refer to Appendix \ref{subsec:HES-ribbon} for details. An important difference between the ribbon and the single-edge case is that we do not have the energy dispersions in closed analytical forms for the ribbon, but instead as the solutions to a cumbersome equation (see Eq.(\ref{eq:implicit-eq-for-energies-for-finite-width})). Nevertheless, the physical consequence of the finite width is clear: A gap opens at the crossing point of the dispersions found for the isolated edge, see Fig.~\ref{fig:dispersion_noR}.\cite{Zhou-PRL-2008} The dispersions for a ribbon have a limiting cusp form for a wide ribbon, i.e.
\begin{align}\label{eq:cusp-form}
E^{e=\pm}_{k_x}\rightarrow E_0\pm\h v |k_x|
\quad \textrm{for}\quad
\  W \rightarrow \infty,
\end{align} 
where $E^{+}_{k_x}$ ($E^{-}_{k_x}$) is the energy dispersion above (below) the gap for $W\leq\infty$. Therefore, the label $e=\pm$ should not be confused with the single-edge case, where $\pm$ often refers to the sign of the velocity.  The velocity $v$ and energy shift $E_0$ are identical to the single-edge case. Noticeably,  $E^{e=\pm}_{k_x}$ are independent of the spin $\sigma$, since equal spins travel in opposite directions on the two edges.  

A ribbon with edges at $y=\pm W/2$ has four HESs without RSOC\cite{Zhou-PRL-2008} $\psi_{k_x\sigma}^{e}(x,y)$, where $e=\pm$ labels the energy $E^{e}_{k_x}$ to which the state belongs. As for an isolated edge, the states have a plane-wave part running along the edges, i.e. $\psi_{k_x\sigma}^{e}\propto e^{ik_xx}$. Only the first (last) two components of the states $\psi_{k_x\ua}^{e}$ ($\psi_{k_x\da}^{e}$) are non-zero, corresponding to the spin-up (spin-down) block of $H_0$. However, in contrast to the single-edge case, the spinors are not constant, but the relative weight of the two components vary with both $k_x$ and $y$. A particular state $\psi_{k_x\sigma}^{e}$ is not always localized on the same edge. Instead, the localization changes gradually from one edge to the other when crossing $k_x=0$. For $k_x>0$, the states
\begin{subequations}\label{eq:localization-of-HESs} 
\begin{align}
\psi_{k_x\ua}^{+},\
\psi_{k_x\da}^{-} \ & \textrm{are localized close to}\ y=W/2\ \textrm{and} \\
\psi_{k_x\ua}^{-},\ 
\psi_{k_x\da}^{+}\ & \textrm{are localized close to}\ y=-W/2, 
\end{align}
\end{subequations} 
and vice versa for $k_x<0$.  

As for the isolated edge, we build an analytical model using only the HESs without RSOC. Since this approach leaves out the overlaps between bulk and edge states, it becomes less good for a narrow ribbon, where bulk and edge states become comparable in spatial extend. Therefore, our analytical results are most reliable for small momenta well within the bulk gap as we shall see.

By including the RSOC in the subspace of the HESs without RSOC, $\psi_{k_x\sigma}^{e}$, the Hamiltonian becomes
\begin{align}
\mathcal{H}=&\mathcal{H}_0+\mathcal{H}_R
=\sum_{\sigma,k_x,e}{E_{k_x}^{e}(c^{e}_{k_x\sigma})^\dagger c_{k_x\sigma}^e}
\nonumber\\
&+ 
\sum_{k_x,k'_x}\sum_{\sigma,\sigma'}\sum_{e,e'}
{\langle\psi_{k_x\sigma}^{e}|H_R| \psi_{k'_x\sigma'}^{e'}\rangle
(c^{e}_{k_x\sigma})^\dagger c_{k'_x\sigma'}^{e'}},
\end{align}
where $(c^{e}_{k_x\sigma})^\dagger$ [$c_{k_x\sigma}^{e}$] creates [annihilates] a particle in the HES $\psi_{k_x\sigma}^{e}$ of energy $E_{k_x}^{e}$. The RSOC Eq.(\ref{Rashba_hamil}) only couples opposite spins and the Hamiltonian is diagonal in $k_x$, since $\langle\psi_{k_x\sigma}^{e}|H_R| \psi_{k'_x\sigma'}^{e'}\rangle\propto\delta_{k^{}_x,k'_x}$. We order the basis as $\{ | \psi_{k_x\ua}^{+}\rangle, \,   | \psi_{k_x\da}^{-}\rangle,\,  | \psi_{k_x\ua}^{-}\rangle,\,| \psi_{k_x\da}^{+}\rangle\}$ such that the first two entries are localized on the opposite edge of the last two as seen in Eq.(\ref{eq:localization-of-HESs}), i.e.
\begin{align}\label{full_hamiltonian}
&\mathcal{H}=\mathcal{H}_0+\mathcal{H}_R
\nonumber\\
&=
\sum_{k_x}
\mathbf{C}^{\dagger}_{k_x}\!
\left(\!
\begin{array}{cccc}
E^{+}_{k_x}&ib&0&id_{+}\\
-ib&E^{-}_{k_x}&id_{-}&0 \\
0&-id_{-}&E^{-}_{k_x}&-ib \\
-id_{+}&0&ib&  E^{+}_{k_x}
\end{array} 
\right)
\!\mathbf{C}_{k_x},
\end{align}
where $\mathbf{C}^{\dagger}_{k_x}=\big[(c_{k_x\ua}^+)^\dagger,(c_{k_x\da}^-)^\dagger,(c_{k_x\ua}^-)^\dagger,(c_{k_x\da}^+)^\dagger\big]$ and we introduced the \emph{inter}-edge matrix elements 
\begin{subequations}\label{eq:inter-matrix-elements} 
\begin{align}
id_{+}=\langle\psi_{k_x\da}^{+} | H_R| \psi_{k_x\ua}^{+}\rangle,\\
id_{-}=\langle\psi_{k_x\ua}^{-} | H_R| \psi_{k_x\da}^{-}\rangle,
\end{align}
\end{subequations} 
and the \emph{intra}-edge matrix element
\begin{align}\label{eq:intra-matrix-element}
ib_{}=\langle\psi_{k_x\da}^{-} | H_R| \psi_{k_x\ua}^{+}\rangle,
\end{align}
which all depend on $k_x$. In Eq.(\ref{full_hamiltonian}), we used that the intra-edge matrix elements on opposite edges are related as $ib=\langle\psi_{k_x\da}^{-} | H_R| \psi_{k_x\ua}^{+}\rangle=-\langle\psi_{k_x\da}^{+} | H_R| \psi_{k_x\ua}^{-}\rangle$, as discussed in Appendix \ref{subsec:ribbon-details}. Thereby, we are left with three matrix elements only, which depend on the implicitly known dispersions relations $E^{\pm}_{k_x}$. The detailed formulas are given in Appendix \ref{subsec:ribbon-details}.

Due to the ordering of the basis, the Hamiltonian (\ref{full_hamiltonian}) has two $2\times2$ blocks in the diagonal, one for each edge. Each $2\times2$ block resembles the Hamiltonian (\ref{eq:full-non-diagonal-H}) found for an isolated edge and an effective \emph{intra}-edge RSOC could be introduced as $b/k_x$ as in Sec.~\ref{subsec:Semi}. However, due to the limiting cusp form of the energy dispersions Eq.(\ref{eq:cusp-form}), one should instead  define the effective intra-edge RSOC as $\alpha_{k_x}^{\textrm{intra}}=-b/|k_x|$. With this definition, $\alpha_{k_x}^{\textrm{intra}}$ corresponds to the effective RSOC for the isolated edge in Eq.(\ref{eq:alpha-ky}) in the wide ribbon limit. However, as the width gets smaller, we find an increased $k_x$-dependence of $\alpha_{k_x}^{\textrm{intra}}$ for small $k_x$. This indicates an increased spin-orientation change for small $k_x$ as $W$ decreases, which we also find below by direct calculation.      

The opposite edges of the ribbon are coupled by the \emph{inter}-edge elements $d_{\pm}$ in the anti-diagonal of $\mathcal{H}$, which vanish for $W\rightarrow \infty$. Finally, we mention that performing unitary transformations of $\mathcal{H}$ to find $k_0$ as in Sec.~\ref{subsec:Semi} is difficult, since we do not have closed formulas for $E_{k_x}^e$.

By diagonalization of the Hamiltonian (\ref{full_hamiltonian}), the dispersion relations including the RSOC become 
\begin{align}\label{eq:energy-rsoc-ribbon}
E^{\textsc{rsoc}}_{k_x,s \tau}
=&
\tau
\frac{1}{2}\sqrt{[s(d_{-}-d_{+})+E_{k_x}^--E_{k_x}^+]^2+4 b^2}
\nonumber\\
&+\frac{1}{2} \left[s(d_{+}+d_{-})+E_{k_x}^++E_{k_x}^-\right]
\end{align}
where  $ s=\pm1$ and $\tau=\pm1$. In the wide ribbon limit, where the inter-edge matrix elements $d_\pm$ are insignificant, these dispersions resemble the isolated-edge dispersions Eq.(\ref{eq:dispersions_single_edge_including_R0}) (disregarding the cusp limit of $E^{\pm}_{k_x}$). For a finite width $W$, however, the inter-edge matrix elements $d_{\pm}$ come into play and create four different dispersions. As shown in the bottom panel of Fig.~\ref{fig:dispersions} and in Fig.~\ref{fig:ribbon-dispersions-anticross}, two gaps arise symmetrically with respect to $k_x=0$. Therefore, we have found that the spin degeneration present for $R_0=0$ between $\psi_{k_x\ua}^{e}$ and $\psi_{k_x\da}^{e}$ is broken by the interplay of RSOC and a finite width, where both ingredients are necessary. A similar effect of SIA combined with finite size have also been found in one dimension higher, namely for the 2D Dirac surface states on a 3D TI.\cite{Shan-NJP-2010}

\begin{figure}
\includegraphics[width=0.95\linewidth]{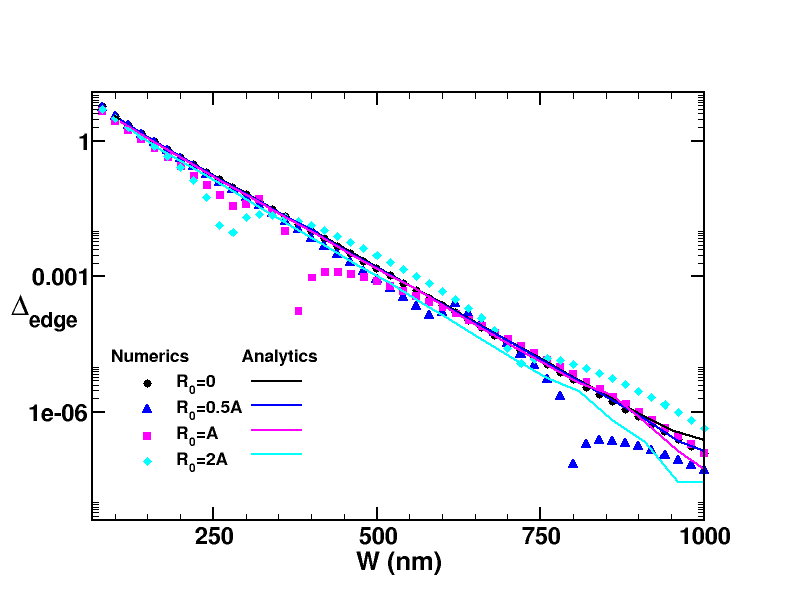}
\includegraphics[width=0.95\linewidth]{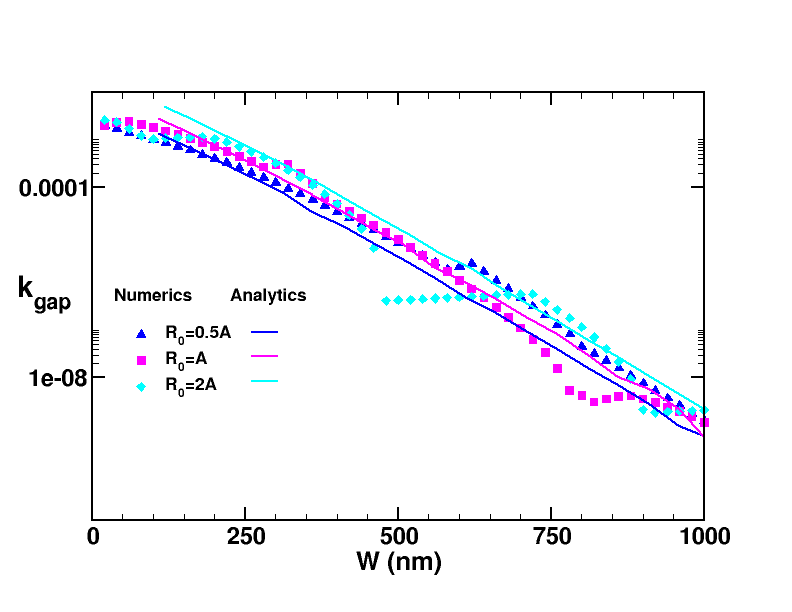}
\caption{\label{fig:gapW}
Comparison of analytical and numerical results for the value of the gap $\Delta_{\textrm{edge}}$ [$meV$] vs.~width of the ribbon (top panel) and position of the gap in momentum $k_{\textrm{gap}}$ [$nm^{-1}$] vs.~width of the ribbon (bottom panel). We use the HgTe parameters in Table \ref{table:parameters}.}
\end{figure}

In Fig.~\ref{fig:gapW}, we show the position $k_{\textrm{gap}}$ and value $\Delta_{\textrm{edge}}$ of the gaps as a function of $W$ for different $R_0$. We compare numerical and analytical calculations on a logarithmic scale. As the width is increased, the position of the gap goes rapidly towards $k_x=0$ and the value of the gap  goes to zero, such that we recover the result for an isolated edge. Interestingly enough, there are several values of the width, depending on the value of $R_0$, where the gap is particularly reduced.  The reason is essentially that the actual transverse wave function including RSOC for an isolated edge has a form similar to an exponential times a sine. This means that the solution for an isolated edge state (including the gapless dispersions) also becomes the solution for a finite ribbon, when the zeros of the transverse wave function match the width. This destructive interference has been studied before both with\cite{Takagaki-PRB-2014} and without\cite{Mehdi-JPCM-2012} RSOC. Similar physics have also been discussed for thin films of 3D TIs.\cite{Linder-PRB-2009,Liu-PRB-2010-3DTI,Lu-PRB-2010} In Ref.~\onlinecite{Takagaki-PRB-2014}, Takagaki showed that the gap vanishes periodically with a period almost inversely proportional to the strength of the RSOC. In these particular values of the width, the coupling between the edges is cancelled without reaching the large width limit. As shown in Fig. \ref{fig:gapW}, this effect is not captured by the analytical theory, although it correctly gives the essential decaying trends of both the gap size $\Delta_{\textrm{edge}}$ and position $k_{\textrm{gap}}$.

The eigenvectors in $k_x$-space in the basis presented above, i.e. $\{ | \psi_{k_x\ua}^{+}\rangle,    | \psi_{k_x\da}^{-}\rangle,  | \psi_{k_x\ua}^{-}\rangle, | \psi_{k_x\da}^{+}\rangle\}$ are: 
\begin{align}\label{two_rashba}
\Psi_{k_x,s\tau}= 
\frac{1}{\sqrt{8b^2+2\zeta^2_{s\tau}}}
\left(
\begin{array}{c}
is 2b\\
-s\zeta_{s\tau}\\
i\zeta_{s\tau}\\
2b
\end{array}
\right)
\end{align}
where $s=\pm$ and $\tau=\pm$ and we defined
\begin{align}
\zeta_{s\tau}= 
&s(d_{+}-d_{-})+E_{k_x}^+-E_{k_x}^-
\nonumber\\
&-\tau\sqrt{\big[s(d_{-}-d_{+})+E_{k_x}^- -E_{k_x}^+\big]^2+4 b^2}.
\end{align}
Here the two first and the two last components of $\Psi_{k_x,s\tau}$ represent spinors localized on opposite edges. The four states $\Psi_{k_x,s\tau}$ are clearly present on both edges, but in a very particular way: The spinor localized on one edge, $\varphi_a\propto (is 2b, -s\zeta_{s\tau})^T$,  is always orthogonal to the spinor $\varphi_b\propto (i\zeta_{s\tau},2b)^T$ localized on the opposite edge, since $\varphi_a^\dag\varphi_b=0$. In other words, the squared projections on the basis states on opposite edges are pairwise identical, i.e. $|\langle\psi_{k_x\ua}^{+}|\Psi_{k_x,s\tau}\rangle|^2=|\langle\psi_{k_x\da}^{+}|\Psi_{k_x,s\tau}\rangle|^2$ and $|\langle\psi_{k_x\da}^{-}|\Psi_{k_x,s\tau}\rangle|^2=|\langle\psi_{k_x\ua}^{-}|\Psi_{k_x,s\tau}\rangle|^2$. Thus, the states always have half of the weight on each edge, i.e. $|\langle\psi_{k_x\ua}^{+}|\Psi_{k_x,s\tau}\rangle|^2+|\langle\psi_{k_x\da}^{-}|\Psi_{k_x,s\tau}\rangle|^2=1/2$ independently of $k_x$.  Moreover, the  Kramers partner of $\Psi_{k_x,\pm\pm}$ is $\Psi_{-k_x,\mp\pm}$, which can be seen by using that $E_{k_x}^\pm$ and $b$ are even in $k_x$ and $d_\pm$ is odd such that $\zeta_{s\tau}(-k_x)=\zeta_{-s\tau}(k_x)$, see Appendix \ref{subsec:ribbon-details}. Furthermore, the dispersions $E^{\textsc{rsoc}}_{k_x,s \tau}$ (\ref{eq:energy-rsoc-ribbon}) and eigenstates (\ref{two_rashba}) depend on both $E_{k_x}^+$ and $E_{k_x}^-$, and therefore only well-defined for momenta $k_x$, where both $E_{k_x}^-$ and $E_{k_x}^+$ are well-defined, see Figs.~\ref{fig:dispersions} and \ref{fig:dispersion_noR}. 

\begin{figure}
 \includegraphics[width=0.95\linewidth]{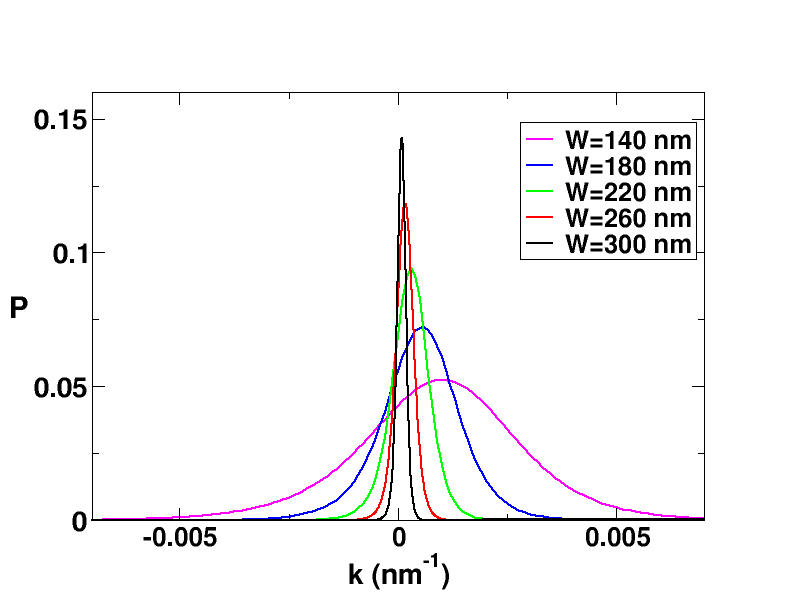}
 \caption{\label{fig:projW} The analytical projection of a ribbon edge state with primarily spin up character (on the lower edge $y=-W/2$ and $k>0$) into the spin down subspace, i.e $P=|\langle\psi_{k_x\da}^{+}|\Psi_{k_x,+-}\rangle|^2$. In other words, the last component squared of $\Psi_{k_x,+-}$ in Eq.(\ref{two_rashba}). The parameters for HgTe in table \ref{table:parameters} and $R_0=A$ are used.}
\end{figure}

Now, we will argue that the eigenstates $\Psi_{k_x,s\tau}$ are GHESs, since their spin orientation on a single edge depends on $k_x$. Due to the structure of $\Psi_{k_x,s\tau}$ discussed above, we observe that the two edges of the ribbon suffer the same --- but opposite --- spin rotation. Due to the coupling between the two edges and the RSOC, the dispersions have two avoided crossings. These avoided crossings induce some particular characteristics of the spin rotation. Fig.~\ref{fig:projW} shows the analytical results for the projection onto the spin down subspace, $P=|\langle\psi_{k_x\da}^{+}|\Psi_{k_x,s=1\tau=-1}\rangle|^2$, as a function of $k_{x}$ for one of the edge states  which asymptotically is more than 99$\%$ spin up for different values of $W$  on lower edge ($y=-W/2$) and $k_x>0$.  We can see that the projection reaches a relatively high value, higher for larger widths, but in a very narrow range of $k_x$, smaller for larger widths. The peaks of the projections are located close to the position of the gap $k_{\textrm{gap}}$. For clarity, we only show a range of $W$ from $140$nm to $300$nm, but the trend goes on indefinitely.  

The avoided crossings of the ribbon dispersions and the associated spin-structure of the GHESs are illustrated on Fig.~\ref{fig:ribbon-dispersions-anticross}. As discussed above, the GHESs $\Psi_{k_x,s\tau}$ Eq.(\ref{two_rashba}) are always equally present on the lower ($y=-W/2$) and the upper ($y=W/2$) edge. Fig.~\ref{fig:ribbon-dispersions-anticross} only shows the spin-structure of the lower edge. We illustrate how to understand this by using the state $\Psi_{k_x,+-}$ as an example. Away from the avoided crossing (the green region), we find 
\begin{align} \label{eq:pure-spin-GHES+-}
\Psi_{k_x,+-}=\frac{1}{\sqrt{2}}(-\psi_{k_x\da}^-+i\psi_{k_x\ua}^-)
\end{align}
with more than $99\%$ accuracy. For $k_x>0$, $\psi_{k_x\ua}^-$ ($\psi_{k_x\da}^-$) is localized near the lower (upper) edge and vice versa for $k_x<0$, see Eq.(\ref{eq:localization-of-HESs}). Thus, in this sense, $\Psi_{k_x,+-}$ is spin $\ua$ for $k_x>0$ (blue region) and spin $\da$ for $k_x<0$ (yellow region) \emph{on the lower edge}, while the upper edge has the opposite spin-structure. In between these regions of almost pure spin $\ua$ or $\da$, the states become genuine GHESs with sizable amounts of both spin $\ua$ and $\da$ present on each edge (the green region). These regions are quantified by the peaks in the projections shown in Fig.~\ref{fig:projW}. Noticeably, the weight of each spin-component in the almost pure spin regions (blue/yellow) of the GHESs is only $1/2$, see e.g.~Eq.(\ref{eq:pure-spin-GHES+-}). Thus, the entire weight of one spin-component on one edge is carried by two different dispersion curves. This is vastly different from the two simple linear dispersions found for an isolated edge. Therefore, it is now clear that our states $\Psi_{k_x,s\tau}$ indeed are GHESs with $k_x$-dependent spin orientation. A remarkable difference to the isolated edge case is that the spin-orientation change is enhanced a great deal by the finite size.

\begin{figure}
\includegraphics[width=0.98\linewidth]{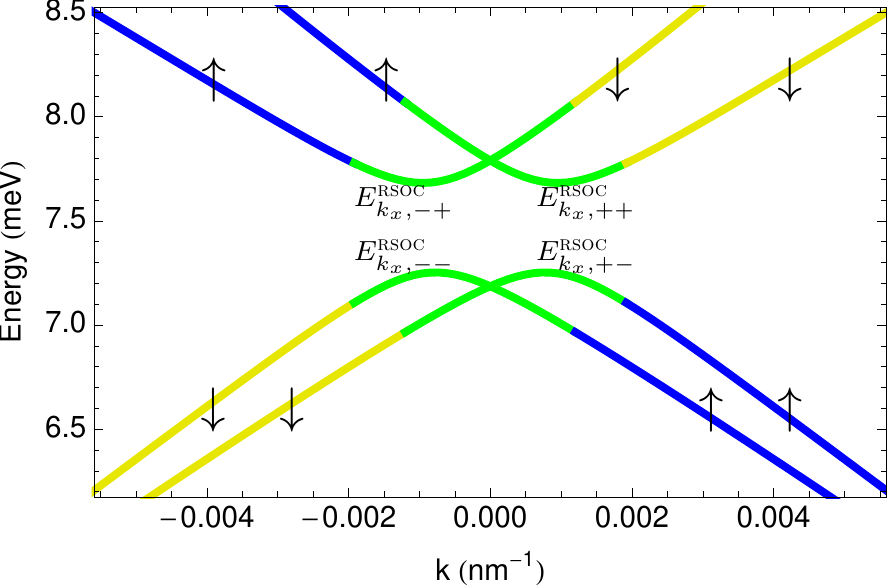}
\caption{\label{fig:ribbon-dispersions-anticross} The ribbon energy dispersions $E^{\textsc{rsoc}}_{k_x,s \tau}$ Eq.(\ref{eq:energy-rsoc-ribbon}) and the spin-structure of the GHESs close to the lower edge at $y=-W/2$. The combination of finite width and RSOC produce splitting in $k_x$-space and energy gaps at $k_{\textrm{gap}}\neq0$. The spin-structure associated to the two avoided crossings is illustrated by the colors: The states are more than 99$\%$ pure spin $\ua$ ($\da$) on the lower edge in the blue (yellow) part of the dispersions, whereas the spin-orientation rotates when all states come close together (green regions). The upper edge at $y=W/2$ has the opposite spin-structure (see the main text). Therefore, the states become true GHESs in the green regions, which coincide with the peaks in the projection seen in Fig.~\ref{fig:projW}. The parameters for HgTe in table \ref{table:parameters} are used together with $R_0=A$ and $W=200$nm.}
\end{figure}

\begin{figure}
 \includegraphics[width=0.95\linewidth]{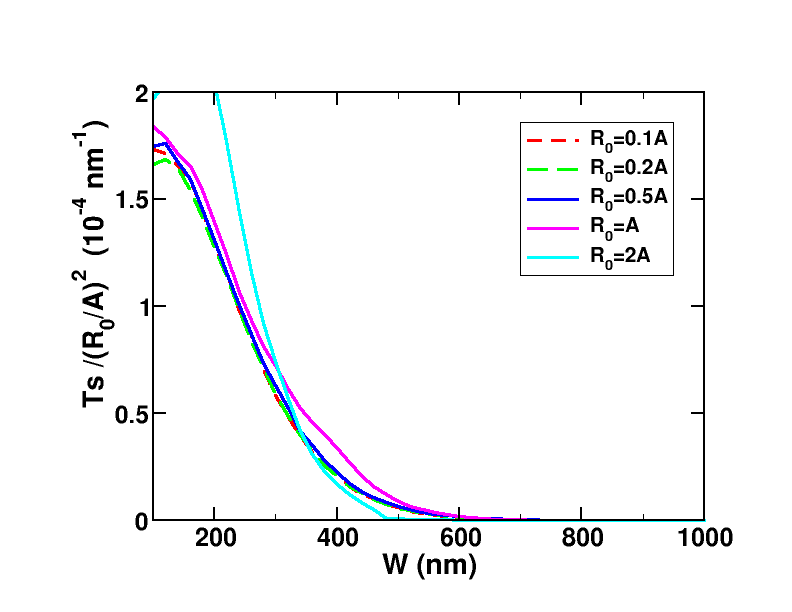}
\caption{\label{fig:integral} Numerical value of the total spin rotation $T_s$  of the GHES $\Psi_{k_x,+-}$ rescaled with $R_0^2$ as a function of the ribbon width for different values of $R_0$.}
\end{figure}

In Fig.~\ref{fig:integral}, we show that the total spin rotation $T_s$ of $\Psi_{k_x,+-}$ scales with $R_0^2$ for not too large values of $R_0$. The total spin rotation is essentially the integral of the projections in Fig.~\ref{fig:projW} due to our choice of $k_1$ in Eq.(\ref{eq:ts}). We only show the numerical results as the analytical states (\ref{two_rashba}) are only available in the small $k_x$ range, where both  $E_{k_x}^+$ and $E_{k_x}^-$ are well-defined. Nevertheless, using the analytical states to find $T_s$ very similar results are obtained for analytically feasible values of $R_0$ and $W$.  Although the maximum of the spin projection in Fig.~\ref{fig:projW} increases, the total value of the integral is reduced, when the ribbon widens to the single edge limit. The scaling with $R_0^2$ works perfectly well for $R_0\lesssim0.5A$ except for very small values of the width. (Note that we only show $W>100$nm in Fig.~\ref{fig:integral}.) For larger values of $R_0$, the scaled total spin rotation $T_s/R_0^2$ increases compared to the values of $R_0\leq0.5A$. However, for very large spin-orbit couplings (like $R_0=2A$ in Fig. 8) and large widths, we obtain smaller $T_s/R_0^2$ probably due to the reduced bulk gap of the system. 

\section{Summary}
\label{conclusions}

We have analyzed the spin-structure of the generic helical edge states appearing at the boundary of 2D TIs without axial spin symmetry. For the usual helical edge states in a 2D TI, the spin and propagation direction are locked in such a way that the spin-orientation is energy independent. However, for the GHESs the spin-orientation varies with energy or equivalently momentum $k$. This is possible in systems without axial spin symmetry, broken for instance by spin-orbit coupling. Importantly, time reversal symmetry still ensures counterpropagating states to be Kramers partners with orthogonal spins, but the spin-orientation of neighbouring states with different energies are not identical. This opens the possibility of inelastic scattering and thereby deviations from quantized conductance.\cite{Schmidt-PRL-2012} 

Our study is focused on the GHESs produced by Rashba spin-orbit coupling within the BHZ model. We use HgTe QWs and InAs/GaSb double QWs as concrete examples. We analyze two situations: (i) a pair of GHESs at an isolated edge and (ii) the two pairs of GHESs in a ribbon with two parallel boundaries.  In both cases, we employ an analytical approach, where the GHESs \emph{with} RSOC are found within a reduced basis consisting of the HESs \emph{without} RSOC. This is a good approximation, since the bulk and edge states are usually well separated spatially --- especially for small $k$ within the bulk energy gap. We also use a numerical tight-binding regularization of the BHZ model including RSOC to verify the analytical approach and, moreover, obtain independent valuable information.   

For an isolated boundary, our analytical approach gives rise to a $2\times2$ Hamiltonian Eq.(\ref{eq:full-non-diagonal-H}), which is formally equivalent to a simple 1D model of a pair of HESs with a phenomenological spin-orbit coupling. From this analogy, we discover that GHESs are produced, when the effective spin-orbit coupling term is nonlinear in the momentum. In contrast, no GHESs appear for a linear effective spin-orbit coupling term within our framework. Moreover, we find the effective RSOC $\al_{k_y}$ in terms of the BHZ parameters. We also obtain the pair of GHESs in Eq.(\ref{eq:eigenstate-single-edge-with_R0}), where the velocity has been renormalized. Using our insights into linear versus nonlinear effective RSOC terms, we are able to provide an explicit expression for the so-called spin-structure parameter $k_0$, which measures the amount of spin-orientation variation for small $k$. The spin-structure parameter $k_0$ was originally deduced by symmetry arguments\cite{Schmidt-PRL-2012} and it is interesting to have an expression in a concrete case. For instance, it shows that $k_0$ depends on the RSOC strength $R_0$  and the Dirac mass $M_0$ in non-trivial ways. Moreover, $1/k_0^2$ vanishes when the particle-hole symmetry parameter $D$ of the BHZ model is zero. This statement is in fact more general: the effective RSOC term becomes exactly linear for $D=0$ such that only ordinary HESs appear in this case. For realistic HgTe and InAs/GaSb TIs, we observe that the spin-orientation of the edge states are quite robust against even large RSOC strengths $R_0$ for the single-edge case. Nevertheless, the spin-orientation does change slightly with energy. Moreover, we find good agreement between the numerical and analytical approaches. 

Now we turn to the case of a ribbon, where the change in the spin-orientation of the GHESs is enhanced substantially for realistic HgTe TIs. The new physical element of the ribbon compared to the isolated edge, is the coupling of the GHESs on opposite edges. This finite size effect --- even without RSOC --- produce a gap in the HES spectrum.\cite{Zhou-PRL-2008} Now combining the finite width and the RSOC, two gaps and two associated avoided crossings arise in the GHESs spectrum symmetrically around $k=0$ as  shown in Fig~\ref{fig:ribbon-dispersions-anticross}. Our analytical approach shows that the inter-edge RSOC is responsible for the avoided crossings to take place at finite momenta, which is evident from the dispersions in Eq.(\ref{eq:energy-rsoc-ribbon}). Moreover, we find the position in momentum of these gaps and their size $\Delta_{\textrm{edge}}$ versus the ribbon width. The analytical and numerical results for these quantities compare well, except at certain widths where the full numerical calculation reveals an interesting destructive interference effect. From our analytical approach, we find the GHESs including the RSOC in Eq.(\ref{two_rashba}). Remarkably, they consist of two orthogonal spinors, one on each side of the ribbon. Thus, the states are equally distributed on the two parallel edges. The states become true GHESs with a sizable variation in the spin-orientation close to the two avoided crossings in the GHES spectrum, where all the states are close in energy. We show in Fig.~\ref{fig:projW} that the region in $k$-space of sizable spin-orientation variation becomes wider, if the ribbon becomes narrower. On the other hand, widening the ribbon increases the maximal value of the projection, which measures the change in spin-orientation. To quantify this further, we find the total spin rotation $T_s$ Eq.(\ref{eq:ts}), which is related to the integral of the spin-orientation variation over the entire region of $k$-space. The numerical calculations show that the total spin rotation decreases with the ribbon width and, moreover, that  $T_s\propto R_0^2$ for values of $R_0 \lesssim 0.5A$. 

Our analytical GHESs for both the isolated edge and the ribbon open the possibility to study other effects in the presence of RSOC. For instance, scattering of magnetic impurities or the nuclear spins in the crystal could be studied. Furthermore, it would be interesting to explore the transport properties of a ribbon, since we found a significant spin-orientation change.

\acknowledgements

This work has been funded by Spanish Government projects: FIS2012-33152, FIS2012-34479, MAT2014-58241-P and CAM research consortium QUITEMAD+ S2013/ICE-2801. AML acknowledges A.~Fernandez Romero and financial support from the Carlsberg Foundation.

\appendix

\section{The time-reversal operator within the BHZ framework}\label{app:TR}

This appendix provides the time-reversal operator $\Theta$ for a wave function expanded in terms of the BHZ basis states, $\ket{E\pm}$ and $\ket{H\pm}$. The time-reversal operator is only defined up to a phase factor and works differently in different bases, so it is important to keep the basis fixed throughout a calculation.\cite{Sakurai-modern-BOOK} Here we use $\Theta=-i\sigma_y K$, where $\sigma_y$ is a Pauli matrix in spin-space and $K$ is the operator for complex conjugation.  With this definition of $\Theta$ and by writing the BHZ basis states within the envelope function approximation, one obtains
\begin{subequations}
\begin{align}
\Theta\ket{E\pm}&=\mp\ket{E\mp},\\
\Theta\ket{H\pm}&=\mp\ket{H\mp},
\end{align}    
\end{subequations}
see Appendix A of Ref.~\onlinecite{Lunde-PRB-2013} for a deviation. Therefore, the Kramers partner of some wave function $\varphi(x,y)$ written in the BHZ basis $\{\ket{E+},\ket{H+},\ket{E-},\ket{H-}\}$ is
\begin{align}\label{eq:TR-operator-def}
\Theta\varphi(x,y)=
\Theta\left(
\begin{array}{c}
\varphi_{E+}^{}(x,y)\\
\varphi_{H+}^{}(x,y)\\
\varphi_{E-}^{}(x,y)\\
\varphi_{H-}^{}(x,y)
\end{array}
\right)=
\left(
\begin{array}{c}
\varphi_{E-}^{\ast}(x,y)\\
\varphi_{H-}^{\ast}(x,y)\\
-\varphi_{E+}^{\ast}(x,y)\\
-\varphi_{H+}^{\ast}(x,y)
\end{array}
\right)
\end{align}    
and we get $\Theta^2\varphi(x,y)=-\varphi(x,y)$ as expected.

\section{The helical edge states without Rashba spin-orbit interaction}\label{appendix:HES}

In this appendix, we provide (i) details on the method used to obtain the HESs without the RSOC within the BHZ model and (ii) the HESs obtained for an isolated edge and a finite width ribbon.

\subsection{On the derivation of the helical edge states}\label{subsec:HES-derivation}

Various methods have been used to study the HESs at the boundary of a TI.\cite{Murakami-PRB-2007,Berry-PRSLA-1987,Zhou-PRL-2008} Here we follow Zhou \emph{et al.}\cite{Zhou-PRL-2008} and simply set the wave function to zero at the boundary of the TI, which is possible despite the Dirac-like nature of the BHZ model due to the second order derivatives.\cite{Flindt-NJP-2009}

Now we provide the overall steps of the derivation in Ref.~\onlinecite{Zhou-PRL-2008}. The block diagonal form of the BHZ hamiltonian (\ref{eq:BHZ-H}) allows one to solve the two blocks separately. Mathematically, each block leads to a homogeneous system of two coupled linear ordinary differential equations with spatially-independent coefficients. The upper block gives the following system of differential equations
\begin{widetext}
\begin{align}\label{eq:diff-system}
\left(\begin{array}{cc}
 M_0-B_+(-\p^2_x+k_y^2) & A(-i\p_x+ik_y) \\
 A(-i\p_x-ik_y) & -[M_0-B_-(-\p^2_x+k_y^2)] 
\end{array}
\right)
\left(\begin{array}{c}
 \varphi_{E+,E}(x,k_y) \\
 \varphi_{H+,E}(x,k_y) 
\end{array}\right)
=E
\left(
\begin{array}{c}
 \varphi_{E+,E}(x,k_y) \\
 \varphi_{H+,E}(x,k_y) 
\end{array}\right),
\end{align}
\end{widetext}
where $B_{\pm}=B\pm D$. For simplicity, we assume translational symmetry along the $y$-axis such that $k_y$ is a good quantum number. In contrast, we use broken translational symmetry along the $x$-axis, so $k_x=-i\p_x$ by the Peierls substitution. In other words, Eq.(\ref{eq:diff-system}) is for one or more edges parallel to the $y$-axis. This can be varied at will to study the HESs of any geometric structure.\cite{Michetti-PRB-2011} Here, the real-space wave function is 
\begin{align}
\varphi_E(x,y)=
e^{ik_yy}\varphi_E(x,k_y)=
e^{ik_yy}
\left(
\begin{array}{c}
 \varphi_{E+,E}(x,k_y) \\
 \varphi_{H+,E}(x,k_y) 
\end{array}
\right),
\nonumber
\end{align}
where the zeros in the two last components of the entire four-vector are implicit, i.e. $\psi_{\ua,E}=[\varphi_E,0,0]^T$. The coupling of the two blocks of $H_0$ is in fact the difficulty that appears by trying to include the RSOC exactly, since four coupled linear differential equations appear. 

The mathematical method to solve this kind of system of differential equations (\ref{eq:diff-system}) is to substitute $\varphi_E(x,k_y)$ by the ansatz $e^{\lambda x}\phi_{\lambda}$ and find all possible values of $\lambda$. Importantly, the vector $\phi_{\lambda}$ is independent of $x$. Since the system of differential equations (\ref{eq:diff-system}) is linear, the general solution is a linear combination of all possible ansatz solutions, $e^{\lambda_i x}\phi_{\lambda_i}$, weighted by $c_i$, i.e.
\begin{align}
\varphi_E(x,k_y)=
\sum_i
c_i
e^{\lambda_i x}\phi_{\lambda_i}. 
\end{align} 

To find all possible $\lambda_i$, the weights $c_i$ and the dispersion relations $E$, we use the boundary condition(s) and the normalization of the wave functions. The wave function is set to zero at the boundaries.\cite{Zhou-PRL-2008} Thus, the boundary condition for the isolated edge of the half-plane $x>0$ is
\begin{subequations}\label{eq:boundary}
\begin{align}
&\varphi_E(x=0,k_y)
= 
\left(
\begin{array}{c}
0\\
0
\end{array} 
\right),
\end{align}
and boundary conditions for the ribbon of width $W$ are
\begin{align}
&
\varphi_E(x=\pm W/2,k_y)
= 
\left(
\begin{array}{c}
0\\
0
\end{array} 
\right).
\end{align}
\end{subequations}
Moreover, the HESs are by definition not extended into the bulk, so we require them to be bounded and normalized in the direction perpendicular to the edge(s), i.e. 
\begin{align}\label{norma}
& \int 
dx |\varphi_E(x,k_y)|^2 =1, 
\end{align}
where the integral goes from $0$ to $\infty$ in the case of an isolated edge and from $-W/2$ to $W/2$ for a ribbon. Therefore, we now have the necessary equations to find all possible $\lambda_i$ and $c_i$ and the corresponding energy dispersions for the HESs at an isolated boundary and for a ribbon.

\subsection{HESs at an isolated boundary}\label{subsec-Appendix:HES-isolated-boundary}

\begin{figure}
\includegraphics[width=0.9\linewidth]{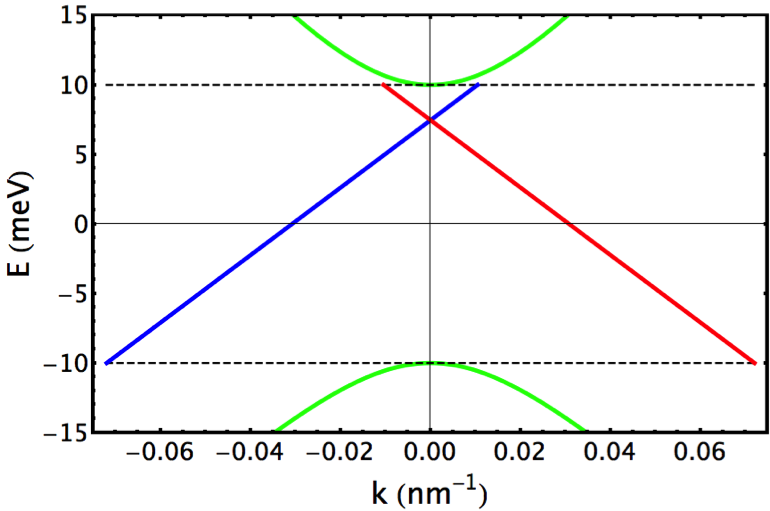}
\includegraphics[width=0.9\linewidth]{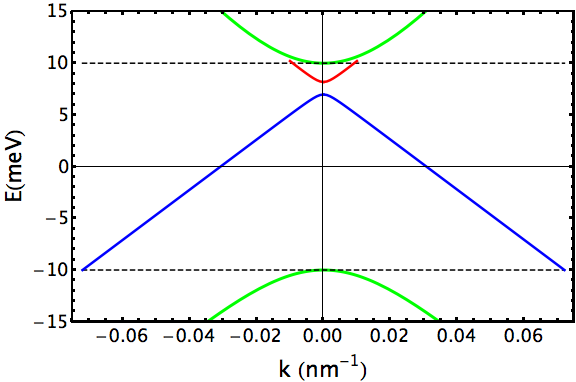}
\caption{
The HES dispersions (red/blue) and bulk bands (green) without the RSOC for an isolated edge (top panel) and for a 200nm wide ribbon (bottom panel). The ribbon dispersions $E^+_{k_x}$ (red) and $E^-_{k_x}$ (blue) are spin-degenerate, since they include the HESs on both sides of the ribbon.  The BHZ parameters for HgTe in Table \ref{table:parameters} have been used.
\label{fig:dispersion_noR}}
\end{figure}

\begin{figure}
\includegraphics[width=0.9\linewidth]{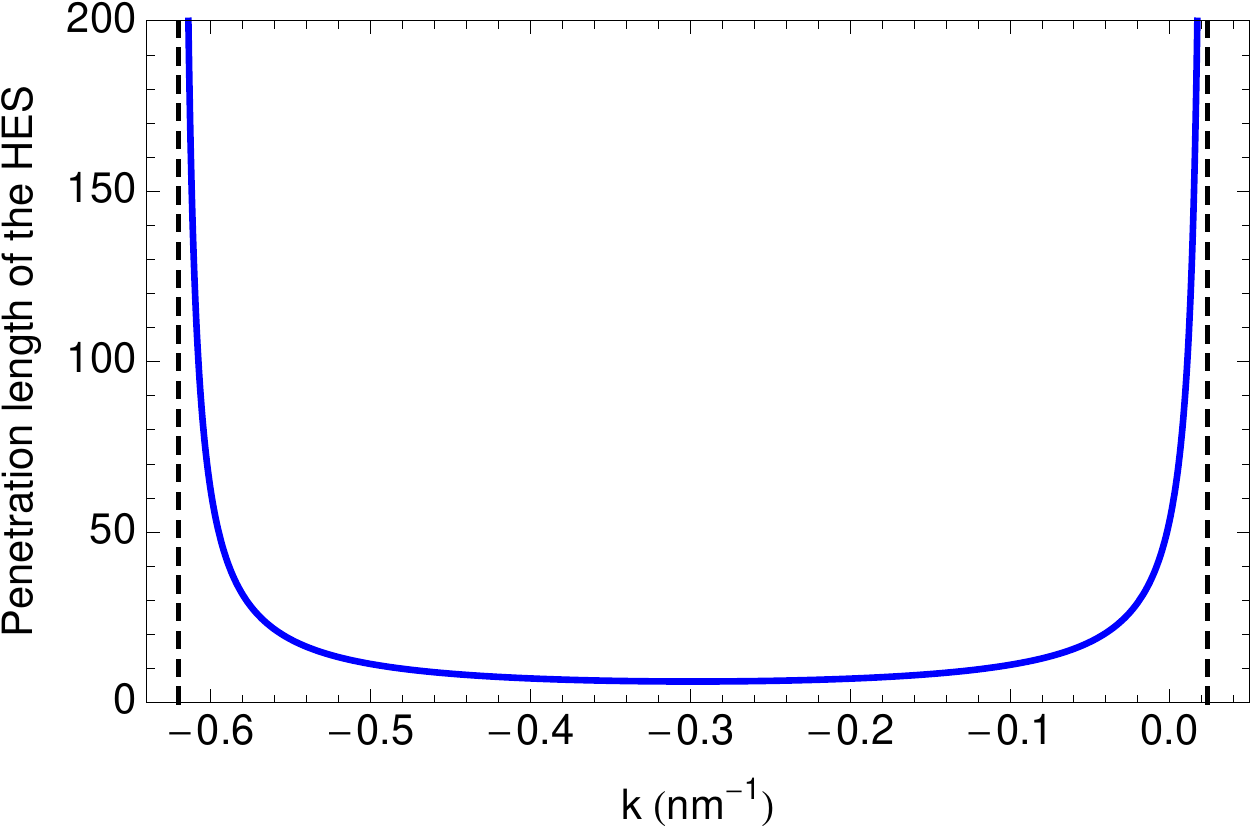}
\includegraphics[width=0.9\linewidth]{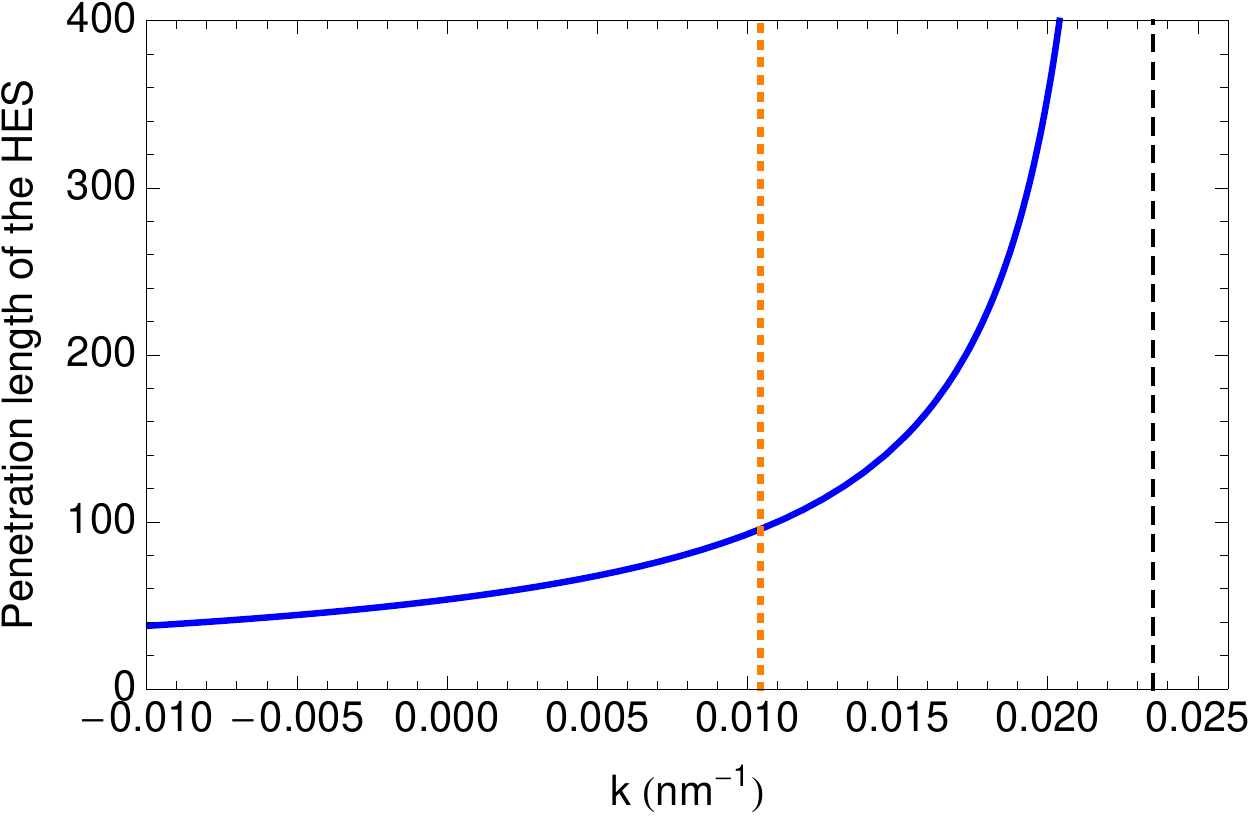}
\caption{The penetration length $1/\la_2$  Eq.(\ref{eq:pene-length}) of the HES $\psi_{k_y\ua}$ into the bulk of the TI. (The penetration length for the other HES $\psi_{k_y\da}$ is found by the replacement $k_y\rightarrow-k_y$.) The upper figure shows the large region of finite penetration length and the points where it diverges, $\la_2=0$, indicated by vertical dashed black lines. The lower figure focus on the region close to the upper edge of the bulk band gap shown by the vertical orange dotted line. A region of coexistence of edge and bulk bands is clearly seen (between the vertical orange dotted and black dashed lines), even though the HESs widens and gradually looses the localization characteristic of an \emph{edge} state. Nevertheless, this facilitates the use of our analytical theory in Sec.~\ref{subsec:Semi} beyond the boundaries of the bulk band gap, since it requires the presence of both HESs without RSOC. The parameters for HgTe in Table \ref{table:parameters} are used.
\label{fig:penetration-length}}
\end{figure}

Applying the method presented above, one can find the pair of HESs without RSOC located at the boundary of the half-plane $x>0$ to be
\begin{subequations}\label{eq:HES_single_edge_app}
\begin{align}
\psi_{k_y\ua}(x,y)
&=
\frac{1}{\sqrt{L}}
e^{ik_yy}
g_{k_y}(x)\hat{\phi}_\ua, 
\\
\psi_{k_y\da}(x,y)
&=
\frac{1}{\sqrt{L}}
e^{ik_yy}
g_{-k_y}(x)
\hat{\phi}_\da, 
\end{align}
\end{subequations}
as presented in Eq.(\ref{eq:HES-R0-zero}) of the main text, however, without the complete specification given below. The energy dispersion relations for the HESs (\ref{eq:HES_single_edge_app}) are
\begin{align}\label{eq:dispersions-isolated-edge-app}
E_{\ua k_y}&=
E_0+\h v k_y
\quad \textrm{and}\quad
E_{\da k_y}&=
E_0-\h v k_y
\end{align}
as seen in Fig.~\ref{fig:dispersion_noR}. Both the velocity $v=-\sqrt{B^2-D^2}\frac{|A|}{\h B}$ and $E_0=-\frac{M_0D}{B}$ are positive for the parameters in Table \ref{table:parameters}. Interestingly, the dispersions are exactly linear for an isolated edge.\cite{Wada-PRB-2011} The $k_y$-independent spinors $\hat{\phi}_{\sigma}$ are
\begin{align}\label{eq:spinors-y-direction}
&\hat{\phi}_{\ua}=
\mathfrak{n}
\left(
\begin{array}{c}
 -i\frac{A}{|A|}\\
 \frac{\sqrt{B_+B_-}}{B_-}\\
 0\\
 0
\end{array}
\right),
&\hat{\phi}_{\da}=
\mathfrak{n}
\left(
\begin{array}{c}
 0\\
 0\\
 +i\frac{A}{|A|}\\
 \frac{\sqrt{B_+B_-}}{B_-}
\end{array}
\right),
\end{align} 
where $B_\pm=B\pm D$ and $\mathfrak{n}=\sqrt{B_-/(2B)}$. The real and normalized transverse wave function $g_{k_y}(x)$ is
\begin{align}\label{eq:transverse-HES-single-edge}
g_{k_y}(x)&=
\sqrt{\frac{2\lambda_1\lambda_2(\lambda_1+\lambda_2)}{(\lambda_1-\lambda_2)^2}}
\left(e^{-\lambda_1 x}-e^{-\lambda_2 x}\right),
\end{align} 
where the length scale $1/\lambda_2$ is the penetration length of the HES into the bulk of the TI. Moreover, the $k_y$-dependence of $g_{k_y}$ is in $\lambda_{1}$ and $\lambda_{2}$ as
\begin{subequations}\label{eq:lambdas-single-edge}
\begin{align} 
\lambda_1
=&\frac{1}{\sqrt{B_-B_+}} \left(\frac{|A|}{2}+\sqrt{Z_{k_y}}\right),
\\
\lambda_2
=&\frac{1}{\sqrt{B_-B_+}} \left(\frac{|A|}{2}-\sqrt{Z_{k_y}}\right),
\label{eq:pene-length}
\end{align} 
\end{subequations}
where we defined
\begin{align}\label{eq:def_af_Z}
Z_{k_y}=&
\left(\frac{A^2}{4}-\frac{M_0}{B}B_+B_-\right)
+\frac{D|A|\sqrt{B_+B_-}}{B}k_y\nonumber\\
&+B_+B_-k^2_y.
\end{align}
The BHZ model only host HESs in the TI regime where $M_0/B>0$. Moreover, the explicit forms of the HESs with real $\lambda_{1,2}$ presented here are found under the assumption that  $0\leq M_0/B\leq A^2/(4B^2)$, which is fulfilled for the parameters in table \ref{table:parameters}. Furthermore, the HESs are well-localized at the boundary within the bulk energy gap with a fairly small penetration length $1/\la_2$ on the order of tens of nm. However, above the upper bulk band gap edge, there is a region of coexistence of edge and bulk states before the penetration length diverges as seen in Fig.~\ref{fig:penetration-length}. Coexistence of bulk and edge states has recently been studied.\cite{Baum-PRL-2015} Finally, we remark that $\psi_{k_y\ua}$ and $\psi_{-k_y\da}$ are Kramers partners, since $\Theta\psi_{k_y\ua}(x,y)=-\psi_{-k_y\da}(x,y)$ by using Eq.(\ref{eq:TR-operator-def}).

In passing, we remark that the HESs along the perpendicular direction are different in a non-trivial way from the ones presented in Eq.(\ref{eq:HES_single_edge_app}). If we consider the HESs running along the $x$-axis instead of the $y$-axis at the boundary of the half-plane $y>0$,  then we find $\psi_{k_x\sigma}(x,y)=\frac{1}{\sqrt{L}}e^{ik_xx}g_{-\mathfrak{s}k_x}(y)\tilde{\phi}_\sigma$ and $E_{\sigma k_x}=E_0-\mathfrak{s}\h v k_x$, where $\mathfrak{s}=+(-)$ for $\sigma=\ua(\da)$. The exchange of $k_y$ by $-k_x$ is natural in order for the states to be connected correctly in e.g. the corner of the TI.\cite{Lunde-PRB-2013} A more interesting fact is that the imaginary unit $i$ disappears from the spinors, i.e.
\begin{align}
&\tilde{\phi}_{\ua}=
\mathfrak{n}
\left(
\begin{array}{c}
 \frac{A}{|A|}\\
 \frac{\sqrt{B_+B_-}}{B_-}\\
 0\\
 0
\end{array}
\right),
&\tilde{\phi}_{\da}=
\mathfrak{n}
\left(
\begin{array}{c}
 0\\
 0\\
 \frac{A}{|A|}\\
 \frac{\sqrt{B_+B_-}}{B_-}
\end{array}
\right),
\end{align} 
where $\mathfrak{n}=\sqrt{B_-/(2B)}$ as in Eq.(\ref{eq:spinors-y-direction}).

\subsection{HESs for a finite width ribbon}\label{subsec:HES-ribbon}

Now we turn to the HESs for a ribbon with edges at $y=\pm W/2$ (i.e.~parallel to the $x$-axis in contrast to the case in Eq.(\ref{eq:HES_single_edge_app}))  as considered by Zhou \emph{et al.}\cite{Zhou-PRL-2008} By use of the ansatz function $e^{\la y}\phi_\la$  and the boundary conditions, the energy dispersions $E^{e=\pm}_{k_x}$ discussed in Sec.~\ref{subsec:twoedges} become the solutions to the following implicit equation
\begin{equation}\label{eq:implicit-eq-for-energies-for-finite-width}
\frac{\tanh\!\big[\frac{\la_1W}{2}\big]}{\tanh\!\big[\frac{\la_2W}{2}\big]}
{+}\frac{\tanh\!\big[\frac{\la_2W}{2}\big]}{\tanh\!\big[\frac{\la_1W}{2}\big]}
= 
\frac{\al_{1}^2\la_{2}^2+\al_{2}^2\la_{1}^2-k_x^2(\al_1{-}\al_2)^2}{\al_{1}\al_{2}\la_{1}\la_{2}},
\end{equation}
where we have introduced
\begin{subequations}\label{eq:lambda-ribbon}
\begin{align}
\la_{1}^2&=
k_x^2+F+\sqrt{F^2-\frac{M_0^2-E^2}{B^2-D^2}},
\\
\la_{2}^2&=
k_x^2+F-\sqrt{F^2-\frac{M_0^2-E^2}{B^2-D^2}},
\\
\alpha_{j}&=E-M_0+B_+(k_x^2-\la_{j}^2)
\quad\textrm{for}\quad j=1,2,
\end{align}
\end{subequations}
and $F=\big[A^2-2(M_0B+ED)\big]/(2B_+B_-)$. As in the previous section, we assume that $\la_{1,2}$ are real and in particular define $\la_{1,2}$ to be the positive root in Eq.(\ref{eq:lambda-ribbon}), i.e.~$\la_{1,2}>0$. Here $\la_{1,2}$ are not identical to the ones for an isolated edge in Eq.(\ref{eq:lambdas-single-edge}), since the energy dispersions differ in the two cases. Moreover, the dispersions for an isolated edge Eq.(\ref{eq:dispersions-isolated-edge-app}) come out correctly in the limit $W\rightarrow\infty$, where the left-hand side of Eq.(\ref{eq:implicit-eq-for-energies-for-finite-width}) is equal to 2. The dispersions $E^\pm_{k_x}$ for a ribbon are seen in Fig.~\ref{fig:dispersion_noR}, where $E^+_{k_x}$ is the upper dispersion ($E^+_{k_x}>E^-_{k_x}$).

The four HESs $\psi^e_{k_x\sigma}$ for the ribbon are all proportional to a plane-wave running along the edges, i.e.~$\psi^e_{k_x\sigma}(x,y)\propto e^{ik_xx}$. Moreover, the spinor and the transverse wave function do not factorize in contrast to the case of an isolated edge in Eq.(\ref{eq:HES_single_edge_app}). For a ribbon, the HESs are 
\begin{subequations}\label{states_two_edges}
\begin{align}
\psi_{k_x\ua}^{+}(x,y)
&=
\tilde{c}_{+}
\frac{e^{ik_xx}}{\sqrt{L}}
\left(
\begin{array}{c}
f_{+}-\gamma^{+}_{k_x}f_{-}\\ 
\gamma^{+}_{k_x}\eta_{2}^{+} f_{+}-\eta_{1}^{+}f_{-}\\
0\\ 
0
\end{array} 
\right),
\\
\psi_{k_x\ua}^{-}(x,y)
&=
\tilde{c}_{-}
\frac{e^{ik_xx}}{\sqrt{L}}
\left(\!
\begin{array}{c}
-\gamma^{-}_{k_x}f_{+}+f_{-}\\
-\eta_{2}^{-}f_{+}+\gamma^{-}_{k_x}\eta_{1}^{-}f_{-}\\
0\\
0
\end{array} 
\right),
\\
\psi_{k_x\da}^{+}(x,y)
&=
-\tilde{c}_{+}
\frac{e^{ik_xx}}{\sqrt{L}}
\left(\!
\begin{array}{c}
0\\ 
0\\ 
f_{+}+\gamma^{+}_{k_x}f_{-}\\ 
-\gamma^{+}_{k_x}\eta_{2}^{+}f_{+}-\eta_{1}^{+}f_{-}
\end{array} 
\!
\right)\!,
\\
\psi_{k_x\da}^{-}(x,y)
&=
-\tilde{c}_{-}
\frac{e^{ik_xx}}{\sqrt{L}}
\left(\!
\begin{array}{c}
0\\ 
0\\
\gamma^{-}_{k_x}f_{+}+f_{-}\\
-\eta_{2}^{-}f_{+}-\gamma^{-}_{k_x}\eta_{1}^{-}f_{-}
\end{array} 
\!\right)\!.
\end{align}
\end{subequations}
All the spatial dependence of $\psi_{k_x\sigma}^{e}(x,y)$ are in the functions $f_{\pm}\equiv f_{\pm}(y,k_x,E)$, where the subscript denotes the parity, i.e.~$f_{\pm}(-y,k_x,E)=\pm f_{\pm}(y,k_x,E)$. These are 
\begin{subequations}\label{eq:def-af-f+-f-}
\begin{align}
f_{+}(y,k_x,E)=\left( \frac{\cosh(\la_{1} y)}{\cosh\big(\frac{\la_1W}{2}\big)}-\frac{\cosh(\la_{2} y)}{\cosh\big(\frac{\la_2W}{2}\big)}\right),
\\
f_{-}(y,k_x,E)=\left( \frac{\sinh(\la_{1} y)}{\sinh\big(\frac{\la_1W}{2}\big)}
-\frac{\sinh(\la_{2} y)}{\sinh\big(\frac{\la_2W}{2}\big)}\right), 
\end{align}
\end{subequations}
which vanish on the boundaries $y=\pm W/2$. Here both $k_x$ and $E$ are explicitly written as variables in $f_\pm$ in order to keep track of which dispersion is used, $E^+_{k_x}$ or $E^-_{k_x}$. The HESs (\ref{states_two_edges}) also include the following space-independent quantities 
\begin{align}
\eta_{1}^{\pm}&=
\frac{\al_{2}-\al_{1}}{A\big(\la_{1}\coth\!\big[\frac{\la_1W}{2}\big]
-\la_{2}\coth\!\big[\frac{\la_2W}{2}\big]\big)}
\Bigg|_{E=E_{k_x}^{\pm}},
\nonumber\\
\eta_{2}^{\pm}&=
\frac{\al_{2}-\al_{1}}{A\big(\la_{1}\tanh\!\big[\frac{\la_1W}{2}\big]-\la_{2}\tanh\!\big[\frac{\la_2W}{2}\big]\big)}
\bigg|_{E=E_{k_x}^{\pm}}\!\!,
\nonumber\\
\gamma^{+}_{k_x}&=
\frac{(\al_{2}-\al_{1})k_x}{\al_{2}\la_{1}\tanh\!\big[\frac{\la_1W}{2}\big]
-\al_{1}\la_{2}\tanh\!\big[\frac{\la_2W}{2}\big]}\frac{\eta_1^{+}}{\eta_2^{+}}
\bigg|_{E=E^+_{k_x}},
\nonumber\\
\gamma^{-}_{k_x}&=
\frac{(\al_{2}-\al_{1})k_x}{\al_{2}\la_{1}\coth\!\big[\frac{\la_1W}{2}\big]
-\al_{1}\la_{2}\coth\!\big[\frac{\la_2W}{2}\big]}\frac{\eta_2^{-}}{\eta_1^{-}}
\bigg|_{E=E^-_{k_x}},
\nonumber
\end{align}
which all depend on $k_x$ and the energy dispersions $E=E^\pm_{k_x}$. We remark that some of the signs in the HESs presented in Eq.(\ref{states_two_edges}) are not identical to the ones found in Ref.~\onlinecite{Zhou-PRL-2008}. The reason is that Zhou \emph{et al.}\cite{Zhou-PRL-2008} have the opposite sign in front of $k_y$ in the BHZ Hamiltonian (\ref{eq:BHZ-H}) compared to one used here and in e.g.~Ref.~\onlinecite{Rothe-NJP-2010}. (In fact, the sign of $k_y$ in the BHZ hamiltonian varies throughout the literature.) 

Finally, we find the normalization constants to be
\begin{align}
\tilde{c}_{+}&=
\frac{1}{\sqrt{\Gamma_{++}^+\big[1+(\gamma_{k_x}^+)^2(\eta_2^+)^2\big]
+\Gamma_{++}^-\big[(\gamma_{k_x}^+)^2+(\eta_1^+)^2\big]}},
\nonumber\\
\tilde{c}_-&=
\frac{1}{\sqrt{\Gamma_{--}^+\big[(\gamma_{k_x}^-)^2+(\eta_2^-)^2\big]
+\Gamma_{--}^-\big[1+(\gamma_{k_x}^-)^2(\eta_1^-)^2\big]}},
\nonumber 
\end{align}
where we introduced
\begin{align}\label{eq:def-Gamma}
\Gamma^{\tau}_{ee'}=\int^{W/2}_{-W/2} dy\;
f_{\tau}(y,k_x,E_{k_x}^e)f_{\tau}(y,k_x,E_{k_x}^{e'}).
\end{align}
The expressions for $\Gamma^+_{\pm\pm}$ and $\Gamma^-_{\pm\pm}$ are given in Eqs.(\ref{eq:Gamma_+_pmpm}) and (\ref{eq:Gamma_-_pmpm}), respectively. 
The integral over two functions of opposite parity is zero, $\int dy f_+f_-=0$, so there is no need to include this possibility in the definition of $\Gamma^{\tau}_{ee'}$.

For the HESs of a ribbon, the Kramers partner of $\psi_{k_x\ua}^\pm$ is $\psi_{-k_x\da}^\pm$, since $\Theta\psi_{k_x\ua}^\pm(x,y)=\psi_{-k_x\da}^\pm(x,y)$ by the help of Eq.(\ref{eq:TR-operator-def}). To find this result, we use that the energy dispersion is even in $k_x$, $E^\pm_{-k_x}=E^\pm_{k_x}$, as seen from Eq.(\ref{eq:implicit-eq-for-energies-for-finite-width}). This in terms leads to $\la_{1,2}$ in Eq.(\ref{eq:lambda-ribbon}) and  $\eta_{1,2}^\pm$ being even in $k_x$ and finally that $\gamma^\pm_{-k_x}=-\gamma^\pm_{k_x}$.

\section{Details of the calculation with RSOC}\label{appendix:Rashba}

In this paper, we find the analytical forms of the GHESs in the presence of RSOC by assuming that the GHESs can be written as combinations of the HESs without RSOC. Therefore, we diagonalized the RSOC $\mathcal{H}_R$ in two bases, namely  $\{ \ket{\psi_{k_x\ua}}, \ket{\psi_{k_x\da}}\}$ for an isolated edge and $\{ | \psi_{k_x\ua}^{+}\rangle, \,   | \psi_{k_x\da}^{-}\rangle,\,  | \psi_{k_x\ua}^{-}\rangle,\,| \psi_{k_x\da}^{+}\rangle\}$ for a ribbon. Here we provide various technical details for these calculations left out in the main text.

\subsection{Details for the case of an isolated edge}\label{subsec:details-RSOC-isolated}

In Eq.(\ref{eq:alpha-ky}) in Sec.~\ref{subsec:Semi}, we only give the effective RSOC $\al_{k_y}$ up to second order in $k_y$. However, the exact result can easily be found to be
\begin{align}\label{eq:alpha-single-edge-exact}
\alpha_{k_y}&=
\frac{\langle\psi_{k_y\ua}|H_{R}|\psi_{k_y\da}\rangle}{k_y}
\nonumber\\
&=
R_0\frac{B-D}{2Bk_y}
\int_{0}^{\infty}\!\! dx
g_{k_y}(x)\big[\p_x+k_y\big] g_{-k_y}(x)
\nonumber\\
&=
R_0\frac{B-D}{2Bk_y}
\Big(
\nu_{k_y}+k_y\xi_{k_y}
\Big),
\end{align}
where we use the transverse wave functions $g_{\pm k_y}(x)$ for an isolated edge Eq.(\ref{eq:transverse-HES-single-edge}) and introduce
\begin{subequations}
\begin{align}\label{eq:def-eta-og-nu-k}
\xi_{k_y} 
&\equiv
\int_{0}^{\infty} d x
g_{k_y}(x)g_{-k_y}(x), \\
\nu_{k_y} 
&\equiv
\int_{0}^{\infty} d x
g_{k_y}(x)\p_x g_{-k_y}(x).
\end{align}
\end{subequations}
Here it is evident that $\alpha_{-k_y}=\alpha_{k_y}$, since $\xi_{-k_y}=\xi_{k_y}$ and $\nu_{-k_y}=-\nu_{k_y}$ by using partial integration. Moreover, we find that $\xi_{k_y=0}=1$ due to the normalization of $g_{k_y}$. The full expressions for  $\xi_{k_y}$  and $\nu_{k_y} $ are
\begin{align}
\xi_{k_y}=h(k_y)\omega(k_y) 
\qquad\textrm{and}\qquad 
\nu_{k_y} =h(k_y)\beta(k_y),
\end{align}
where 
\begin{align}
h(k_y)=&
2\sqrt{\frac{\lambda_{1}^{-}\lambda_{2}^{-} (\lambda_{1}^{-}+\lambda_{2}^{-})}{(\lambda_{1}^{-}-\lambda_{2}^{-})^2}}
\sqrt{\frac{\lambda_{1}^{+}\lambda_{2}^{+} (\lambda_{1}^{+}+\lambda_{2}^+)}{(\lambda_{1}^{+}-\lambda_{2}^{+})^2}}, 
\label{eq-def-h}
\\
\omega(k_y)=&
\frac{1}{\lambda_{1}^{-}+\lambda_{1}^{+}}
-\frac{1}{\lambda_{2}^{-}+\lambda_{1}^{+}}
-\frac{1}{\lambda_{1}^{-}+\lambda_{2}^{+}}
+\frac{1}{\lambda_{2}^{-}+\lambda_{2}^{+}},
\nonumber\\
\beta(k_y)=&
\lambda_{1}^{-} 
\left(
\frac{1}{\lambda_{1}^{-}+\lambda_{2}^{+}}
-\frac{1}{\lambda_{1}^{-}+\lambda_{1}^{+}}
\right)
\nonumber\\
&+\lambda_{2}^{-} 
\left(\frac{1}{\lambda_{1}^{+}+\lambda_{2}^{-}}
-\frac{1}{\lambda_{2}^{-}+\lambda_{2}^{+}}
\right).\nonumber
\end{align}
Here we used the shorthand notation $\lambda_i^{\pm}=\lambda_i(\pm k_y)$, where $\la_{1,2}$ are given in Eq.(\ref{eq:lambdas-single-edge}). Note that $\la_{1,2}$ are not even functions of $k_y$ for an isolated edge in contrast to $\la_{1,2}$ for a ribbon. The reason is that the dispersions are not even in $k_y$ for an isolated edge as they are for a ribbon. For the parameters in Table \ref{table:parameters}, the exact result for $\alpha_{k_y}$ Eq.(\ref{eq:alpha-single-edge-exact}) is very well approximated by the second order expansion given in Eq.(\ref{eq:alpha-ky}) in the main text. 

In the limit of a particle-hole symmetric BHZ Hamiltonian, i.e.~$D=0$, we have $g_{k_y}(x)=g_{-k_y}(x)$, since $Z^{(D=0)}_{k_y}$ Eq.(\ref{eq:def_af_Z}) and thereby also $\la_{1,2}^{(D=0)}$ Eq.(\ref{eq:lambdas-single-edge}) become even. This means that $\xi^{D=0}_{k_y}=1$ for all $k_y$ by normalization of $g_{k_y}$. Moreover, $g_{k_y}(x)=g_{-k_y}(x)$ forces $\nu^{(D=0)}_{k_y}$ to be even, but we already found $\nu_{k_y}$ to be generally odd, so we have to conclude that $\nu^{(D=0)}_{k_y}=0$. Therefore, we are left with the exact result 
\begin{align}\label{eq:alpha-D_equal_0}
\alpha_{k_y}^{(D=0)}=R_0/2.
\end{align} 
As discussed in the main text (below Eq.(\ref{eq:k0-general})), such an $k_y$-independent $\alpha_{k_y}$ means that the spin-orientation remain fixed for $R_0\neq0$, so no GHESs appear for $D=0$. 

\subsection{Details for the ribbon calculation}\label{subsec:ribbon-details}

In this section, we provide the matrix elements of the RSOC $H_R$ between the HESs $\psi^e_{k_x\sigma}$ for a ribbon, which were used --- but not given explicitly --- in Sec.~\ref{subsec:twoedges}. 

Using the HESs $\psi^e_{k_x\sigma}$ Eq.(\ref{states_two_edges}), we find the matrix elements to be diagonal in $k_x$ due to the plane-waves running along the edges, $\psi^e_{k_x\sigma}\propto e^{ik_xx}$. Moreover,  $H_R$ Eq.(\ref{Rashba_hamil}) only couples opposite spins, i.e.
\begin{align}
&\langle\psi_{k_x\sigma}^{e} | H_R| \psi_{k_x'\sigma'}^{e'}\rangle=
\nonumber\\
&
\int^{W/2}_{-W/2} \!\!\!\!\!\!\!\! dy  \int^{L/2}_{-L/2} \!\!\!\!\!\!\!dx \;
[\psi_{k_x\sigma}^{e}(x,y)]^\dag H_R \psi_{k_x'\sigma'}^{e'}(x,y)
\propto \delta_{k^{}_x,k'_x} \delta_{\overline{\sigma}\sigma'}, 
\nonumber
\end{align}
where $\bar{\sigma}$ denote the opposite spin of $\sigma$. Therefore, out of the 10 possible matrix elements of $H_R$ in Eq.(\ref{full_hamiltonian}) (accounting for the hermiticity of $H_R$), we are now left with only four different non-zero matrix elements. Moreover, we find below that $\langle\psi_{k_x\da}^{-} | H_R| \psi_{k_x\ua}^{+}\rangle=-\langle\psi_{k_x\da}^{+} | H_R| \psi_{k_x\ua}^{-}\rangle$ such that only three different non-zero matrix elements of $H_R$ remain in Eq.(\ref{full_hamiltonian}).

The \emph{inter}-edge matrix elements Eq.(\ref{eq:inter-matrix-elements}) involving HESs localized on opposite edges are
\begin{subequations}\label{eq:inter-matrix-elements-appendix}
\begin{align}
id_{+}&=
\langle\psi_{k_x\da}^{+} | H_R| \psi_{k_x\ua}^{+}\rangle 
\\
&=
-iR_0 (\tilde{c}_+)^2
\Big[ k_x\big[\Gamma^{+}_{++} -(\gamma^{+}_{k_x})^2\Gamma^{-}_{++}\big]
-2\gamma^{+}_{k_x}\Omega^{+-}_{++}
\Big],
\nonumber\\
id_{-}&=
\langle\psi_{k_x\ua}^{-} |H_R| \psi_{k_x\da}^{-}\rangle 
\\
&=
+iR_0 (\tilde{c}_-)^2
\Big[k_x\big[\Gamma^{-}_{--}-(\gamma^{-}_{k_x})^2\Gamma^{+}_{--}\big]
+2\gamma^{-}_{k_x}\Omega^{+-}_{--}\Big],
\nonumber
\end{align}
\end{subequations}
while the \emph{intra}-edge matrix element Eq.(\ref{eq:intra-matrix-element})  between HESs on the same edge is
\begin{align}\label{eq:ib-matrix-element-explicit}
&ib=
\langle\psi_{k_x\da}^{-} |H_R| \psi_{k_x\ua}^{+}\rangle 
\\
&{=}iR_0 \tilde{c}_+\tilde{c}_-
\Big[k_x(\gamma^{+}_{k_x} \Gamma^{-}_{-+}-\gamma^{-}_{k_x}\Gamma^{+}_{-+})
-\Omega^{-+}_{-+}
+\gamma^{-}_{k_x}\gamma^{+}_{k_x}\Omega^{+-}_{-+}
\Big].
\nonumber
\end{align}
These elements are written in terms of the quantities appearing in the ribbon HESs given in Sec.~\ref{subsec:HES-ribbon} and integrals involving $f_{\pm}$ in Eq.(\ref{eq:def-af-f+-f-}), namely  $\Gamma^\tau_{ee'}$ in Eq.(\ref{eq:def-Gamma}) and 
\begin{align}\label{eq:def-Omega}
\Omega^{\tau\tau'}_{ee'}=
\int^{W/2}_{-W/2}  dy\; f_{\tau}(y,k_x,E_{k_x}^e)
\partial_yf_{\tau'}(y,k_x,E_{k_x}^{e'}).
\end{align}
The parity of $f_\pm$ leads to $\Omega^{++}_{ee'}=\Omega^{--}_{ee'}=0$ and partial integration (with zero boundary term) gives $\Omega^{\tau\tau'}_{ee'}=-\Omega^{\tau'\tau}_{e'e}$.

Now we turn to the intra-edge matrix element between the HESs localized on the opposite edge compared to the HESs in $ib=\langle\psi_{k_x\da}^{-} |H_R| \psi_{k_x\ua}^{+}\rangle $ and show that we find the same result. As discussed before Eq.(\ref{eq:localization-of-HESs}), the HESs $\{\psi_{k_x\da}^{+}, \psi_{k_x\ua}^{-}\}$ are localized on the opposite edge of the HESs $\{\psi_{k_x\da}^{-}, \psi_{k_x\ua}^{+}\}$, which are used in $ib$. Direct calculation gives
\begin{align}
&\langle\psi_{k_x\da}^{+} | H_R| \psi_{k_x\ua}^{-}\rangle= 
\nonumber\\
&\  iR_0 \tilde{c}_+\tilde{c}_-
\left[k_x(\gamma^{-}_{k_x}\Gamma^{+}_{+-}-\gamma^{+}_{k_x} \Gamma^{-}_{+-})
-\Omega^{+-}_{+-}
+\gamma^{-}_{k_x}\gamma^{+}_{k_x}\Omega^{-+}_{+-}
\right]\!.\nonumber
\end{align}
Therefore, by comparing to $ib$ in  Eq.(\ref{eq:ib-matrix-element-explicit}) and using that $\Omega^{\tau\tau'}_{ee'}=-\Omega^{\tau'\tau}_{e'e}$ and $\Gamma^{\tau}_{ee'}=\Gamma^{\tau}_{e'e}$, we find
\begin{align}
\langle\psi_{k_x\da}^{+} | H_R| \psi_{k_x\ua}^{-}\rangle= 
-\langle\psi_{k_x\da}^{-} |H_R| \psi_{k_x\ua}^{+}\rangle=-ib.  
\end{align}
This is a physically sound result, since the two edges are physically equivalent.

Finally, we give the integrals $ \Gamma^{\tau}_{ee'}$ in Eq.(\ref{eq:def-Gamma}) and $\Omega^{\tau\tau'}_{ee'}$ in Eq.(\ref{eq:def-Omega}). There are six non-zero $ \Gamma^{\tau}_{ee'}$. These are
\begin{widetext}
\begin{align}\label{eq:Gamma_+_pmpm}
\Gamma^+_{\pm\pm}&=
\frac{W}{2}
\left[
\frac{1}{\cosh^2\!\Big[\frac{W\la_1^{\pm}}{2}\Big]}
+\frac{1}{\cosh^2\!\Big[\frac{W\la_2^{\pm}}{2}\Big]}
\right]
+
\frac{\la_1^\pm[3(\la_2^{\pm})^2 + (\la_1^{\pm})^2]
\tanh\!\Big[\frac{W\la_2^{\pm}}{2}\Big]
-\la_2^\pm[3(\la_1^{\pm})^2 + (\la_2^{\pm})^2]
\tanh\!\Big[\frac{W\la_1^{\pm}}{2}\Big]}
{\la_1^{\pm}\la_2^{\pm}[(\la_1^{\pm})^2 - (\la_2^{\pm})^2]}
\end{align}
where $\la_{1,2}^{\pm}=\la_{1,2}(E_{k_x}^{\pm})$ in Eq.(\ref{eq:lambda-ribbon}). Note that the notation $\la_{1,2}^\pm$ is not identical to the one used in Sec.~\ref{subsec:details-RSOC-isolated} for an isolated edge. Moreover, we obtain 
\begin{align}\label{eq:Gamma_-_pmpm}
\Gamma^-_{\pm\pm}&=
-\frac{W}{2}
\left[
\frac{1}{\sinh^2\!\Big[\frac{W\la_1^{\pm}}{2}\Big]}
+\frac{1}{\sinh^2\!\Big[\frac{W\la_2^{\pm}}{2}\Big]}
\right]
+
\frac{\la_1^\pm[3(\la_2^{\pm})^2 + (\la_1^{\pm})^2]
\coth\!\Big[\frac{W\la_2^{\pm}}{2}\Big]
-\la_2^\pm[3(\la_1^{\pm})^2 + (\la_2^{\pm})^2]
\coth\!\Big[\frac{W\la_1^{\pm}}{2}\Big]}
{\la_1^{\pm}\la_2^{\pm}[(\la_1^{\pm})^2 - (\la_2^{\pm})^2]},
\end{align}
which is related to $\Gamma_{\pm\pm}^{+}$ by exchanging $\cosh^2(W\la_i^\tau/2)\rightarrow-\sinh^2(W\la_i^\tau/2)$ and $\tanh(W\la_i^\tau/2)\rightarrow\coth(W\la_i^\tau/2)=1/\tanh(W\la_i^\tau/2)$. We remark that $\Gamma^\tau_{\pm\pm}$ are invariant under interchange of $\la_1^{\pm}$ and $\la_2^{\pm}$. The $\Gamma_{ee'}^{\tau}$ with $e\neq e'$ are
\begin{align}
\Gamma^+_{-+}=
\Gamma^+_{+-}&=
\frac{2\la^-_1\tanh\!\Big[\frac{W\la^-_1}{2}\Big]-2\la^+_1\tanh\!\Big[\frac{W\la^+_1}{2}\Big]}{(\la_1^-)^2-(\la_1^+)^2}
+
\frac{2\la^-_2\tanh\!\Big[\frac{W\la^-_2}{2}\Big]-2\la^+_2\tanh\!\Big[\frac{W\la^+_2}{2}\Big]}{(\la_2^-)^2-(\la_2^+)^2}
\nonumber\\
&\ -
\frac{2\la^+_1\tanh\!\Big[\frac{W\la^+_1}{2}\Big]-2\la^-_2\tanh\!\Big[\frac{W\la^-_2}{2}\Big]}{(\la_1^+)^2-(\la_2^-)^2}
-
\frac{2\la^-_1\tanh\!\Big[\frac{W\la^-_1}{2}\Big]-2\la^+_2\tanh\!\Big[\frac{W\la^+_2}{2}\Big]}{(\la_1^-)^2-(\la_2^+)^2}
\end{align}
and 
\begin{align}
\Gamma^-_{-+}=
\Gamma^-_{+-}&=
\frac{2\la^-_1\coth\!\Big[\frac{W\la^-_1}{2}\Big]-2\la^+_1\coth\!\Big[\frac{W\la^+_1}{2}\Big]}{(\la_1^-)^2-(\la_1^+)^2}
+
\frac{2\la^-_2\coth\!\Big[\frac{W\la^-_2}{2}\Big]-2\la^+_2\coth\!\Big[\frac{W\la^+_2}{2}\Big]}{(\la_2^-)^2-(\la_2^+)^2}
\nonumber\\
&\ -
\frac{2\la^+_1\coth\!\Big[\frac{W\la^+_1}{2}\Big]-2\la^-_2\coth\!\Big[\frac{W\la^-_2}{2}\Big]}{(\la_1^+)^2-(\la_2^-)^2}
-
\frac{2\la^-_1\coth\!\Big[\frac{W\la^-_1}{2}\Big]-2\la^+_2\coth\!\Big[\frac{W\la^+_2}{2}\Big]}{(\la_1^-)^2-(\la_2^+)^2},
\end{align}
where $\Gamma_{ee'}^\tau=\Gamma_{e'e}^\tau$ follows from the definition of $\Gamma_{ee'}^\tau$ in Eq.(\ref{eq:def-Gamma}). Here we observe that $\Gamma_{\mp\pm}^-$ is related to $\Gamma_{\mp\pm}^+$ by interchanging $\tanh(W\la_i^\tau/2)$ and $\coth(W\la_i^\tau/2)$. 

Furthermore, there are four different non-zero $\Omega^{\tau\tau'}_{ee'}$ (remembering that $\Omega^{\tau\tau'}_{ee'}=-\Omega^{\tau'\tau}_{e'e}$ and $\Omega^{\pm\pm}_{ee'}=0$). These are
\begin{align}
\Omega^{+-}_{\pm\pm}
&=
W
\left[
\frac{\la_1^\pm}{\sinh[W\la_1^{\pm}]}
+\frac{\la_2^\pm}{\sinh[W\la_2^{\pm}]}
\right]
+
\frac{2\la_1^\pm\la_2^\pm}{(\la_1^\pm)^2-(\la_2^\pm)^2}
\left[
\frac{\tanh\!\Big[\frac{W\la_2^\pm}{2}\Big]}{\tanh\!\Big[\frac{W\la_1^\pm}{2}\Big]}
-\frac{\tanh\!\Big[\frac{W\la_1^\pm}{2}\Big]}{\tanh\!\Big[\frac{W\la_2^\pm}{2}\Big]}
\right],
\end{align}
and 
\begin{align}\label{eq_Omega^-+_-+}
\Omega^{-+}_{-+}
=
-2 
\Bigg\{
&
-\frac{(\la_1^+)^2}{(\la_1^+)^2-(\la_1^-)^2}
+\frac{(\la_1^+)^2}{(\la_1^+)^2-(\la_2^-)^2}
-\frac{(\la_1^-)^2}{(\la_1^-)^2-(\la_2^+)^2}
+\frac{(\la_2^-)^2}{(\la_2^-)^2-(\la_2^+)^2}
\nonumber\\
&
-\frac{\la_1^- \la_1^+}{(\la_1^-)^2-(\la_1^+)^2}
\frac{\tanh \left[\frac{\la_1^+ W}{2}\right]}{\tanh \left[\frac{\la_1^- W}{2}\right]} 
+\frac{\la_1^- \la_2^+}{(\la_1^-)^2-(\la_2^+)^2}
\frac{\tanh \left[\frac{\la_2^+ W}{2}\right]}{\tanh \left[\frac{\la_1^- W}{2}\right]}
\nonumber\\
&
+\frac{\la_1^+ \la_2^-}{(\la_2^-)^2-(\la_1^+)^2}
\frac{\tanh \left[\frac{\la_1^+ W}{2}\right]}{\tanh \left[\frac{\la_2^- W}{2}\right]}
-\frac{\la_2^- \la_2^+}{(\la_2^-)^2-(\la_2^+)^2}
\frac{\tanh \left[\frac{\la_2^+ W}{2}\right]}{\tanh\left[\frac{\la_2^- W}{2}\right]}
\Bigg\}
\end{align}
and 
\begin{align}
\Omega^{+-}_{-+}
=
+2 
\Bigg\{
&
-\frac{(\la_1^-)^2}{(\la_1^-)^2-(\la_1^+)^2}
+\frac{(\la_1^-)^2}{(\la_1^-)^2-(\la_2^+)^2}
-\frac{(\la_1^+)^2}{(\la_1^+)^2-(\la_2^-)^2}
+\frac{(\la_2^+)^2}{(\la_2^+)^2-(\la_2^-)^2}
\nonumber\\
&
-\frac{\la_1^+ \la_1^-}{(\la_1^+)^2-(\la_1^-)^2}
\frac{\tanh \left[\frac{\la_1^- W}{2}\right]}{\tanh \left[\frac{\la_1^+ W}{2}\right]} 
+\frac{\la_1^+ \la_2^-}{(\la_1^+)^2-(\la_2^-)^2}
\frac{\tanh \left[\frac{\la_2^- W}{2}\right]}{\tanh \left[\frac{\la_1^+ W}{2}\right]}
\nonumber\\
&
+\frac{\la_1^- \la_2^+}{(\la_2^+)^2-(\la_1^-)^2}
\frac{\tanh \left[\frac{\la_1^- W}{2}\right]}{\tanh \left[\frac{\la_2^+ W}{2}\right]}
-\frac{\la_2^+ \la_2^-}{(\la_2^+)^2-(\la_2^-)^2}
\frac{\tanh \left[\frac{\la_2^- W}{2}\right]}{\tanh\left[\frac{\la_2^+ W}{2}\right]}
\Bigg\}.
\end{align}
We note that $\Omega^{+-}_{-+}=-\Omega^{-+}_{+-}$ and that $\Omega^{-+}_{+-}$ can be found by interchanging $E^+_{k_x}$ and $E^-_{k_x}$ in $\Omega^{-+}_{-+}$ in Eq.(\ref{eq_Omega^-+_-+}) (i.e. interchanging $\la_{i}^+$ and $\la_{i}^-$ for $i=1,2$). We also remark that all the integrals $\Gamma_{ee'}^\tau$ and  $\Omega^{\tau\tau'}_{ee'}$ are even in $k_x$, since $\la_{1,2}^{\pm}=\la_{1,2}(E_{k_x}^{\pm})$ is even in $k_x$. We now have all the integrals $\Gamma_{ee'}^\tau$ and  $\Omega^{\tau\tau'}_{ee'}$ appearing in the matrix elements $id_\pm$ Eq.(\ref{eq:inter-matrix-elements-appendix}) and $ib$ Eq.(\ref{eq:ib-matrix-element-explicit}). 
\end{widetext}

\section{On the numerical tight-binding formulation}\label{app:tight-binding}

In this appendix, we briefly discuss the lattice regularization of the BHZ model and its formulation for the ribbon geometry. 

In order to map a continuous model to a tight-binding model, we use the standard tight-binding regularization procedure. For a  2D square lattice, this consists in making the replacements
\begin{eqnarray}
k_i & \rightarrow &\frac{1}{a}\sin{k_i a}, \\
k_i^2 & \rightarrow & \frac{2}{a^2} \left(1-\cos{k_i a}\right),
\end{eqnarray}
with $a$ being the lattice constant and $i=x,y$. The two quantities are equal only in the long wavelength limit, $k_ia \rightarrow 0$. This tight-binding regularization has been extensively used in the literature to study the BHZ Hamiltonian.\cite{book_Shen,Imura10} The fermion doubling problem that usually occurs when discretizing massless Dirac particles does not directly affect our calculations as the RSOC term breaks chiral symmetry, which is one of the conditions for the fullfillment of the no-go theorem by Nielsen and Ninomiya.\cite{Nielsen81-1,Nielsen81-2} Moreover, we have checked that the topological properties of the Hamiltonian are unchanged in the tight-binding version for the parametric regimes that we have explored.

To transfer the Hamiltonian from momentum space onto a real-space lattice, we perform a Fourier transformation. The chosen form of the tight-binding regularization implies that the hopping terms in the lattice model exist only between nearest neighbor sites. For the calculations presented in the paper, we have a ribbon of finite width in one direction ($y$) and periodic boundary conditions in the orthogonal direction ($x$). Thus, we perform a Fourier transformation only in the direction of finite width ($y$) and obtain a $k_x$-dependent Hamiltonian:
\begin{widetext}
\begin{align}
\mathcal{H}(k_x)&=
\sum_j \mathcal{H}_{jj}c^{\dagger}_{j}c_{j}+\left(\mathcal{H}_{jj+1}c^{\dagger}_{j}c_{j+1}+H.c.\right), \\
\mathcal{H}_{jj}
&=
\left(\!\!
\begin{array}{cccc}
\ M-2B_+(2-\cos k_x)&A\sin k_x&-iR_0\sin k_x&0 \\A\sin k_x& -M+2B_-(2-\cos k_x)&0&0 \\iR_0\sin k_x&0&M-2B_+(2-\cos k_x)&\ -A\sin k_x \\ 0&0&-A\sin k_x&\ -M+2B_-(2-\cos k_x)
\end{array} 
\!\right)
\!,
 \\
\mathcal{H}_{jj+1}
&=
\frac{1}{2}\left(\begin{array}{cccc}
\ 2B_+&+A& +iR_0&0 \\-A& -2B_-&0&0 \\+iR_0&0&2B_+&\ +A \\ 0&0&-A&\ -2B_-\end{array} \right).
\end{align}
\end{widetext}
The number of sites in the simulation varied between $200$ and $2000$. We set the value of the lattice spacing $a$ such that we obtain the required width for the particular case under study. Changing the number of sites and $a$ help us to make sure that the results for the relevant values of $k$ did not depend on the details of the tight-binding regularization.

\section{Higher order Rashba spin-orbit couplings}\label{app:higher-order-Rashba}

Rothe \emph{et al.}\cite{Rothe-NJP-2010} derived the RSOC in the BHZ basis up to third order in the momentum. The calculations in the main text only include the first order term as seen in Eq.(\ref{Rashba_hamil}). Here we discuss the effects of the second and third order terms on a pair of HESs at an isolated boundary.  

The RSOC Hamiltonian in the BHZ basis to third order is\cite{Rothe-NJP-2010} 
\begin{align}
H_R&=
H_R^{(1)}+H_R^{(2)}+H_R^{(3)} 
\nonumber\\
&=
\left(\begin{array}{cccc}
 0 & 0 & -iR_0k_- & -S_0k_-^2 \\
 0 & 0 & S_0k_-^2 & i T_0k_-^3 \\
 iR_0k_+ & S_0k_+^2 & 0 & 0 \\
 -S_0k_+^2 & -i T_0k_+^3 & 0 & 0
\end{array}\right),
\label{eq:RSOC-to-3-order-BHZ-basis}
\end{align}
where $k_\pm=k_x\pm ik_y$ and the superscript $n$ on $H_R^{(n)}$ indicate the order of the momentum that it represents, i.e. the main text only use $H_R=H_R^{(1)}$ in Eq.(\ref{Rashba_hamil}). Here each order in momentum has its own constant prefactor, namely $R_0$, $S_0$ and $T_0$. 

To incorporate the higher order RSOC terms $H_R^{(2)}$ and $H_R^{(3)}$ into the $2\times2$ Hamiltonian (\ref{eq:full-non-diagonal-H}) for a pair of HESs at an isolated edge, we need to include the matrix elements $\langle\psi_{k_y\ua}|H_R^{(2)}|\psi_{k_y\da}\rangle$ and $\langle\psi_{k_y\ua}|H_R^{(3)}|\psi_{k_y\da}\rangle$ into the effective RSOC $\alpha_{k_y}$.  Here $\psi_{k_y\sigma}$ for $\sigma = \da,\ua$ are the HESs in Eq.(\ref{eq:HES_single_edge_app}). 

We begin by noticing that 
\begin{align} 
\langle\psi_{k_y\ua}|&H_R^{(2)}|\psi_{k_y\da}\rangle=0, 
\end{align}
i.e.~the second order RSOC term $H_R^{(2)}$ does not contribute to the RSOC for a pair of HESs at an isolated edge within our analytical approach. The same kind of cancellation was found by considering the BIA term in Eq.(\ref{eq:BIA}). In fact, both cancellations stem from the alternating signs in the anti-diagonal of $H_{BIA}$ and $H_R^{(2)}$, respectively. Hence the cancellation is independent of the details of the transversal wave function of the HESs.

\begin{figure}
\includegraphics[width=0.96\linewidth]{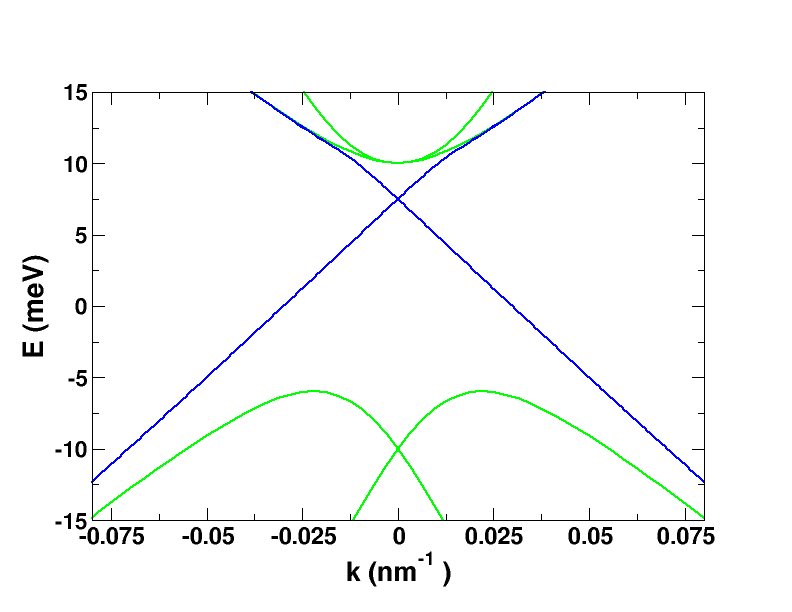}
\caption{\label{fig:first-and-third-order-RSOC-band-structure} The edge state dispersions including both the first and third order RSOC terms, $H_R^{(1)}$ and $H_R^{(3)}$, found by a numerical tight-binding calculation. The parameters are the same as in Fig.~\ref{fig:dispersions} (top panel). Thus, we observe that the third order term only introduce minor changes in the edge state spectrum. The prefactor $T_0$ of the third order RSOC term is chosen such that $T_0=0.57R_0$ according to the values given by Rothe \emph{et al.}\cite{Rothe-NJP-2010}}
\end{figure}

Now we turn to the third order RSOC $H_R^{(3)}$, where we will see how our analytical approach actually fails to give the correct answer. A straightforward calculation gives 
\begin{align} 
\langle&\psi_{k_y\ua}|H_R^{(3)}|\psi_{k_y\da}\rangle=
i T_0 \frac{B_+}{2B} 
\int_{0}^{\infty} dx 
g_{k_y}(x)(k_-)^3
g_{-k_y}(x)
\nonumber \\
&=
-T_0
\frac{B_+}{2B} 
\left[
a_{k_y}^{(3)} 
+3k_y a_{k_y}^{(2)}
+3k_y^2 a_{k_y}^{(1)}
+k_y^3 a_{k_y}^{(0)}
\right],
\label{eq:error-element}
\end{align}
where $k_-=-i(\p_x +k_y)$ and we introduced
\begin{align} 
&a_{k_y}^{(n)}=
\int_{0}^{\infty}\! dx\; 
g_{k_y}(x)\p_x^n
g_{-k_y}(x)
\nonumber\\
&=
h(k_y) 
\Bigg[
(-1)^n (\lambda_1^-)^n 
\left(\frac{1}{\lambda_1^++\lambda_1^-}-\frac{1}{\lambda_1^-+\lambda_2^+}\right)
\nonumber\\
&
\hspace{11mm} +
(-1)^n (\lambda_2^-)^n 
\left(\frac{1}{\lambda_2^++\lambda_2^-}-\frac{1}{\lambda_1^++\lambda_2^-}\right)
\Bigg].
\end{align}
Here we use $h(k_y)$ from Eq.(\ref{eq-def-h}) and the shorthand notation $\lambda_i^{\pm}=\lambda_i(\pm k_y)$ for the two inverse length scales in Eq.(\ref{eq:lambdas-single-edge}) that appear in the transversal wave functions of the HESs $g_{\pm k_y}(x)$ in Eq.(\ref{eq:transverse-HES-single-edge}). \emph{We emphasize that even though the calculus leading to this result is correct, the result itself is not correct.} For instance, it leads to a non-zero matrix element at $k_y=0$, which in turn introduces a gap in the edge state spectrum of the Hamiltonian (\ref{eq:full-non-diagonal-H}). This is obviously not correct since the RSOC is invariant under time reversal symmetry and therefore \emph{no} gap should be opened by $H_R^{(3)}$. To emphasise this point, we have performed a numerical tight-binding calculation including both $H_R^{(1)}$ and $H_R^{(3)}$, which indeed shows that no gap appears in the edge state spectrum, see Fig.~\ref{fig:first-and-third-order-RSOC-band-structure}. Now, to pinpoint the origin of the problem further,  we consider the matrix element (\ref{eq:error-element}) at $k_y=0$, which erroneously was found to be non-zero. By using integration by parts repeatedly, we obtain 
\begin{align}
\langle\psi_{k_y=0\ua}|H_R^{(3)}|\psi_{k_y=0\da}\rangle&=
-\frac{B_+}{2B} T_0
\int_{0}^{\infty} dx 
g_{0}(x)\p_x^3
g_{0}(x)
\nonumber\\
&=
-\frac{B_+}{4B} T_0
\Big[\p_xg_0(x=0)\Big]^2. 
\end{align}
Hence, the matrix element is proportional to the square of the derivative of the transversal wave function of the HES, $[\p_xg_0(x)]^2$, at the boundary $x=0$. We found the transverse wave function $g_{k_y}(x)$  analytically using the simple hard-wall boundary condition that $g_{k_y}(x=0)=0$ (see Appendix \ref{subsec-Appendix:HES-isolated-boundary} and Refs.~\onlinecite{Zhou-PRL-2008,Wada-PRB-2011}). This produce an artificial discontinuity in the derivative of $g_{k_y}(x)$ at the boundary $x=0$ --- just as for the textbook example of an infinitely deep square well. It is this discontinuity that gives the incorrect non-zero matrix element at $k_y=0$. For any smooth boundary potential (or a finite step potential), $\p_xg_0(x)$ would be zero at the boundary of the integral (not necessarily at $x=0$) and thereby give the correct result. Unfortunately, it is hard to obtain analytical wave functions for these potentials. In other words, we seem to get a non-zero result due to our crude approximation for the transversal wave function. However, for integrals involving only the first order derivatives of $g_{k_y}$ as in the main text, we can still use the HESs in Eq.~(\ref{eq:HES_single_edge_app}) in our analytical approach.

A similar situation is found in the use of $\mathbf{k}\cdot\mathbf{p}$ theory to describe confined structures by the envelope function approximation. Here hard-wall boundaries are often used to describe structures in accordance with experimental observations, even though the envelope function approximation in principle requires smooth potentials. This has been justified in some case, but remains a problematic issue for other cases (see Sec.~4.1 in Ref.~\onlinecite{Winkler-BOOK-2003} for a discussion).


\begin{thebibliography}{65}%
\makeatletter
\providecommand \@ifxundefined [1]{%
 \@ifx{#1\undefined}
}%
\providecommand \@ifnum [1]{%
 \ifnum #1\expandafter \@firstoftwo
 \else \expandafter \@secondoftwo
 \fi
}%
\providecommand \@ifx [1]{%
 \ifx #1\expandafter \@firstoftwo
 \else \expandafter \@secondoftwo
 \fi
}%
\providecommand \natexlab [1]{#1}%
\providecommand \enquote  [1]{``#1''}%
\providecommand \bibnamefont  [1]{#1}%
\providecommand \bibfnamefont [1]{#1}%
\providecommand \citenamefont [1]{#1}%
\providecommand \href@noop [0]{\@secondoftwo}%
\providecommand \href [0]{\begingroup \@sanitize@url \@href}%
\providecommand \@href[1]{\@@startlink{#1}\@@href}%
\providecommand \@@href[1]{\endgroup#1\@@endlink}%
\providecommand \@sanitize@url [0]{\catcode `\\12\catcode `\$12\catcode
  `\&12\catcode `\#12\catcode `\^12\catcode `\_12\catcode `\%12\relax}%
\providecommand \@@startlink[1]{}%
\providecommand \@@endlink[0]{}%
\providecommand \url  [0]{\begingroup\@sanitize@url \@url }%
\providecommand \@url [1]{\endgroup\@href {#1}{\urlprefix }}%
\providecommand \urlprefix  [0]{URL }%
\providecommand \Eprint [0]{\href }%
\providecommand \doibase [0]{http://dx.doi.org/}%
\providecommand \selectlanguage [0]{\@gobble}%
\providecommand \bibinfo  [0]{\@secondoftwo}%
\providecommand \bibfield  [0]{\@secondoftwo}%
\providecommand \translation [1]{[#1]}%
\providecommand \BibitemOpen [0]{}%
\providecommand \bibitemStop [0]{}%
\providecommand \bibitemNoStop [0]{.\EOS\space}%
\providecommand \EOS [0]{\spacefactor3000\relax}%
\providecommand \BibitemShut  [1]{\csname bibitem#1\endcsname}%
\let\auto@bib@innerbib\@empty
\bibitem [{\citenamefont {Kane}\ and\ \citenamefont
  {Mele}(2005{\natexlab{a}})}]{Kane-PRL-2005a}%
  \BibitemOpen
  \bibfield  {author} {\bibinfo {author} {\bibfnamefont {C.~L.}\ \bibnamefont
  {Kane}}\ and\ \bibinfo {author} {\bibfnamefont {E.~J.}\ \bibnamefont
  {Mele}},\ }\href {\doibase 10.1103/PhysRevLett.95.226801} {\bibfield
  {journal} {\bibinfo  {journal} {Phys. Rev. Lett.}\ }\textbf {\bibinfo
  {volume} {95}},\ \bibinfo {pages} {226801} (\bibinfo {year}
  {2005}{\natexlab{a}})}\BibitemShut {NoStop}%
\bibitem [{\citenamefont {Kane}\ and\ \citenamefont
  {Mele}(2005{\natexlab{b}})}]{Kane-PRL-2005b}%
  \BibitemOpen
  \bibfield  {author} {\bibinfo {author} {\bibfnamefont {C.~L.}\ \bibnamefont
  {Kane}}\ and\ \bibinfo {author} {\bibfnamefont {E.~J.}\ \bibnamefont
  {Mele}},\ }\href {\doibase 10.1103/PhysRevLett.95.146802} {\bibfield
  {journal} {\bibinfo  {journal} {Phys. Rev. Lett.}\ }\textbf {\bibinfo
  {volume} {95}},\ \bibinfo {pages} {146802} (\bibinfo {year}
  {2005}{\natexlab{b}})}\BibitemShut {NoStop}%
\bibitem [{\citenamefont {Qi}\ and\ \citenamefont {Zhang}(2011)}]{Qi-RMP-2011}%
  \BibitemOpen
  \bibfield  {author} {\bibinfo {author} {\bibfnamefont {X.-L.}\ \bibnamefont
  {Qi}}\ and\ \bibinfo {author} {\bibfnamefont {S.-C.}\ \bibnamefont {Zhang}},\
  }\href {\doibase 10.1103/RevModPhys.83.1057} {\bibfield  {journal} {\bibinfo
  {journal} {Rev. Mod. Phys.}\ }\textbf {\bibinfo {volume} {83}},\ \bibinfo
  {pages} {1057} (\bibinfo {year} {2011})}\BibitemShut {NoStop}%
\bibitem [{\citenamefont {Y.Ando}(2013)}]{Ando-JPSJ-2013}%
  \BibitemOpen
  \bibfield  {author} {\bibinfo {author} {\bibnamefont {Y.Ando}},\ }\href@noop
  {} {\bibfield  {journal} {\bibinfo  {journal} {Journal of The Physical
  Society of Japan}\ }\textbf {\bibinfo {volume} {82}},\ \bibinfo {pages}
  {102001} (\bibinfo {year} {2013})}\BibitemShut {NoStop}%
\bibitem [{\citenamefont {Xu}\ and\ \citenamefont {Moore}(2006)}]{Xu-PRB-2006}%
  \BibitemOpen
  \bibfield  {author} {\bibinfo {author} {\bibfnamefont {C.}~\bibnamefont
  {Xu}}\ and\ \bibinfo {author} {\bibfnamefont {J.~E.}\ \bibnamefont {Moore}},\
  }\href {\doibase 10.1103/PhysRevB.73.045322} {\bibfield  {journal} {\bibinfo
  {journal} {Phys. Rev. B}\ }\textbf {\bibinfo {volume} {73}},\ \bibinfo
  {pages} {045322} (\bibinfo {year} {2006})}\BibitemShut {NoStop}%
\bibitem [{\citenamefont {K\"{o}nig}\ \emph {et~al.}(2007)\citenamefont
  {K\"{o}nig}, \citenamefont {Wiedmann}, \citenamefont {Br\"{u}ne},
  \citenamefont {Roth}, \citenamefont {Buhmann}, \citenamefont {Molenkamp},
  \citenamefont {Qi},\ and\ \citenamefont {Zhang}}]{Konig-Science-2007}%
  \BibitemOpen
  \bibfield  {author} {\bibinfo {author} {\bibfnamefont {M.}~\bibnamefont
  {K\"{o}nig}}, \bibinfo {author} {\bibfnamefont {S.}~\bibnamefont {Wiedmann}},
  \bibinfo {author} {\bibfnamefont {C.}~\bibnamefont {Br\"{u}ne}}, \bibinfo
  {author} {\bibfnamefont {A.}~\bibnamefont {Roth}}, \bibinfo {author}
  {\bibfnamefont {H.}~\bibnamefont {Buhmann}}, \bibinfo {author} {\bibfnamefont
  {L.~W.}\ \bibnamefont {Molenkamp}}, \bibinfo {author} {\bibfnamefont {X.-L.}\
  \bibnamefont {Qi}}, \ and\ \bibinfo {author} {\bibfnamefont {S.-C.}\
  \bibnamefont {Zhang}},\ }\href {\doibase 10.1126/science.1148047} {\bibfield
  {journal} {\bibinfo  {journal} {Science}\ }\textbf {\bibinfo {volume}
  {318}},\ \bibinfo {pages} {766} (\bibinfo {year} {2007})}\BibitemShut
  {NoStop}%
\bibitem [{\citenamefont {Roth}\ \emph {et~al.}(2009)\citenamefont {Roth},
  \citenamefont {Br\"{u}ne}, \citenamefont {Buhmann}, \citenamefont
  {Molenkamp}, \citenamefont {Maciejko}, \citenamefont {Qi},\ and\
  \citenamefont {Zhang}}]{Roth-Science-2009}%
  \BibitemOpen
  \bibfield  {author} {\bibinfo {author} {\bibfnamefont {A.}~\bibnamefont
  {Roth}}, \bibinfo {author} {\bibfnamefont {C.}~\bibnamefont {Br\"{u}ne}},
  \bibinfo {author} {\bibfnamefont {H.}~\bibnamefont {Buhmann}}, \bibinfo
  {author} {\bibfnamefont {L.~W.}\ \bibnamefont {Molenkamp}}, \bibinfo {author}
  {\bibfnamefont {J.}~\bibnamefont {Maciejko}}, \bibinfo {author}
  {\bibfnamefont {X.-L.}\ \bibnamefont {Qi}}, \ and\ \bibinfo {author}
  {\bibfnamefont {S.-C.}\ \bibnamefont {Zhang}},\ }\href {\doibase
  10.1126/science.1174736} {\bibfield  {journal} {\bibinfo  {journal}
  {Science}\ }\textbf {\bibinfo {volume} {325}},\ \bibinfo {pages} {294}
  (\bibinfo {year} {2009})}\BibitemShut {NoStop}%
\bibitem [{\citenamefont {Br\"{u}ne}\ \emph {et~al.}(2012)\citenamefont
  {Br\"{u}ne}, \citenamefont {Roth}, \citenamefont {Buhmann}, \citenamefont
  {Hankiewicz}, \citenamefont {Molenkamp}, \citenamefont {Maciejko},
  \citenamefont {Qi},\ and\ \citenamefont {Zhang}}]{Brune-nature-phys-2012}%
  \BibitemOpen
  \bibfield  {author} {\bibinfo {author} {\bibfnamefont {C.}~\bibnamefont
  {Br\"{u}ne}}, \bibinfo {author} {\bibfnamefont {A.}~\bibnamefont {Roth}},
  \bibinfo {author} {\bibfnamefont {H.}~\bibnamefont {Buhmann}}, \bibinfo
  {author} {\bibfnamefont {E.~M.}\ \bibnamefont {Hankiewicz}}, \bibinfo
  {author} {\bibfnamefont {L.~W.}\ \bibnamefont {Molenkamp}}, \bibinfo {author}
  {\bibfnamefont {J.}~\bibnamefont {Maciejko}}, \bibinfo {author}
  {\bibfnamefont {X.-L.}\ \bibnamefont {Qi}}, \ and\ \bibinfo {author}
  {\bibfnamefont {S.-C.}\ \bibnamefont {Zhang}},\ }\href {\doibase
  10.1038/nphys2322} {\bibfield  {journal} {\bibinfo  {journal} {Nature
  Physics}\ }\textbf {\bibinfo {volume} {8}},\ \bibinfo {pages} {485} (\bibinfo
  {year} {2012})}\BibitemShut {NoStop}%
\bibitem [{\citenamefont {K\"onig}\ \emph {et~al.}(2013)\citenamefont
  {K\"onig}, \citenamefont {Baenninger}, \citenamefont {Garcia}, \citenamefont
  {Harjee}, \citenamefont {Pruitt}, \citenamefont {Ames}, \citenamefont
  {Leubner}, \citenamefont {Br\"une}, \citenamefont {Buhmann}, \citenamefont
  {Molenkamp},\ and\ \citenamefont {Goldhaber-Gordon}}]{Konig-PRX-2013}%
  \BibitemOpen
  \bibfield  {author} {\bibinfo {author} {\bibfnamefont {M.}~\bibnamefont
  {K\"onig}}, \bibinfo {author} {\bibfnamefont {M.}~\bibnamefont {Baenninger}},
  \bibinfo {author} {\bibfnamefont {A.~G.~F.}\ \bibnamefont {Garcia}}, \bibinfo
  {author} {\bibfnamefont {N.}~\bibnamefont {Harjee}}, \bibinfo {author}
  {\bibfnamefont {B.~L.}\ \bibnamefont {Pruitt}}, \bibinfo {author}
  {\bibfnamefont {C.}~\bibnamefont {Ames}}, \bibinfo {author} {\bibfnamefont
  {P.}~\bibnamefont {Leubner}}, \bibinfo {author} {\bibfnamefont
  {C.}~\bibnamefont {Br\"une}}, \bibinfo {author} {\bibfnamefont
  {H.}~\bibnamefont {Buhmann}}, \bibinfo {author} {\bibfnamefont {L.~W.}\
  \bibnamefont {Molenkamp}}, \ and\ \bibinfo {author} {\bibfnamefont
  {D.}~\bibnamefont {Goldhaber-Gordon}},\ }\href {\doibase
  10.1103/PhysRevX.3.021003} {\bibfield  {journal} {\bibinfo  {journal} {Phys.
  Rev. X}\ }\textbf {\bibinfo {volume} {3}},\ \bibinfo {pages} {021003}
  (\bibinfo {year} {2013})}\BibitemShut {NoStop}%
\bibitem [{\citenamefont {Bernevig}\ \emph {et~al.}(2006)\citenamefont
  {Bernevig}, \citenamefont {Hughes},\ and\ \citenamefont
  {Zhang}}]{Bernevig-Science-2006}%
  \BibitemOpen
  \bibfield  {author} {\bibinfo {author} {\bibfnamefont {B.~A.}\ \bibnamefont
  {Bernevig}}, \bibinfo {author} {\bibfnamefont {T.~L.}\ \bibnamefont
  {Hughes}}, \ and\ \bibinfo {author} {\bibfnamefont {S.-C.}\ \bibnamefont
  {Zhang}},\ }\href {\doibase 10.1126/science.1133734} {\bibfield  {journal}
  {\bibinfo  {journal} {Science}\ }\textbf {\bibinfo {volume} {314}},\ \bibinfo
  {pages} {1757} (\bibinfo {year} {2006})}\BibitemShut {NoStop}%
\bibitem [{\citenamefont {Liu}\ \emph {et~al.}(2008)\citenamefont {Liu},
  \citenamefont {Hughes}, \citenamefont {Qi}, \citenamefont {Wang},\ and\
  \citenamefont {Zhang}}]{Liu-PRL-2008}%
  \BibitemOpen
  \bibfield  {author} {\bibinfo {author} {\bibfnamefont {C.}~\bibnamefont
  {Liu}}, \bibinfo {author} {\bibfnamefont {T.~L.}\ \bibnamefont {Hughes}},
  \bibinfo {author} {\bibfnamefont {X.-L.}\ \bibnamefont {Qi}}, \bibinfo
  {author} {\bibfnamefont {K.}~\bibnamefont {Wang}}, \ and\ \bibinfo {author}
  {\bibfnamefont {S.-C.}\ \bibnamefont {Zhang}},\ }\href {\doibase
  10.1103/PhysRevLett.100.236601} {\bibfield  {journal} {\bibinfo  {journal}
  {Phys. Rev. Lett.}\ }\textbf {\bibinfo {volume} {100}},\ \bibinfo {pages}
  {236601} (\bibinfo {year} {2008})}\BibitemShut {NoStop}%
\bibitem [{\citenamefont {Knez}\ \emph {et~al.}(2011)\citenamefont {Knez},
  \citenamefont {Du},\ and\ \citenamefont {Sullivan}}]{Knez-PRL-2011}%
  \BibitemOpen
  \bibfield  {author} {\bibinfo {author} {\bibfnamefont {I.}~\bibnamefont
  {Knez}}, \bibinfo {author} {\bibfnamefont {R.-R.}\ \bibnamefont {Du}}, \ and\
  \bibinfo {author} {\bibfnamefont {G.}~\bibnamefont {Sullivan}},\ }\href
  {\doibase 10.1103/PhysRevLett.107.136603} {\bibfield  {journal} {\bibinfo
  {journal} {Phys. Rev. Lett.}\ }\textbf {\bibinfo {volume} {107}},\ \bibinfo
  {pages} {136603} (\bibinfo {year} {2011})}\BibitemShut {NoStop}%
\bibitem [{\citenamefont {Suzuki}\ \emph {et~al.}(2013)\citenamefont {Suzuki},
  \citenamefont {Harada}, \citenamefont {Onomitsu},\ and\ \citenamefont
  {Muraki}}]{Suzuki-PRB-2013}%
  \BibitemOpen
  \bibfield  {author} {\bibinfo {author} {\bibfnamefont {K.}~\bibnamefont
  {Suzuki}}, \bibinfo {author} {\bibfnamefont {Y.}~\bibnamefont {Harada}},
  \bibinfo {author} {\bibfnamefont {K.}~\bibnamefont {Onomitsu}}, \ and\
  \bibinfo {author} {\bibfnamefont {K.}~\bibnamefont {Muraki}},\ }\href
  {\doibase 10.1103/PhysRevB.87.235311} {\bibfield  {journal} {\bibinfo
  {journal} {Phys. Rev. B}\ }\textbf {\bibinfo {volume} {87}},\ \bibinfo
  {pages} {235311} (\bibinfo {year} {2013})}\BibitemShut {NoStop}%
\bibitem [{\citenamefont {Knez}\ \emph {et~al.}(2014)\citenamefont {Knez},
  \citenamefont {Rettner}, \citenamefont {Yang}, \citenamefont {Parkin},
  \citenamefont {Du}, \citenamefont {Du},\ and\ \citenamefont
  {Sullivan}}]{Knez-PRL-2014}%
  \BibitemOpen
  \bibfield  {author} {\bibinfo {author} {\bibfnamefont {I.}~\bibnamefont
  {Knez}}, \bibinfo {author} {\bibfnamefont {C.~T.}\ \bibnamefont {Rettner}},
  \bibinfo {author} {\bibfnamefont {S.-H.}\ \bibnamefont {Yang}}, \bibinfo
  {author} {\bibfnamefont {S.~S.~P.}\ \bibnamefont {Parkin}}, \bibinfo {author}
  {\bibfnamefont {L.}~\bibnamefont {Du}}, \bibinfo {author} {\bibfnamefont
  {R.-R.}\ \bibnamefont {Du}}, \ and\ \bibinfo {author} {\bibfnamefont
  {G.}~\bibnamefont {Sullivan}},\ }\href@noop {} {\bibfield  {journal}
  {\bibinfo  {journal} {Phys. Rev. Lett.}\ }\textbf {\bibinfo {volume} {112}},\
  \bibinfo {pages} {026602} (\bibinfo {year} {2014})}\BibitemShut {NoStop}%
\bibitem [{\citenamefont {Du}\ \emph {et~al.}(2015)\citenamefont {Du},
  \citenamefont {Knez}, \citenamefont {Sullivan},\ and\ \citenamefont
  {Du}}]{Du-PRL-2015}%
  \BibitemOpen
  \bibfield  {author} {\bibinfo {author} {\bibfnamefont {L.}~\bibnamefont
  {Du}}, \bibinfo {author} {\bibfnamefont {I.}~\bibnamefont {Knez}}, \bibinfo
  {author} {\bibfnamefont {G.}~\bibnamefont {Sullivan}}, \ and\ \bibinfo
  {author} {\bibfnamefont {R.-R.}\ \bibnamefont {Du}},\ }\href {\doibase
  10.1103/PhysRevLett.114.096802} {\bibfield  {journal} {\bibinfo  {journal}
  {Phys. Rev. Lett.}\ }\textbf {\bibinfo {volume} {114}},\ \bibinfo {pages}
  {096802} (\bibinfo {year} {2015})}\BibitemShut {NoStop}%
\bibitem [{\citenamefont {Nichele}\ \emph {et~al.}(2015)\citenamefont
  {Nichele}, \citenamefont {Suominen}, \citenamefont {Kjaergaard},
  \citenamefont {Marcus}, \citenamefont {Sajadi}, \citenamefont {Folk},
  \citenamefont {Qu}, \citenamefont {Beukman}, \citenamefont {de~Vries},
  \citenamefont {van Veen}, \citenamefont {Nadj-Perge}, \citenamefont
  {Kouwenhoven}, \citenamefont {Nguyen}, \citenamefont {Kiselev}, \citenamefont
  {Wei~Yi}, \citenamefont {Manfra}, \citenamefont {Spanton},\ and\
  \citenamefont {Moler}}]{Nichele-et-al-2015}%
  \BibitemOpen
  \bibfield  {author} {\bibinfo {author} {\bibfnamefont {F.}~\bibnamefont
  {Nichele}}, \bibinfo {author} {\bibfnamefont {H.~J.}\ \bibnamefont
  {Suominen}}, \bibinfo {author} {\bibfnamefont {M.}~\bibnamefont
  {Kjaergaard}}, \bibinfo {author} {\bibfnamefont {C.~M.}\ \bibnamefont
  {Marcus}}, \bibinfo {author} {\bibfnamefont {E.}~\bibnamefont {Sajadi}},
  \bibinfo {author} {\bibfnamefont {J.~A.}\ \bibnamefont {Folk}}, \bibinfo
  {author} {\bibfnamefont {F.}~\bibnamefont {Qu}}, \bibinfo {author}
  {\bibfnamefont {A.~J.}\ \bibnamefont {Beukman}}, \bibinfo {author}
  {\bibfnamefont {F.~K.}\ \bibnamefont {de~Vries}}, \bibinfo {author}
  {\bibfnamefont {J.}~\bibnamefont {van Veen}}, \bibinfo {author}
  {\bibfnamefont {S.}~\bibnamefont {Nadj-Perge}}, \bibinfo {author}
  {\bibfnamefont {L.~P.}\ \bibnamefont {Kouwenhoven}}, \bibinfo {author}
  {\bibfnamefont {B.-M.}\ \bibnamefont {Nguyen}}, \bibinfo {author}
  {\bibfnamefont {A.~A.}\ \bibnamefont {Kiselev}}, \bibinfo {author}
  {\bibfnamefont {M.~S.}\ \bibnamefont {Wei~Yi}}, \bibinfo {author}
  {\bibfnamefont {M.~J.}\ \bibnamefont {Manfra}}, \bibinfo {author}
  {\bibfnamefont {E.~M.}\ \bibnamefont {Spanton}}, \ and\ \bibinfo {author}
  {\bibfnamefont {K.~A.}\ \bibnamefont {Moler}},\ }\href@noop {} {\bibfield
  {journal} {\bibinfo  {journal} {arXiv:1511.01728}\ } (\bibinfo {year}
  {2015})}\BibitemShut {NoStop}%
\bibitem [{\citenamefont {Gusev}\ \emph {et~al.}(2011)\citenamefont {Gusev},
  \citenamefont {Kvon}, \citenamefont {Shegai}, \citenamefont {Mikhailov},
  \citenamefont {Dvoretsky},\ and\ \citenamefont {Portal}}]{Gusev-PRB-2011}%
  \BibitemOpen
  \bibfield  {author} {\bibinfo {author} {\bibfnamefont {G.~M.}\ \bibnamefont
  {Gusev}}, \bibinfo {author} {\bibfnamefont {Z.~D.}\ \bibnamefont {Kvon}},
  \bibinfo {author} {\bibfnamefont {O.~A.}\ \bibnamefont {Shegai}}, \bibinfo
  {author} {\bibfnamefont {N.~N.}\ \bibnamefont {Mikhailov}}, \bibinfo {author}
  {\bibfnamefont {S.~A.}\ \bibnamefont {Dvoretsky}}, \ and\ \bibinfo {author}
  {\bibfnamefont {J.~C.}\ \bibnamefont {Portal}},\ }\href {\doibase
  10.1103/PhysRevB.84.121302} {\bibfield  {journal} {\bibinfo  {journal} {Phys.
  Rev. B}\ }\textbf {\bibinfo {volume} {84}},\ \bibinfo {pages} {121302}
  (\bibinfo {year} {2011})}\BibitemShut {NoStop}%
\bibitem [{\citenamefont {Grabecki}\ \emph {et~al.}(2013)\citenamefont
  {Grabecki}, \citenamefont {Wr\'obel}, \citenamefont {Czapkiewicz},
  \citenamefont {Cywi\ifmmode~\acute{n}\else \'{n}\fi{}ski}, \citenamefont
  {Giera\l{}towska}, \citenamefont {Guziewicz}, \citenamefont {Zholudev},
  \citenamefont {Gavrilenko}, \citenamefont {Mikhailov}, \citenamefont
  {Dvoretski}, \citenamefont {Teppe}, \citenamefont {Knap},\ and\ \citenamefont
  {Dietl}}]{Grabecki-PRB-2013}%
  \BibitemOpen
  \bibfield  {author} {\bibinfo {author} {\bibfnamefont {G.}~\bibnamefont
  {Grabecki}}, \bibinfo {author} {\bibfnamefont {J.}~\bibnamefont {Wr\'obel}},
  \bibinfo {author} {\bibfnamefont {M.}~\bibnamefont {Czapkiewicz}}, \bibinfo
  {author} {\bibfnamefont {L.}~\bibnamefont {Cywi\ifmmode~\acute{n}\else
  \'{n}\fi{}ski}}, \bibinfo {author} {\bibfnamefont {S.}~\bibnamefont
  {Giera\l{}towska}}, \bibinfo {author} {\bibfnamefont {E.}~\bibnamefont
  {Guziewicz}}, \bibinfo {author} {\bibfnamefont {M.}~\bibnamefont {Zholudev}},
  \bibinfo {author} {\bibfnamefont {V.}~\bibnamefont {Gavrilenko}}, \bibinfo
  {author} {\bibfnamefont {N.~N.}\ \bibnamefont {Mikhailov}}, \bibinfo {author}
  {\bibfnamefont {S.~A.}\ \bibnamefont {Dvoretski}}, \bibinfo {author}
  {\bibfnamefont {F.}~\bibnamefont {Teppe}}, \bibinfo {author} {\bibfnamefont
  {W.}~\bibnamefont {Knap}}, \ and\ \bibinfo {author} {\bibfnamefont
  {T.}~\bibnamefont {Dietl}},\ }\href {\doibase 10.1103/PhysRevB.88.165309}
  {\bibfield  {journal} {\bibinfo  {journal} {Phys. Rev. B}\ }\textbf {\bibinfo
  {volume} {88}},\ \bibinfo {pages} {165309} (\bibinfo {year}
  {2013})}\BibitemShut {NoStop}%
\bibitem [{\citenamefont {Gusev}\ \emph {et~al.}(2014)\citenamefont {Gusev},
  \citenamefont {Kvon}, \citenamefont {Olshanetsky}, \citenamefont {Levin},
  \citenamefont {Krupko}, \citenamefont {Portal}, \citenamefont {Mikhailov},\
  and\ \citenamefont {Dvoretsky}}]{Gusev-et-al-PRB-2014}%
  \BibitemOpen
  \bibfield  {author} {\bibinfo {author} {\bibfnamefont {G.~M.}\ \bibnamefont
  {Gusev}}, \bibinfo {author} {\bibfnamefont {Z.~D.}\ \bibnamefont {Kvon}},
  \bibinfo {author} {\bibfnamefont {E.~B.}\ \bibnamefont {Olshanetsky}},
  \bibinfo {author} {\bibfnamefont {A.~D.}\ \bibnamefont {Levin}}, \bibinfo
  {author} {\bibfnamefont {Y.}~\bibnamefont {Krupko}}, \bibinfo {author}
  {\bibfnamefont {J.~C.}\ \bibnamefont {Portal}}, \bibinfo {author}
  {\bibfnamefont {N.~N.}\ \bibnamefont {Mikhailov}}, \ and\ \bibinfo {author}
  {\bibfnamefont {S.~A.}\ \bibnamefont {Dvoretsky}},\ }\href {\doibase
  10.1103/PhysRevB.89.125305} {\bibfield  {journal} {\bibinfo  {journal} {Phys.
  Rev. B}\ }\textbf {\bibinfo {volume} {89}},\ \bibinfo {pages} {125305}
  (\bibinfo {year} {2014})}\BibitemShut {NoStop}%
\bibitem [{\citenamefont {Spanton}\ \emph {et~al.}(2014)\citenamefont
  {Spanton}, \citenamefont {Nowack}, \citenamefont {Du}, \citenamefont
  {Sullivan}, \citenamefont {Du},\ and\ \citenamefont
  {Moler}}]{Spanton-PRL-2014}%
  \BibitemOpen
  \bibfield  {author} {\bibinfo {author} {\bibfnamefont {E.~M.}\ \bibnamefont
  {Spanton}}, \bibinfo {author} {\bibfnamefont {K.~C.}\ \bibnamefont {Nowack}},
  \bibinfo {author} {\bibfnamefont {L.}~\bibnamefont {Du}}, \bibinfo {author}
  {\bibfnamefont {G.}~\bibnamefont {Sullivan}}, \bibinfo {author}
  {\bibfnamefont {R.-R.}\ \bibnamefont {Du}}, \ and\ \bibinfo {author}
  {\bibfnamefont {K.~A.}\ \bibnamefont {Moler}},\ }\href@noop {} {\bibfield
  {journal} {\bibinfo  {journal} {Phys. Rev. Lett.}\ }\textbf {\bibinfo
  {volume} {113}},\ \bibinfo {pages} {026804} (\bibinfo {year}
  {2014})}\BibitemShut {NoStop}%
\bibitem [{\citenamefont {Schmidt}\ \emph {et~al.}(2012)\citenamefont
  {Schmidt}, \citenamefont {Rachel}, \citenamefont {von Oppen},\ and\
  \citenamefont {Glazman}}]{Schmidt-PRL-2012}%
  \BibitemOpen
  \bibfield  {author} {\bibinfo {author} {\bibfnamefont {T.~L.}\ \bibnamefont
  {Schmidt}}, \bibinfo {author} {\bibfnamefont {S.}~\bibnamefont {Rachel}},
  \bibinfo {author} {\bibfnamefont {F.}~\bibnamefont {von Oppen}}, \ and\
  \bibinfo {author} {\bibfnamefont {L.~I.}\ \bibnamefont {Glazman}},\ }\href
  {\doibase 10.1103/PhysRevLett.108.156402} {\bibfield  {journal} {\bibinfo
  {journal} {Phys. Rev. Lett.}\ }\textbf {\bibinfo {volume} {108}},\ \bibinfo
  {pages} {156402} (\bibinfo {year} {2012})}\BibitemShut {NoStop}%
\bibitem [{\citenamefont {Str\"om}\ \emph {et~al.}(2010)\citenamefont
  {Str\"om}, \citenamefont {Johannesson},\ and\ \citenamefont
  {Japaridze}}]{Strom-PRL-2010}%
  \BibitemOpen
  \bibfield  {author} {\bibinfo {author} {\bibfnamefont {A.}~\bibnamefont
  {Str\"om}}, \bibinfo {author} {\bibfnamefont {H.}~\bibnamefont
  {Johannesson}}, \ and\ \bibinfo {author} {\bibfnamefont {G.~I.}\ \bibnamefont
  {Japaridze}},\ }\href {\doibase 10.1103/PhysRevLett.104.256804} {\bibfield
  {journal} {\bibinfo  {journal} {Phys. Rev. Lett.}\ }\textbf {\bibinfo
  {volume} {104}},\ \bibinfo {pages} {256804} (\bibinfo {year}
  {2010})}\BibitemShut {NoStop}%
\bibitem [{\citenamefont {Budich}\ \emph {et~al.}(2012)\citenamefont {Budich},
  \citenamefont {Dolcini}, \citenamefont {Recher},\ and\ \citenamefont
  {Trauzettel}}]{Budich-PRL-2012}%
  \BibitemOpen
  \bibfield  {author} {\bibinfo {author} {\bibfnamefont {J.~C.}\ \bibnamefont
  {Budich}}, \bibinfo {author} {\bibfnamefont {F.}~\bibnamefont {Dolcini}},
  \bibinfo {author} {\bibfnamefont {P.}~\bibnamefont {Recher}}, \ and\ \bibinfo
  {author} {\bibfnamefont {B.}~\bibnamefont {Trauzettel}},\ }\href {\doibase
  10.1103/PhysRevLett.108.086602} {\bibfield  {journal} {\bibinfo  {journal}
  {Phys. Rev. Lett.}\ }\textbf {\bibinfo {volume} {108}},\ \bibinfo {pages}
  {086602} (\bibinfo {year} {2012})}\BibitemShut {NoStop}%
\bibitem [{\citenamefont {Cr\'{e}pin}\ \emph {et~al.}(2012)\citenamefont
  {Cr\'{e}pin}, \citenamefont {Budich}, \citenamefont {Dolcini}, \citenamefont
  {Recher},\ and\ \citenamefont {Trauzettel}}]{Crepin-PRB-2012}%
  \BibitemOpen
  \bibfield  {author} {\bibinfo {author} {\bibfnamefont {F.}~\bibnamefont
  {Cr\'{e}pin}}, \bibinfo {author} {\bibfnamefont {J.~C.}\ \bibnamefont
  {Budich}}, \bibinfo {author} {\bibfnamefont {F.}~\bibnamefont {Dolcini}},
  \bibinfo {author} {\bibfnamefont {P.}~\bibnamefont {Recher}}, \ and\ \bibinfo
  {author} {\bibfnamefont {B.}~\bibnamefont {Trauzettel}},\ }\href {\doibase
  10.1103/PhysRevB.86.121106} {\bibfield  {journal} {\bibinfo  {journal} {Phys.
  Rev. B}\ }\textbf {\bibinfo {volume} {86}},\ \bibinfo {pages} {121106}
  (\bibinfo {year} {2012})}\BibitemShut {NoStop}%
\bibitem [{\citenamefont {Lezmy}\ \emph {et~al.}(2012)\citenamefont {Lezmy},
  \citenamefont {Oreg},\ and\ \citenamefont {Berkooz}}]{Lezmy-PRB-2012}%
  \BibitemOpen
  \bibfield  {author} {\bibinfo {author} {\bibfnamefont {N.}~\bibnamefont
  {Lezmy}}, \bibinfo {author} {\bibfnamefont {Y.}~\bibnamefont {Oreg}}, \ and\
  \bibinfo {author} {\bibfnamefont {M.}~\bibnamefont {Berkooz}},\ }\href
  {\doibase 10.1103/PhysRevB.85.235304} {\bibfield  {journal} {\bibinfo
  {journal} {Phys. Rev. B}\ }\textbf {\bibinfo {volume} {85}},\ \bibinfo
  {pages} {235304} (\bibinfo {year} {2012})}\BibitemShut {NoStop}%
\bibitem [{\citenamefont {Geissler}\ \emph {et~al.}(2014)\citenamefont
  {Geissler}, \citenamefont {Cr\'epin},\ and\ \citenamefont
  {Trauzettel}}]{Geissler-PRB-2014}%
  \BibitemOpen
  \bibfield  {author} {\bibinfo {author} {\bibfnamefont {F.}~\bibnamefont
  {Geissler}}, \bibinfo {author} {\bibfnamefont {F.}~\bibnamefont {Cr\'epin}},
  \ and\ \bibinfo {author} {\bibfnamefont {B.}~\bibnamefont {Trauzettel}},\
  }\href@noop {} {\bibfield  {journal} {\bibinfo  {journal} {Phys. Rev. B}\
  }\textbf {\bibinfo {volume} {89}},\ \bibinfo {pages} {235136} (\bibinfo
  {year} {2014})}\BibitemShut {NoStop}%
\bibitem [{\citenamefont {Kainaris}\ \emph {et~al.}(2014)\citenamefont
  {Kainaris}, \citenamefont {Gornyi}, \citenamefont {Carr},\ and\ \citenamefont
  {Mirlin}}]{Kainaris-PRB-2014}%
  \BibitemOpen
  \bibfield  {author} {\bibinfo {author} {\bibfnamefont {N.}~\bibnamefont
  {Kainaris}}, \bibinfo {author} {\bibfnamefont {I.~V.}\ \bibnamefont
  {Gornyi}}, \bibinfo {author} {\bibfnamefont {S.~T.}\ \bibnamefont {Carr}}, \
  and\ \bibinfo {author} {\bibfnamefont {A.~D.}\ \bibnamefont {Mirlin}},\
  }\href {\doibase 10.1103/PhysRevB.90.075118} {\bibfield  {journal} {\bibinfo
  {journal} {Phys. Rev. B}\ }\textbf {\bibinfo {volume} {90}},\ \bibinfo
  {pages} {075118} (\bibinfo {year} {2014})}\BibitemShut {NoStop}%
\bibitem [{\citenamefont {V\"ayrynen}\ \emph {et~al.}(2013)\citenamefont
  {V\"ayrynen}, \citenamefont {Goldstein},\ and\ \citenamefont
  {Glazman}}]{Vayrynen-PRL-2013}%
  \BibitemOpen
  \bibfield  {author} {\bibinfo {author} {\bibfnamefont {J.~I.}\ \bibnamefont
  {V\"ayrynen}}, \bibinfo {author} {\bibfnamefont {M.}~\bibnamefont
  {Goldstein}}, \ and\ \bibinfo {author} {\bibfnamefont {L.~I.}\ \bibnamefont
  {Glazman}},\ }\href {\doibase 10.1103/PhysRevLett.110.216402} {\bibfield
  {journal} {\bibinfo  {journal} {Phys. Rev. Lett.}\ }\textbf {\bibinfo
  {volume} {110}},\ \bibinfo {pages} {216402} (\bibinfo {year}
  {2013})}\BibitemShut {NoStop}%
\bibitem [{\citenamefont {Tanaka}\ \emph {et~al.}(2011)\citenamefont {Tanaka},
  \citenamefont {Furusaki},\ and\ \citenamefont {Matveev}}]{Tanaka-PRL-2011}%
  \BibitemOpen
  \bibfield  {author} {\bibinfo {author} {\bibfnamefont {Y.}~\bibnamefont
  {Tanaka}}, \bibinfo {author} {\bibfnamefont {A.}~\bibnamefont {Furusaki}}, \
  and\ \bibinfo {author} {\bibfnamefont {K.~A.}\ \bibnamefont {Matveev}},\
  }\href {\doibase 10.1103/PhysRevLett.106.236402} {\bibfield  {journal}
  {\bibinfo  {journal} {Phys. Rev. Lett.}\ }\textbf {\bibinfo {volume} {106}},\
  \bibinfo {pages} {236402} (\bibinfo {year} {2011})}\BibitemShut {NoStop}%
\bibitem [{\citenamefont {Lunde}\ and\ \citenamefont
  {Platero}(2012)}]{Lunde-PRB-2012}%
  \BibitemOpen
  \bibfield  {author} {\bibinfo {author} {\bibfnamefont {A.~M.}\ \bibnamefont
  {Lunde}}\ and\ \bibinfo {author} {\bibfnamefont {G.}~\bibnamefont
  {Platero}},\ }\href {\doibase 10.1103/PhysRevB.86.035112} {\bibfield
  {journal} {\bibinfo  {journal} {Phys. Rev. B}\ }\textbf {\bibinfo {volume}
  {86}},\ \bibinfo {pages} {035112} (\bibinfo {year} {2012})}\BibitemShut
  {NoStop}%
\bibitem [{\citenamefont {Eriksson}\ \emph {et~al.}(2012)\citenamefont
  {Eriksson}, \citenamefont {Str\"om}, \citenamefont {Sharma},\ and\
  \citenamefont {Johannesson}}]{Eriksson-PRB-2012}%
  \BibitemOpen
  \bibfield  {author} {\bibinfo {author} {\bibfnamefont {E.}~\bibnamefont
  {Eriksson}}, \bibinfo {author} {\bibfnamefont {A.}~\bibnamefont {Str\"om}},
  \bibinfo {author} {\bibfnamefont {G.}~\bibnamefont {Sharma}}, \ and\ \bibinfo
  {author} {\bibfnamefont {H.}~\bibnamefont {Johannesson}},\ }\href {\doibase
  10.1103/PhysRevB.86.161103} {\bibfield  {journal} {\bibinfo  {journal} {Phys.
  Rev. B}\ }\textbf {\bibinfo {volume} {86}},\ \bibinfo {pages} {161103}
  (\bibinfo {year} {2012})}\BibitemShut {NoStop}%
\bibitem [{\citenamefont {Eriksson}(2013)}]{Eriksson-PRB-2013}%
  \BibitemOpen
  \bibfield  {author} {\bibinfo {author} {\bibfnamefont {E.}~\bibnamefont
  {Eriksson}},\ }\href {\doibase 10.1103/PhysRevB.87.235414} {\bibfield
  {journal} {\bibinfo  {journal} {Phys. Rev. B}\ }\textbf {\bibinfo {volume}
  {87}},\ \bibinfo {pages} {235414} (\bibinfo {year} {2013})}\BibitemShut
  {NoStop}%
\bibitem [{\citenamefont {Probst}\ \emph {et~al.}(2014)\citenamefont {Probst},
  \citenamefont {Virtanen},\ and\ \citenamefont {Recher}}]{Probst-arxiv-2014}%
  \BibitemOpen
  \bibfield  {author} {\bibinfo {author} {\bibfnamefont {B.}~\bibnamefont
  {Probst}}, \bibinfo {author} {\bibfnamefont {P.}~\bibnamefont {Virtanen}}, \
  and\ \bibinfo {author} {\bibfnamefont {P.}~\bibnamefont {Recher}},\
  }\href@noop {} {\bibfield  {journal} {\bibinfo  {journal} {arXiv:1407.3253}\
  } (\bibinfo {year} {2014})}\BibitemShut {NoStop}%
\bibitem [{\citenamefont {Lunde}\ and\ \citenamefont
  {Platero}(2013)}]{Lunde-PRB-2013}%
  \BibitemOpen
  \bibfield  {author} {\bibinfo {author} {\bibfnamefont {A.~M.}\ \bibnamefont
  {Lunde}}\ and\ \bibinfo {author} {\bibfnamefont {G.}~\bibnamefont
  {Platero}},\ }\href {\doibase 10.1103/PhysRevB.88.115411} {\bibfield
  {journal} {\bibinfo  {journal} {Phys. Rev. B}\ }\textbf {\bibinfo {volume}
  {88}},\ \bibinfo {pages} {115411} (\bibinfo {year} {2013})}\BibitemShut
  {NoStop}%
\bibitem [{\citenamefont {Rothe}\ \emph {et~al.}(2010)\citenamefont {Rothe},
  \citenamefont {Reinthaler}, \citenamefont {Liu}, \citenamefont {Molenkamp},
  \citenamefont {Zhang},\ and\ \citenamefont {Hankiewicz}}]{Rothe-NJP-2010}%
  \BibitemOpen
  \bibfield  {author} {\bibinfo {author} {\bibfnamefont {D.~G.}\ \bibnamefont
  {Rothe}}, \bibinfo {author} {\bibfnamefont {R.~W.}\ \bibnamefont
  {Reinthaler}}, \bibinfo {author} {\bibfnamefont {C.-X.}\ \bibnamefont {Liu}},
  \bibinfo {author} {\bibfnamefont {L.~W.}\ \bibnamefont {Molenkamp}}, \bibinfo
  {author} {\bibfnamefont {S.-C.}\ \bibnamefont {Zhang}}, \ and\ \bibinfo
  {author} {\bibfnamefont {E.~M.}\ \bibnamefont {Hankiewicz}},\ }\href
  {\doibase doi:10.1088/1367-2630/12/6/065012} {\bibfield  {journal} {\bibinfo
  {journal} {New Journal of Physics}\ }\textbf {\bibinfo {volume} {12}},\
  \bibinfo {pages} {065012} (\bibinfo {year} {2010})}\BibitemShut {NoStop}%
\bibitem [{\citenamefont {Virtanen}\ and\ \citenamefont
  {Recher}(2012)}]{Virtanen-PRB-2012}%
  \BibitemOpen
  \bibfield  {author} {\bibinfo {author} {\bibfnamefont {P.}~\bibnamefont
  {Virtanen}}\ and\ \bibinfo {author} {\bibfnamefont {P.}~\bibnamefont
  {Recher}},\ }\href {\doibase 10.1103/PhysRevB.85.035310} {\bibfield
  {journal} {\bibinfo  {journal} {Phys. Rev. B}\ }\textbf {\bibinfo {volume}
  {85}},\ \bibinfo {pages} {035310} (\bibinfo {year} {2012})}\BibitemShut
  {NoStop}%
\bibitem [{\citenamefont {Rothe}\ and\ \citenamefont
  {Hankiewicz}(2014)}]{Rothe-PRB-2014}%
  \BibitemOpen
  \bibfield  {author} {\bibinfo {author} {\bibfnamefont {D.~G.}\ \bibnamefont
  {Rothe}}\ and\ \bibinfo {author} {\bibfnamefont {E.~M.}\ \bibnamefont
  {Hankiewicz}},\ }\href {\doibase 10.1103/PhysRevB.89.035418} {\bibfield
  {journal} {\bibinfo  {journal} {Phys. Rev. B}\ }\textbf {\bibinfo {volume}
  {89}},\ \bibinfo {pages} {035418} (\bibinfo {year} {2014})}\BibitemShut
  {NoStop}%
\bibitem [{\citenamefont {K\"{o}nig}\ \emph {et~al.}(2008)\citenamefont
  {K\"{o}nig}, \citenamefont {Buhmann}, \citenamefont {Molenkamp},
  \citenamefont {Hughes}, \citenamefont {Liu}, \citenamefont {Qi},\ and\
  \citenamefont {Zhang}}]{Konig-JPSJ-2008}%
  \BibitemOpen
  \bibfield  {author} {\bibinfo {author} {\bibfnamefont {M.}~\bibnamefont
  {K\"{o}nig}}, \bibinfo {author} {\bibfnamefont {H.}~\bibnamefont {Buhmann}},
  \bibinfo {author} {\bibfnamefont {L.~W.}\ \bibnamefont {Molenkamp}}, \bibinfo
  {author} {\bibfnamefont {T.~L.}\ \bibnamefont {Hughes}}, \bibinfo {author}
  {\bibfnamefont {C.-X.}\ \bibnamefont {Liu}}, \bibinfo {author} {\bibfnamefont
  {X.~L.}\ \bibnamefont {Qi}}, \ and\ \bibinfo {author} {\bibfnamefont {S.~C.}\
  \bibnamefont {Zhang}},\ }\href {\doibase 10.1143/JPSJ.77.031007} {\bibfield
  {journal} {\bibinfo  {journal} {J. Phys. Soc. Jpn.}\ }\textbf {\bibinfo
  {volume} {77}},\ \bibinfo {pages} {031007} (\bibinfo {year}
  {2008})}\BibitemShut {NoStop}%
\bibitem [{\citenamefont {Ostrovsky}\ \emph {et~al.}(2012)\citenamefont
  {Ostrovsky}, \citenamefont {Gornyi},\ and\ \citenamefont
  {Mirlin}}]{Ostrovsky-PRB-2012}%
  \BibitemOpen
  \bibfield  {author} {\bibinfo {author} {\bibfnamefont {P.~M.}\ \bibnamefont
  {Ostrovsky}}, \bibinfo {author} {\bibfnamefont {I.~V.}\ \bibnamefont
  {Gornyi}}, \ and\ \bibinfo {author} {\bibfnamefont {A.~D.}\ \bibnamefont
  {Mirlin}},\ }\href {\doibase 10.1103/PhysRevB.86.125323} {\bibfield
  {journal} {\bibinfo  {journal} {Phys. Rev. B}\ }\textbf {\bibinfo {volume}
  {86}},\ \bibinfo {pages} {125323} (\bibinfo {year} {2012})}\BibitemShut
  {NoStop}%
\bibitem [{\citenamefont {Orth}\ \emph {et~al.}(2013)\citenamefont {Orth},
  \citenamefont {Str\"ubi},\ and\ \citenamefont {Schmidt}}]{Orth-PRB-2013}%
  \BibitemOpen
  \bibfield  {author} {\bibinfo {author} {\bibfnamefont {C.~P.}\ \bibnamefont
  {Orth}}, \bibinfo {author} {\bibfnamefont {G.}~\bibnamefont {Str\"ubi}}, \
  and\ \bibinfo {author} {\bibfnamefont {T.~L.}\ \bibnamefont {Schmidt}},\
  }\href {\doibase 10.1103/PhysRevB.88.165315} {\bibfield  {journal} {\bibinfo
  {journal} {Phys. Rev. B}\ }\textbf {\bibinfo {volume} {88}},\ \bibinfo
  {pages} {165315} (\bibinfo {year} {2013})}\BibitemShut {NoStop}%
\bibitem [{\citenamefont {Orth}\ \emph {et~al.}(2015)\citenamefont {Orth},
  \citenamefont {Tiwari}, \citenamefont {Meng},\ and\ \citenamefont
  {Schmidt}}]{Orth-PRB-2015}%
  \BibitemOpen
  \bibfield  {author} {\bibinfo {author} {\bibfnamefont {C.~P.}\ \bibnamefont
  {Orth}}, \bibinfo {author} {\bibfnamefont {R.~P.}\ \bibnamefont {Tiwari}},
  \bibinfo {author} {\bibfnamefont {T.}~\bibnamefont {Meng}}, \ and\ \bibinfo
  {author} {\bibfnamefont {T.~L.}\ \bibnamefont {Schmidt}},\ }\href {\doibase
  10.1103/PhysRevB.91.081406} {\bibfield  {journal} {\bibinfo  {journal} {Phys.
  Rev. B}\ }\textbf {\bibinfo {volume} {91}},\ \bibinfo {pages} {081406}
  (\bibinfo {year} {2015})}\BibitemShut {NoStop}%
\bibitem [{\citenamefont {Rod}\ \emph {et~al.}(2015)\citenamefont {Rod},
  \citenamefont {Schmidt},\ and\ \citenamefont {Rachel}}]{Rod-PRB-2015}%
  \BibitemOpen
  \bibfield  {author} {\bibinfo {author} {\bibfnamefont {A.}~\bibnamefont
  {Rod}}, \bibinfo {author} {\bibfnamefont {T.~L.}\ \bibnamefont {Schmidt}}, \
  and\ \bibinfo {author} {\bibfnamefont {S.}~\bibnamefont {Rachel}},\ }\href
  {\doibase 10.1103/PhysRevB.91.245112} {\bibfield  {journal} {\bibinfo
  {journal} {Phys. Rev. B}\ }\textbf {\bibinfo {volume} {91}},\ \bibinfo
  {pages} {245112} (\bibinfo {year} {2015})}\BibitemShut {NoStop}%
\bibitem [{\citenamefont {Zhou}\ \emph {et~al.}(2008)\citenamefont {Zhou},
  \citenamefont {Lu}, \citenamefont {Chu}, \citenamefont {Shen},\ and\
  \citenamefont {Niu}}]{Zhou-PRL-2008}%
  \BibitemOpen
  \bibfield  {author} {\bibinfo {author} {\bibfnamefont {B.}~\bibnamefont
  {Zhou}}, \bibinfo {author} {\bibfnamefont {H.-Z.}\ \bibnamefont {Lu}},
  \bibinfo {author} {\bibfnamefont {R.-L.}\ \bibnamefont {Chu}}, \bibinfo
  {author} {\bibfnamefont {S.-Q.}\ \bibnamefont {Shen}}, \ and\ \bibinfo
  {author} {\bibfnamefont {Q.}~\bibnamefont {Niu}},\ }\href {\doibase
  10.1103/PhysRevLett.101.246807} {\bibfield  {journal} {\bibinfo  {journal}
  {Phys. Rev. Lett.}\ }\textbf {\bibinfo {volume} {101}},\ \bibinfo {pages}
  {246807} (\bibinfo {year} {2008})}\BibitemShut {NoStop}%
\bibitem [{\citenamefont {Liu}\ and\ \citenamefont
  {Zhang}(2013)}]{ZHANG-BOOK-2013}%
  \BibitemOpen
  \bibfield  {author} {\bibinfo {author} {\bibfnamefont {C.}~\bibnamefont
  {Liu}}\ and\ \bibinfo {author} {\bibfnamefont {S.-C.}\ \bibnamefont
  {Zhang}},\ }\href@noop {} {\emph {\bibinfo {title} {Models and materials for
  topological Insulators}}},\ edited by\ \bibinfo {editor} {\bibfnamefont
  {M.}~\bibnamefont {Franz}}\ and\ \bibinfo {editor} {\bibfnamefont
  {L.}~\bibnamefont {Molenkamp}},\ Vol.\ \bibinfo {volume} {Ch. 7 of
  Topological Insulators}\ (\bibinfo  {publisher} {Springer},\ \bibinfo {year}
  {(2013)})\BibitemShut {NoStop}%
\bibitem [{\citenamefont {Shen}(2011)}]{book_Shen}%
  \BibitemOpen
  \bibfield  {author} {\bibinfo {author} {\bibfnamefont {S.-Q.}\ \bibnamefont
  {Shen}},\ }\href@noop {} {\emph {\bibinfo {title} {Topological Insulators}}}\
  (\bibinfo  {publisher} {Springer-Verlag},\ \bibinfo {year}
  {2011})\BibitemShut {NoStop}%
\bibitem [{\citenamefont {Wada}\ \emph {et~al.}(2011)\citenamefont {Wada},
  \citenamefont {Murakami}, \citenamefont {Freimuth},\ and\ \citenamefont
  {Bihlmayer}}]{Wada-PRB-2011}%
  \BibitemOpen
  \bibfield  {author} {\bibinfo {author} {\bibfnamefont {M.}~\bibnamefont
  {Wada}}, \bibinfo {author} {\bibfnamefont {S.}~\bibnamefont {Murakami}},
  \bibinfo {author} {\bibfnamefont {F.}~\bibnamefont {Freimuth}}, \ and\
  \bibinfo {author} {\bibfnamefont {G.}~\bibnamefont {Bihlmayer}},\ }\href
  {\doibase 10.1103/PhysRevB.83.121310} {\bibfield  {journal} {\bibinfo
  {journal} {Phys. Rev. B}\ }\textbf {\bibinfo {volume} {83}},\ \bibinfo
  {pages} {121310} (\bibinfo {year} {2011})}\BibitemShut {NoStop}%
\bibitem [{\citenamefont {Bernevig}\ and\ \citenamefont {with
  T.~L.~Hughes}(2013)}]{book_BH}%
  \BibitemOpen
  \bibfield  {author} {\bibinfo {author} {\bibfnamefont {B.~A.}\ \bibnamefont
  {Bernevig}}\ and\ \bibinfo {author} {\bibnamefont {with T.~L.~Hughes}},\
  }\href@noop {} {\emph {\bibinfo {title} {Topological Insulators and
  Topological Superconductors}}}\ (\bibinfo  {publisher} {Princeton University
  Press},\ \bibinfo {year} {2013})\BibitemShut {NoStop}%
\bibitem [{\citenamefont {Michetti}\ \emph {et~al.}(2012)\citenamefont
  {Michetti}, \citenamefont {Penteado}, \citenamefont {Egues},\ and\
  \citenamefont {Recher}}]{Michetti-Semicond-Sci-Tech-2012}%
  \BibitemOpen
  \bibfield  {author} {\bibinfo {author} {\bibfnamefont {P.}~\bibnamefont
  {Michetti}}, \bibinfo {author} {\bibfnamefont {P.~H.}\ \bibnamefont
  {Penteado}}, \bibinfo {author} {\bibfnamefont {J.~C.}\ \bibnamefont {Egues}},
  \ and\ \bibinfo {author} {\bibfnamefont {P.}~\bibnamefont {Recher}},\ }\href
  {\doibase doi:10.1088/0268-1242/27/12/124007} {\bibfield  {journal} {\bibinfo
   {journal} {Semiconductor Science and Technology}\ }\textbf {\bibinfo
  {volume} {27}},\ \bibinfo {pages} {124007} (\bibinfo {year}
  {2012})}\BibitemShut {NoStop}%
\bibitem [{\citenamefont {Krueckl}\ and\ \citenamefont
  {Richter}(2011)}]{Krueckl-PRL-2011}%
  \BibitemOpen
  \bibfield  {author} {\bibinfo {author} {\bibfnamefont {V.}~\bibnamefont
  {Krueckl}}\ and\ \bibinfo {author} {\bibfnamefont {K.}~\bibnamefont
  {Richter}},\ }\href {\doibase 10.1103/PhysRevLett.107.086803} {\bibfield
  {journal} {\bibinfo  {journal} {Phys. Rev. Lett.}\ }\textbf {\bibinfo
  {volume} {107}},\ \bibinfo {pages} {086803} (\bibinfo {year}
  {2011})}\BibitemShut {NoStop}%
\bibitem [{\citenamefont {Shan}\ \emph {et~al.}(2010)\citenamefont {Shan},
  \citenamefont {Lu},\ and\ \citenamefont {Shen}}]{Shan-NJP-2010}%
  \BibitemOpen
  \bibfield  {author} {\bibinfo {author} {\bibfnamefont {W.-Y.}\ \bibnamefont
  {Shan}}, \bibinfo {author} {\bibfnamefont {H.-Z.}\ \bibnamefont {Lu}}, \ and\
  \bibinfo {author} {\bibfnamefont {S.-Q.}\ \bibnamefont {Shen}},\ }\href
  {http://stacks.iop.org/1367-2630/12/i=4/a=043048} {\bibfield  {journal}
  {\bibinfo  {journal} {New Journal of Physics}\ }\textbf {\bibinfo {volume}
  {12}},\ \bibinfo {pages} {043048} (\bibinfo {year} {2010})}\BibitemShut
  {NoStop}%
\bibitem [{\citenamefont {Takagaki}(2014)}]{Takagaki-PRB-2014}%
  \BibitemOpen
  \bibfield  {author} {\bibinfo {author} {\bibfnamefont {Y.}~\bibnamefont
  {Takagaki}},\ }\href {\doibase 10.1103/PhysRevB.90.165305} {\bibfield
  {journal} {\bibinfo  {journal} {Phys. Rev. B}\ }\textbf {\bibinfo {volume}
  {90}},\ \bibinfo {pages} {165305} (\bibinfo {year} {2014})}\BibitemShut
  {NoStop}%
\bibitem [{\citenamefont {Medhi}\ and\ \citenamefont
  {Shenoy}(2012)}]{Mehdi-JPCM-2012}%
  \BibitemOpen
  \bibfield  {author} {\bibinfo {author} {\bibfnamefont {A.}~\bibnamefont
  {Medhi}}\ and\ \bibinfo {author} {\bibfnamefont {V.~B.}\ \bibnamefont
  {Shenoy}},\ }\href {http://stacks.iop.org/0953-8984/24/i=35/a=355001}
  {\bibfield  {journal} {\bibinfo  {journal} {Journal of Physics: Condensed
  Matter}\ }\textbf {\bibinfo {volume} {24}},\ \bibinfo {pages} {355001}
  (\bibinfo {year} {2012})}\BibitemShut {NoStop}%
\bibitem [{\citenamefont {Linder}\ \emph {et~al.}(2009)\citenamefont {Linder},
  \citenamefont {Yokoyama},\ and\ \citenamefont {Sudb\o{}}}]{Linder-PRB-2009}%
  \BibitemOpen
  \bibfield  {author} {\bibinfo {author} {\bibfnamefont {J.}~\bibnamefont
  {Linder}}, \bibinfo {author} {\bibfnamefont {T.}~\bibnamefont {Yokoyama}}, \
  and\ \bibinfo {author} {\bibfnamefont {A.}~\bibnamefont {Sudb\o{}}},\ }\href
  {\doibase 10.1103/PhysRevB.80.205401} {\bibfield  {journal} {\bibinfo
  {journal} {Phys. Rev. B}\ }\textbf {\bibinfo {volume} {80}},\ \bibinfo
  {pages} {205401} (\bibinfo {year} {2009})}\BibitemShut {NoStop}%
\bibitem [{\citenamefont {Liu}\ \emph {et~al.}(2010)\citenamefont {Liu},
  \citenamefont {Zhang}, \citenamefont {Yan}, \citenamefont {Qi}, \citenamefont
  {Frauenheim}, \citenamefont {Dai}, \citenamefont {Fang},\ and\ \citenamefont
  {Zhang}}]{Liu-PRB-2010-3DTI}%
  \BibitemOpen
  \bibfield  {author} {\bibinfo {author} {\bibfnamefont {C.-X.}\ \bibnamefont
  {Liu}}, \bibinfo {author} {\bibfnamefont {H.~J.}\ \bibnamefont {Zhang}},
  \bibinfo {author} {\bibfnamefont {B.}~\bibnamefont {Yan}}, \bibinfo {author}
  {\bibfnamefont {X.-L.}\ \bibnamefont {Qi}}, \bibinfo {author} {\bibfnamefont
  {T.}~\bibnamefont {Frauenheim}}, \bibinfo {author} {\bibfnamefont
  {X.}~\bibnamefont {Dai}}, \bibinfo {author} {\bibfnamefont {Z.}~\bibnamefont
  {Fang}}, \ and\ \bibinfo {author} {\bibfnamefont {S.-C.}\ \bibnamefont
  {Zhang}},\ }\href {\doibase 10.1103/PhysRevB.81.041307} {\bibfield  {journal}
  {\bibinfo  {journal} {Phys. Rev. B}\ }\textbf {\bibinfo {volume} {81}},\
  \bibinfo {pages} {041307} (\bibinfo {year} {2010})}\BibitemShut {NoStop}%
\bibitem [{\citenamefont {Lu}\ \emph {et~al.}(2010)\citenamefont {Lu},
  \citenamefont {Shan}, \citenamefont {Yao}, \citenamefont {Niu},\ and\
  \citenamefont {Shen}}]{Lu-PRB-2010}%
  \BibitemOpen
  \bibfield  {author} {\bibinfo {author} {\bibfnamefont {H.-Z.}\ \bibnamefont
  {Lu}}, \bibinfo {author} {\bibfnamefont {W.-Y.}\ \bibnamefont {Shan}},
  \bibinfo {author} {\bibfnamefont {W.}~\bibnamefont {Yao}}, \bibinfo {author}
  {\bibfnamefont {Q.}~\bibnamefont {Niu}}, \ and\ \bibinfo {author}
  {\bibfnamefont {S.-Q.}\ \bibnamefont {Shen}},\ }\href {\doibase
  10.1103/PhysRevB.81.115407} {\bibfield  {journal} {\bibinfo  {journal} {Phys.
  Rev. B}\ }\textbf {\bibinfo {volume} {81}},\ \bibinfo {pages} {115407}
  (\bibinfo {year} {2010})}\BibitemShut {NoStop}%
\bibitem [{\citenamefont {Sakurai}(1993)}]{Sakurai-modern-BOOK}%
  \BibitemOpen
  \bibfield  {author} {\bibinfo {author} {\bibfnamefont {J.~J.}\ \bibnamefont
  {Sakurai}},\ }\href@noop {} {\emph {\bibinfo {title} {Modern Quantum
  Mechanics}}}\ (\bibinfo  {publisher} {Addison Wesley},\ \bibinfo {year}
  {1993})\BibitemShut {NoStop}%
\bibitem [{\citenamefont {Murakami}\ \emph {et~al.}(2007)\citenamefont
  {Murakami}, \citenamefont {Iso}, \citenamefont {Avishai}, \citenamefont
  {Onoda},\ and\ \citenamefont {Nagaosa}}]{Murakami-PRB-2007}%
  \BibitemOpen
  \bibfield  {author} {\bibinfo {author} {\bibfnamefont {S.}~\bibnamefont
  {Murakami}}, \bibinfo {author} {\bibfnamefont {S.}~\bibnamefont {Iso}},
  \bibinfo {author} {\bibfnamefont {Y.}~\bibnamefont {Avishai}}, \bibinfo
  {author} {\bibfnamefont {M.}~\bibnamefont {Onoda}}, \ and\ \bibinfo {author}
  {\bibfnamefont {N.}~\bibnamefont {Nagaosa}},\ }\href@noop {} {\bibfield
  {journal} {\bibinfo  {journal} {Phys. Rev. B}\ }\textbf {\bibinfo {volume}
  {76}},\ \bibinfo {pages} {205304} (\bibinfo {year} {2007})}\BibitemShut
  {NoStop}%
\bibitem [{\citenamefont {Berry}\ and\ \citenamefont
  {Mondragon}(1987)}]{Berry-PRSLA-1987}%
  \BibitemOpen
  \bibfield  {author} {\bibinfo {author} {\bibfnamefont {M.}~\bibnamefont
  {Berry}}\ and\ \bibinfo {author} {\bibfnamefont {R.~J.}\ \bibnamefont
  {Mondragon}},\ }\href@noop {} {\bibfield  {journal} {\bibinfo  {journal}
  {Proc. R. Soc. Lond. A}\ }\textbf {\bibinfo {volume} {412}},\ \bibinfo
  {pages} {53} (\bibinfo {year} {1987})}\BibitemShut {NoStop}%
\bibitem [{\citenamefont {F\"{u}rst}\ \emph {et~al.}(2009)\citenamefont
  {F\"{u}rst}, \citenamefont {Pedersen}, \citenamefont {Flindt}, \citenamefont
  {Mortensen}, \citenamefont {Brandbyge}, \citenamefont {Pedersen},\ and\
  \citenamefont {Jauho}}]{Flindt-NJP-2009}%
  \BibitemOpen
  \bibfield  {author} {\bibinfo {author} {\bibfnamefont {J.~A.}\ \bibnamefont
  {F\"{u}rst}}, \bibinfo {author} {\bibfnamefont {J.~G.}\ \bibnamefont
  {Pedersen}}, \bibinfo {author} {\bibfnamefont {C.}~\bibnamefont {Flindt}},
  \bibinfo {author} {\bibfnamefont {N.~A.}\ \bibnamefont {Mortensen}}, \bibinfo
  {author} {\bibfnamefont {M.}~\bibnamefont {Brandbyge}}, \bibinfo {author}
  {\bibfnamefont {T.~G.}\ \bibnamefont {Pedersen}}, \ and\ \bibinfo {author}
  {\bibfnamefont {A.-P.}\ \bibnamefont {Jauho}},\ }\href {\doibase
  doi:10.1088/1367-2630/11/9/095020} {\bibfield  {journal} {\bibinfo  {journal}
  {New Journal of Physics}\ }\textbf {\bibinfo {volume} {11}},\ \bibinfo
  {pages} {095020} (\bibinfo {year} {2009})}\BibitemShut {NoStop}%
\bibitem [{\citenamefont {Michetti}\ and\ \citenamefont
  {Recher}(2011)}]{Michetti-PRB-2011}%
  \BibitemOpen
  \bibfield  {author} {\bibinfo {author} {\bibfnamefont {P.}~\bibnamefont
  {Michetti}}\ and\ \bibinfo {author} {\bibfnamefont {P.}~\bibnamefont
  {Recher}},\ }\href {\doibase 10.1103/PhysRevB.83.125420} {\bibfield
  {journal} {\bibinfo  {journal} {Phys. Rev. B}\ }\textbf {\bibinfo {volume}
  {83}},\ \bibinfo {pages} {125420} (\bibinfo {year} {2011})}\BibitemShut
  {NoStop}%
\bibitem [{\citenamefont {Baum}\ \emph {et~al.}(2015)\citenamefont {Baum},
  \citenamefont {Posske}, \citenamefont {Fulga}, \citenamefont {Trauzettel},\
  and\ \citenamefont {Stern}}]{Baum-PRL-2015}%
  \BibitemOpen
  \bibfield  {author} {\bibinfo {author} {\bibfnamefont {Y.}~\bibnamefont
  {Baum}}, \bibinfo {author} {\bibfnamefont {T.}~\bibnamefont {Posske}},
  \bibinfo {author} {\bibfnamefont {I.~C.}\ \bibnamefont {Fulga}}, \bibinfo
  {author} {\bibfnamefont {B.}~\bibnamefont {Trauzettel}}, \ and\ \bibinfo
  {author} {\bibfnamefont {A.}~\bibnamefont {Stern}},\ }\href {\doibase
  10.1103/PhysRevLett.114.136801} {\bibfield  {journal} {\bibinfo  {journal}
  {Phys. Rev. Lett.}\ }\textbf {\bibinfo {volume} {114}},\ \bibinfo {pages}
  {136801} (\bibinfo {year} {2015})}\BibitemShut {NoStop}%
\bibitem [{\citenamefont {Imura}\ \emph {et~al.}(2010)\citenamefont {Imura},
  \citenamefont {Yamakage}, \citenamefont {Mao}, \citenamefont {Hotta},\ and\
  \citenamefont {Kuramoto}}]{Imura10}%
  \BibitemOpen
  \bibfield  {author} {\bibinfo {author} {\bibfnamefont {K.-I.}\ \bibnamefont
  {Imura}}, \bibinfo {author} {\bibfnamefont {A.}~\bibnamefont {Yamakage}},
  \bibinfo {author} {\bibfnamefont {S.}~\bibnamefont {Mao}}, \bibinfo {author}
  {\bibfnamefont {A.}~\bibnamefont {Hotta}}, \ and\ \bibinfo {author}
  {\bibfnamefont {Y.}~\bibnamefont {Kuramoto}},\ }\href {\doibase
  10.1103/PhysRevB.82.085118} {\bibfield  {journal} {\bibinfo  {journal} {Phys.
  Rev. B}\ }\textbf {\bibinfo {volume} {82}},\ \bibinfo {pages} {085118}
  (\bibinfo {year} {2010})}\BibitemShut {NoStop}%
\bibitem [{\citenamefont {Nielsen}\ and\ \citenamefont
  {Ninomiya}(1981{\natexlab{a}})}]{Nielsen81-1}%
  \BibitemOpen
  \bibfield  {author} {\bibinfo {author} {\bibfnamefont {H.}~\bibnamefont
  {Nielsen}}\ and\ \bibinfo {author} {\bibfnamefont {M.}~\bibnamefont
  {Ninomiya}},\ }\href {\doibase
  http://dx.doi.org/10.1016/0550-3213(81)90361-8} {\bibfield  {journal}
  {\bibinfo  {journal} {Nucl. Phys. B}\ }\textbf {\bibinfo {volume} {185}},\
  \bibinfo {pages} {20 } (\bibinfo {year} {1981}{\natexlab{a}})}\BibitemShut
  {NoStop}%
\bibitem [{\citenamefont {Nielsen}\ and\ \citenamefont
  {Ninomiya}(1981{\natexlab{b}})}]{Nielsen81-2}%
  \BibitemOpen
  \bibfield  {author} {\bibinfo {author} {\bibfnamefont {H.}~\bibnamefont
  {Nielsen}}\ and\ \bibinfo {author} {\bibfnamefont {M.}~\bibnamefont
  {Ninomiya}},\ }\href {\doibase
  http://dx.doi.org/10.1016/0550-3213(81)90524-1} {\bibfield  {journal}
  {\bibinfo  {journal} {Nucl. Phys. B}\ }\textbf {\bibinfo {volume} {193}},\
  \bibinfo {pages} {173 } (\bibinfo {year} {1981}{\natexlab{b}})}\BibitemShut
  {NoStop}%
\bibitem [{\citenamefont {Winkler}(2003)}]{Winkler-BOOK-2003}%
  \BibitemOpen
  \bibfield  {author} {\bibinfo {author} {\bibfnamefont {R.}~\bibnamefont
  {Winkler}},\ }\href@noop {} {\emph {\bibinfo {title} {Spin-orbit Coupling
  Effects in Two-Dimensional Electron and Hole Systems (Springer Tracts in
  Modern Physics)}}}\ (\bibinfo  {publisher} {Springer},\ \bibinfo {year}
  {2003})\BibitemShut {NoStop}%
\end{thebibliography}

%

\end{document}